\documentclass[aps,prb,notitlepage,showkeys,showpacs,superscriptaddress,nofootinbib,longbibliography,preprintnumbers]{revtex4-2}

\usepackage[utf8]{inputenc}

\usepackage{here}
\usepackage{ulem}

\usepackage{url}
\usepackage{amscd}

\usepackage[top=30truemm,bottom=30truemm,left=25truemm,right=25truemm]{geometry}

\usepackage{spectralsequences}
\usepackage{tikz}

\usepackage{xcolor}

\newcommand{\zmod}[1]{\mathbb{Z}/#1\mathbb{Z}}

\newcommand{\tr}[1]{\mathrm{tr}\left(#1\right)}

\newcommand{\cohoZ}[2]{\mathrm{H}^{#1}(#2;\mathbb{Z})}

\usepackage[all]{xy}
\usetikzlibrary{cd}

\usepackage[colorlinks=true,linktoc=page,citecolor=red,linkcolor=blue]{hyperref}	
\usepackage[whole]{bxcjkjatype} 

\usepackage{slashed}

\usepackage{amsthm}
\usepackage{thmtools}
\usepackage{blindtext}
\usepackage{braket}

\declaretheoremstyle[
       shaded={bgcolor=\color{rgb}{0.9,0.9,0.9}}  
]{theorem}

\declaretheoremstyle[
       shaded={bgcolor=\color{rgb}{0.9,0.9,0.9}}
]{question}

\declaretheoremstyle[
       shaded={bgcolor=\color{rgb}{0.9,0.9,0.9}}  
]{remark}

\declaretheoremstyle[
       shaded={bgcolor=\color{rgb}{0.9,0.9,0.9}}  
]{proposition}

\declaretheoremstyle[
       shaded={bgcolor=\color{rgb}{0.9,0.9,0.9}}  
]{definition}

\declaretheoremstyle[
       shaded={bgcolor=\color{rgb}{0.9,0.9,0.9}}  
]{assumption}

\declaretheoremstyle[
       shaded={bgcolor=\color{rgb}{0.9,0.9,0.9}}  
]{conjecture}

\declaretheoremstyle[
       shaded={bgcolor=\color{rgb}{0.9,0.9,0.9}}  
]{corrorary}

\declaretheoremstyle[
       shaded={bgcolor=\color{rgb}{0.9,0.9,0.9}}  
]{axiom}

\declaretheoremstyle[
       shaded={bgcolor=\color{rgb}{0.9,0.9,0.9}}  
]{lemma}

\def\>{{\geq }}
\def\<{{\leq }}

\usepackage{mathrsfs}
\usepackage{amsmath,amssymb}

\usepackage{esvect}

\begin{document}

\title{Higher Berry Connection for Matrix Product States}

\author{Shuhei Ohyama}
\email{shuhei.ohyama@riken.jp}
\affiliation{
RIKEN Center for Emergent Matter Science, Wako, Saitama, 351-0198, Japan}
\author{Shinsei Ryu}
\email{shinseir@princeton.edu}
\affiliation{Department of Physics, Princeton University, Princeton, New Jersey 08544, USA}

\date{\today} 

\begin{abstract}
In one spatial dimension, families of short-range entangled many-body quantum states,
parameterized over some parameter space, 
can be topologically distinguished and classified by 
topological invariants built from the higher Berry phase
-- a many-body generalization of the Berry phase.
Previous works identified 
the underlying mathematical structure (the gerbe structure) 
and introduced a multi-wavefunction overlap,
a generalization of the inner product in quantum mechanics,
which allows for the extraction of the higher Berry phase 
and topological invariants.
In this paper, 
building on these works, 
we introduce a connection, the higher Berry 
connection, for a family of parameterized 
Matrix Product States (MPS) over 
a parameter space.
We demonstrate the use of our formula for simple non-trivial models.
\end{abstract}

\maketitle

\setcounter{tocdepth}{3}
\tableofcontents

\section{Introduction}

The higher Berry phase is a many-body generalization of
the regular Berry phase, and has been actively studied recently.
See, e.g., Refs.\ \cite{
Kitaev13,
KS20-1,KS20-2,
Hsin_2020,
Cordova_2020a, Cordova_2020b,
shiozaki21,
Choi_2022,
Wen_2023,
OTS23,
beaudry2023homotopical,
qi2023charting,
shiozaki2023higher,
artymowicz2023quantization,
yao2024modulating}.
Notably, the Berry phase and Berry curvature 
have played crucial roles in topological phases of matter, 
in particular in understanding the topological properties of Bloch electrons.  
For example, the quantized Hall conductance is the canonical example
\cite{TKNN82, Kohomoto85}. 
Beyond characterizing the topological properties of ground states, the Berry phase also plays an important role in characterizing adiabatic processes, such as the quantized charge transport in the Thouless pump
\cite{Thouless83}.
These examples can fully be understood in terms of non-interacting fermions,
and in terms of the Berry phase/curvature of finite-dimensional quantum mechanical systems in the parameter spaces, i.e., Bloch Hamiltonians. In recent years, it has been recognized that the Berry phase/curvature in quantum many-body systems has different and more interesting structures. For example, the Berry connection for many-body quantum states in $d = 1$ spatial dimension is expected to be a higher-form (two-form), as opposed to the one-form Berry connection in regular cases. 
The formulas for the higher Berry phase and curvature have been proposed and utilized to study specific examples  
\cite{KS20-1,OR23,shiozaki2023higher}.
In Ref.\ \cite{OR23},
using matrix product states (MPS), we introduced a generalized inner product, what we call the triple inner product or the multi-wavefunction overlap. The multi-wavefunction overlap assigns a complex number for more than two states.
Just like the regular Berry phase can be extracted from the inner product of two states, the phase part of the multi-wavefunction overlap 
is relevant to the higher Berry phase. 
For the case of $d=1$ quantum many-body states, we need an inner product for three states.

The previous works above studied the higher Berry phase and curvature, without discussing a connection. This paper will fill this gap and construct a connection for the higher Berry phase.
The summary of our results and the organization of the paper are as follows. 
\begin{itemize}
\item
In Sec.\ \ref{Preliminaries},
we start our discussion by going through the necessary preliminaries, 
such as the infinite MPS formalism,
and the gerbe structure that underlies 
the higher Berry phase for parameterized MPS.  
In particular, the consistency conditions satisfied by 
1-form and 2-form connections of a gerbe, 
and how they are transformed under gauge transformations are reviewed.

\item
Upshots of our construction are the formula for the higher Berry connection summarized in
Sec.\ \ref{summary},
and further explained in  
Sec.\ \ref{Construction of an MPS gerbe connection}.
These include
the formulas for 1-form and 2-form connections, 
\eqref{sum: 1-form conn} and
\eqref{sum: 2-form conn a}-\eqref{sum: 2-form conn b},
respectively.
For more details, 
see also Eqs.\ \eqref{1-form conn}, 
\eqref{eq:def_of_b} 
and \eqref{2-form conn}. 
These formulas, in principle, can directly be applicable to 
any parameter-dependent MPS.
In formulating higher-Berry connections, 
it is important to discuss parameterized MPS whose rank 
is not globally constant over the parameter space,
since only for such cases the higher-Berry phase 
(more precisely, the higher Berry curvature and the integral topological invariant) can be non-trivial
\footnote{
While parameterized constant-rank MPS cannot 
be non-trivial in the free part of the higher Berry class, 
i.e., 
cannot realize non-trivial integral topological invariants,
they can still be non-trivial in
the torsion part of the higher Berry class. 
See Ref.\ \cite{OTS23} for nontrivial examples.}. 
The case of non-constant rank MPS is discussed in Sec.\ \ref{General MPS gerbe}.


\item
In Sec.\ \ref{Examples}, we discuss various examples. We calculate the topological invariant, the integrated 3-form higher Berry curvature, and confirm that it is properly quantized for the examples. This provides further support for our formulas.

\item 
Finally in Sec.\ \ref{Discussion},
we close by discussing open problems.
In particular, we discuss a link between
the multi-wavefunction overlap (triple inner product) introduced in \cite{OTS23} and our formulas for connections.
Namely, we follow closely 
the spirit of the work of Berry \cite{Berry1984} that relates the regular 
quantum mechanical inner product (wavefunction overlap)
and a connection of a fibre bundle (complex line bundle).
For those who wonder where our formulas in Sec.\ \ref{Construction of an MPS gerbe connection}
come from,
this discussion may serve as a motivation.

\item 
In two Appendices
\ref{Wen's model parameterized over X=S3}
and
\ref{A model parametrized S1xS2},
we provide the details of our examples,
including their MPS representations.
In addition, we calculate their topological invariants,
the Dixmier-Douady class,
without using (the integral of) the higher Berry curvature.
We confirm the agreement 
with the calculations
in Sec.\ \ref{Discussion}.
\end{itemize}
%
%
%
%

%

\section{Preliminaries}
\label{Preliminaries}

\subsection{Infinite MPS}\label{sec:infinite}

Throughout this paper, we use the infinite MPS representation of 
short-range entangled ground states
(invertible states) of (1+1)d many-body systems. 
We begin by recalling some of its preliminary properties
\cite{Cirac_2021}.
Let us first consider a normal MPS represented by 
$\mathsf{D} \times \mathsf{D}$ normal matrices $\left\{A^s\right\}_{s=1,\cdots, \mathsf{D}}$.
Here, $s=1,\cdots, \mathsf{D}$ represents 
the "physical" index, e.g., spin degrees of freedom in the local Hilbert space, and $\mathsf{D}$ is the dimension of the local Hilbert space. While not explicitly displayed, 
$A^s$ also carries "virtual" indices.  
An MPS representation has a gauge redundancy
under
\begin{align}
   A^s \to e^{i\theta} g A^s 
   g^{\dagger}.
\end{align}
Remark that, when we take the MPS representation $\{A^s\}$, we implicitly fix the gauge of MPS.

From an MPS $\{A^s\}$,
we define its transfer matrix as
$T_A=\sum_s A^{s *} \otimes A^s$. 
For normal MPS, its transfer matrix has unique right and left eigenvectors 
$\Lambda^R_A$ and $\Lambda^L_A$ 
with eigenvalue $\lambda$. 
I.e., 
$
T_{A}\cdot \Lambda^R_A 
= \sum_s A^s \Lambda^R_A A^{s\dagger}
=
\lambda \Lambda^R_A
$
and
$\Lambda^L_A\cdot T_{A} = \sum_s 
A^{s \dagger} \Lambda^L_A A^{s}
=
\lambda \Lambda^L_A$.
Throughout the paper, unless stated otherwise, we will work with 
the right canonical form in which
$\lambda=1$ and 
$\Lambda^R_A=1_\mathsf{D}$. 
We represent the transfer matrix and 
its eigenvectors (fixed points)
diagrammatically 
as
\begin{gather}
T_A=
  \begin{tikzpicture}[line cap=round,line join=round,x=1.0cm,y=1.0cm, scale=0.25, baseline={([yshift=-.4ex]current bounding box.center)}, thick, shift={(0,0)}, scale=0.7]
  \def\tate{2} 
  \def\yoko{1} 
  \def\sen{7} 
  \def\gaisen{2} 
  \draw (0,\tate) -- (\yoko+\yoko,\tate);
  \draw (0,-\tate) -- (\yoko + \yoko,-\tate);
  \draw (\yoko,-\tate) -- (\yoko,\tate);
\end{tikzpicture}\, ,
\quad 
\Lambda^R_A=
  \begin{tikzpicture}[line cap=round,line join=round,x=1.0cm,y=1.0cm, scale=0.25, baseline={([yshift=-.4ex]current bounding box.center)}, thick, shift={(0,0)}, scale=0.7]
  \def\tate{2} 
  \def\yoko{1} 
  \def\sen{7} 
  \def\gaisen{2} 
  \draw (\yoko,\tate) arc (90:-90:\tate);
\end{tikzpicture}\, ,
\quad 
\Lambda^L_A=
  \begin{tikzpicture}[line cap=round,line join=round,x=1.0cm,y=1.0cm, scale=0.25, baseline={([yshift=-.4ex]current bounding box.center)}, thick, shift={(0,0)}, scale=0.7]
  \def\tate{2} 
  \def\yoko{1} 
  \def\sen{7} 
  \def\gaisen{2} 
  \draw (3,\tate) arc (90:270:\tate);
\end{tikzpicture}\, . 
\end{gather}
Note that in our notation, we do not
draw boxes explicitly for MPS matrices and other tensors. MPS matrices are represented 
simply by trivalent vertices. 
Similarly, the eigen equations are represented as 
\begin{gather}
\label{eig eq}
  \begin{tikzpicture}[line cap=round,line join=round,x=1.0cm,y=1.0cm, scale=0.25, baseline={([yshift=-.4ex]current bounding box.center)}, thick, shift={(0,0)}, scale=0.7]
  \def\tate{2} 
  \def\yoko{1} 
  \def\sen{7} 
  \def\gaisen{2} 
  \draw (0,\tate) -- (\yoko,\tate);
  \draw (0,-\tate) -- (\yoko,-\tate);
  \draw (\yoko,\tate) arc (90:-90:\tate);
  \draw (\yoko,-\tate) -- (\yoko,\tate);
    \node[anchor=west] at (3,0) 
  {$=$};
  \draw (6,\tate) arc (90:-90:\tate);
\end{tikzpicture}\, ,
\quad 
  \begin{tikzpicture}[line cap=round,line join=round,x=1.0cm,y=1.0cm, scale=0.25, baseline={([yshift=-.4ex]current bounding box.center)}, thick, shift={(0,0)}, scale=0.7]
  \def\tate{2} 
  \def\yoko{1} 
  \def\sen{7} 
  \def\gaisen{2} 
  \draw (0,\tate) -- (\yoko,\tate);
  \draw (0,-\tate) -- (\yoko,-\tate);
  \draw (0,\tate) arc (90:270:\tate);
  \draw (0,-\tate) -- (0,\tate);
    \node[anchor=west] at (1,0) 
  {$=$};
  \draw (6,\tate) arc (90:270:\tate);
\end{tikzpicture}\, .
\end{gather}

An infinite MPS is an MPS with $\Lambda^L_A$ and $1_\mathsf{D}$ imposed as left and right boundary conditions. 
This allows us to calculate the expectation values of 
observables in the thermodynamic limit. For example, 
the expectation value of the identity operator, i.e., 
the normalization, is given by
\begin{align}
\label{norm}
  \begin{tikzpicture}[line cap=round,line join=round,x=1.0cm,y=1.0cm, scale=0.25, baseline={([yshift=-.4ex]current bounding box.center)}, thick, shift={(0,0)}, scale=0.7]
  \def\tate{2} 
  \def\yoko{8} 
  \def\sen{7} 
  \def\gaisen{2} 
  \draw (0,\tate) -- (\yoko,\tate);
  \draw (0,-\tate) -- (\yoko,-\tate);
  \draw (0,\tate) arc (90:270:\tate);
  \draw (\yoko,\tate) arc (90:-90:\tate);
  \node[anchor=east] at (-\tate,0) 
  {$\Lambda_{A}^{L}$};
  \draw (0,-\tate) -- (0,\tate);
  \draw (1,-\tate) -- (1,\tate);
  \node[anchor=east] at (6,0) 
  {$\cdots$};
  \draw (7,-\tate) -- (7,\tate);
  \draw (8,-\tate) -- (8,\tate);
    \node[anchor=east] at (13.5,0) 
  {$1_{\mathsf{D}}$};
\end{tikzpicture}
=
\begin{tikzpicture}[line cap=round,line join=round,x=1.0cm,y=1.0cm, scale=0.3, baseline={([yshift=-.4ex]current bounding box.center)}, thick, shift={(0,0)}, scale=0.7]
  \def\tate{2} 
  \def\yoko{10} 
  \def\sen{7} 
  \def\gaisen{2} 
  \draw (0,\tate) arc (90:270:\tate);
  \draw (0,\tate) arc (90:-90:\tate);
\end{tikzpicture}
=
\mathrm{tr}\, \Lambda^L_{A}.
\end{align}
In the following, we choose the normalization
$\mathrm{tr}\, \Lambda^L_{A}=1$.
%


We are interested in 
parameterized families of 
(1+1)d many-body states in the infinite MPS representation.
To set the stage, we fix an open covering 
$\{U_{\alpha}\}$ of $X$ 
and consider 
a parameterized family of MPSs
$\{A^s_{\alpha}(x)\}$
on each $U_{\alpha}$.
In general, the rank of the MPS is different for each
patch. We denote the rank of MPS matrices on $U_\alpha$ as $\mathsf{D}_{\alpha}$.
On the intersection 
$U_{\alpha\beta}=U_{\alpha}\cap U_{\beta}$, 
we have two MPS  
$\{A^s_{\alpha} \}$
and 
$\{A^s_{\beta} \}$
representing the same physical state.
By the fundamental theorem, 
two MPSs 
$\{A^s_{\alpha}(x)\}$
and
$\{A^s_{\beta}(x)\}$
are related 
by the MPS gauge transformation, 
\begin{align}
\label{MPS gauge}
  A^s_{\alpha}(x) = 
  e^{i \theta_{\alpha\beta}(x)}
  g^{\ }_{\alpha\beta}(x) 
  A^s_{\beta}(x) 
  g_{\alpha\beta}^{\dag}(x),
\end{align}
where  
the $U(1)$ phase 
$e^{i \theta_{\alpha\beta}(x)}$
is related to the usual phase ambiguity of 
quantum mechanical wave functions, 
while 
$g^{\ }_{\alpha\beta}(x)$
is 
related to the higher Berry phase. 

Here, we remark that, from the fundamental theorem, $g_{\alpha\beta}$ is determined only up to $U(1)$ phase ambiguity. There are two ways to handle $g_{\alpha\beta}$. One is to treat $g_{\alpha\beta}$ abstractly as an element of the projective unitary group. The other is to treat 
$g_{\alpha\beta}$ as a unitary matrix and 
take into account its behavior under 
redefinition of $U(1)$ phases. 
The former is mathematically analogous to considering a quantum state as a ray in Hilbert space, while the latter corresponds to treating a quantum state as an element of Hilbert space and the phase indeterminacy of the state as a gauge redundancy. 
Here, we will follow the latter approach 
and treat $g_{\alpha\beta}$ as a unitary matrix. 
In this case, it is necessary to 
keep track of the behavior of 
any object defined 
in terms of $g_{\alpha\beta}$ 
under phase redefinitions. 
Indeed, we will define a gerbe connection using $g_{\alpha\beta}$ in Secs.\ \ref{Construction of an MPS gerbe connection} and \ref{sec:MPSgerbe}, and we will discuss 
in Sec.\ \ref{Gauge redundancy} that the change in the connection under the phase redefinition of $g_{\alpha\beta}$ can be absorbed as a gauge transformation of the gerbe connection.

The mixed transfer matrix,
defined by
\begin{eqnarray}
    T_{\alpha\beta}:= \sum_s A^{s*}_{\beta}\otimes A^s_{\alpha},
\end{eqnarray}
plays a central role 
in the formulation of the higher Berry phase in Ref.\ \cite{OR23}.
We can take
the right and left eigenstates
of $T_{\alpha\beta}$ with eigenvalue $1$
as
\begin{eqnarray}
\label{normalization mixed fixed pt}
\Lambda_{\alpha\beta}^{R}:=\Lambda_{\alpha}^{R}g_{\alpha\beta}=g_{\alpha\beta}\Lambda_{\beta}^{R},
\quad
\Lambda_{\alpha\beta}^{L}:=g_{\beta\alpha}\Lambda_{\alpha}^{L}=\Lambda_{\beta}^{L}g_{\beta\alpha},
\end{eqnarray}
so that 
$\mathrm{tr}\, (\Lambda^R_{\alpha\beta} 
\Lambda^L_{\alpha\beta})
=1$.
\footnote{Here, the phase ambiguity of $g_{\alpha\beta}$ is reinterpreted as that of the eigenvector.}
We note that,
under 
the MPS gauge transformations
on $U_{\alpha}$ and $U_{\beta}$,
\begin{align}
   A^s_{\alpha} \to e^{i\theta_{\alpha}} g_{\alpha} A^i_{\alpha} 
   g^{\dagger}_{\alpha},
   \quad
   A^s_{\beta} \to e^{i\theta_{\beta}} g_{\beta} A^s_{\beta} 
   g^{\dagger}_{\beta},
\end{align}
the left and right fixed points of 
$T_{\beta\alpha}$ 
transform as
\begin{align}
\label{gauge trsf mixed}
\Lambda^R_{\beta\alpha} \to
g_{\beta}\Lambda^R_{\beta\alpha} g^{\dagger}_{\alpha},
\quad
\Lambda^L_{\beta\alpha} \to
g_{\alpha}\Lambda^L_{\beta\alpha} g^{\dagger}_{\beta}.
\end{align}
Note that 
the phases $e^{i\theta_{\alpha,\beta}}$ do not
affect $\Lambda^{L,R}_{\beta\alpha}$,
while they affect the eigenvalues of the 
mixed transfer matrix.
We also note that,
under the transformation, 
the regular inner product
of the two MPS wavefunctions
undergoes the change
$\mathrm{tr}\, [ \Lambda^{L}_{\beta\alpha}  \Lambda^{R}_{\beta\alpha} ]
\to 
\mathrm{tr}\, [ 
g_{\alpha} \Lambda^L_{\beta\alpha} g^{\dagger}_{\beta}
g_{\beta} \Lambda^R_{\beta\alpha} g^{\dagger}_{\alpha}]
$.
Thus, $g_{\alpha}$ and $g_{\beta}$ 
do not affect the regular inner product (as expected).
%

\subsection{MPS gerbe}\label{sec:MPSgerbe}

Just like a complex line bundle
provides the mathematical structure 
to describe the regular Berry phase, 
a gerbe serves as the underlying 
mathematical structure for the higher Berry phase.
Let $X$ be a topological space. 
Generically, 
a gerbe on a topological space $X$ is defined by 
the data 
$(\{U_\alpha\},\{L_{\alpha \beta}\},
\{\sigma_{\alpha \beta \gamma}\})$
\cite{GT10}. 
Here, 
$\{U_\alpha\}$ is an open covering of a base space $X$, $L_{\alpha \beta}$ is a complex vector bundle over
$U_{\alpha \beta}=U_\alpha \cap U_\beta$, and $\sigma_{\alpha \beta \gamma}: L_{\alpha \beta} \otimes L_{\beta \gamma} \rightarrow L_{\alpha \gamma}$ is an isomorphism between complex vector bundles. 
They satisfy a commutative diagram
\begin{eqnarray}
\label{eq:consistency}
  \begin{CD}
     L_{\alpha\beta}\otimes L_{\beta\gamma}\otimes L_{\gamma\delta} @>{1\otimes\sigma_{\beta\gamma\delta}}>> L_{\alpha\beta}\otimes L_{\beta\delta} \\
  @V{\sigma_{\alpha\beta\gamma}}VV    @V{\sigma_{\alpha\beta\delta}\otimes1}VV \\
     L_{\alpha\gamma}\otimes L_{\gamma\delta}   @>{\sigma_{\alpha\gamma\delta}}>>  L_{\alpha\delta}.
  \end{CD}.
\end{eqnarray}

The data $\sigma_{\alpha\beta\gamma}$ is a little bit abstract. 
To describe this concretely, let's take a section of each line bundle, i.e., $\ket{\psi_{\alpha\beta}}$ such that $L_{\alpha\beta}=\mathbb{C}\ket{\psi_{\alpha\beta}}$. For simplicity, we normalize this section, i.e., $\braket{\psi_{\alpha\beta}|\psi_{\alpha\beta}}=1$. Then, an isomorphism between two line bundles is nothing but a multiplication of some $U(1)$-valued scalar. Thus, we can extract a $U(1)$-valued function $c_{\alpha\beta\gamma}^{(0)}$\footnote{
The superscript "$(0)$" in $c^{(0)}_{\alpha\beta\gamma}$
indicates that $c_{\alpha\beta\gamma}$ 
is a zero form. We use similar notations 
throughout the paper. 
When there is no confusion, 
we omit the superscripts.} from $\sigma_{\alpha\beta\gamma}$ as follows:
\begin{eqnarray}
    \sigma_{\alpha\beta\gamma}:\ket{\psi_{\alpha\beta}}\otimes\ket{\psi_{\beta\gamma}}\mapsto c_{\alpha\beta\gamma}^{(0)}\ket{\psi_{\alpha\gamma}}. 
\end{eqnarray}
Here, $c_{\alpha\beta\gamma}^{(0)}$ is a $U(1)$-valued function on triple intersection $U_{\alpha\beta\gamma}$, and this is a higher 
analogue of the transition function of a line bundle. 
Similar to  
the Chern class
in the case of a line bundle, $c_{\alpha\beta\gamma}^{(0)}$ 
has topological information of the gerbe, and we can construct a topological invariant that takes its values in $\cohoZ{3}{X}$.
$[c_{\alpha\beta\gamma}^{(0)}]$ 
is called the Dixmier-Douady class.

To introduce a gerbe structure for a family of MPS,
following the above generalities, 
we need to specify the data 
$(\{U_\alpha\},\{L_{\alpha \beta}\},
\{\sigma_{\alpha \beta \gamma}\})$
in terms of MPS.
Following
Ref.\ \cite{OR23,QSWSPBH23},
we introduce $\ket{\psi_{\alpha\beta}}$ as the fixed point $\ket{\psi_{\alpha\beta}^{\rm MPS}}:=\Lambda_{\alpha\beta}^{R}$ of the mixed transfer matrix $T_{\alpha\beta}$, and an isomorphism $\sigma_{\alpha\beta\gamma}$ is given by the matrix multiplication
\begin{eqnarray}
    \sigma_{\alpha\beta\gamma}^{\rm MPS}:\Lambda_{\alpha\beta}^{R}\otimes\Lambda_{\beta\gamma}^{R}\mapsto\Lambda_{\alpha\beta}^{R}\Lambda_{\beta\gamma}^{R}.
\end{eqnarray}
Then, $(\{U_{\alpha}\},\{\mathbb{C}\ket{\psi_{\alpha\beta}^{\rm MPS}}\},\{\sigma_{\alpha\beta\gamma}^{\rm MPS}\})$ gives a gerbe over the parameter space. 
See Refs.\ \cite{OR23,QSWSPBH23} for the details. 
Note that the Dixmier-Douady class can be computed as the triple inner product of three MPS.

\subsection{Gerbe connections}

The subject of our central interest in this paper is
to construct a connection on an MPS gerbe.
In this section, we will go through some basics of gerbe connections in general.
\footnote{
Generically,
for a gerbe over $X$,
$\mathcal{G}=
(\{U_\alpha\},\{L_{\alpha \beta}\},
\{\sigma_{\alpha \beta \gamma}\})$,
a connection on ${\cal G}$ is 
the data 
$(\{B_\alpha\},
\{\nabla_{\alpha\beta}\})$
\cite{GT10}. 
Here, $\nabla_{\alpha \beta}$ is a covariant derivative on the line bundle $L_{\alpha\beta}\to U_{\alpha\beta}$
(i.e., a connection of the complex line bundle $L_{\alpha \beta}$),
and $B_\alpha$ is a 2-form on a patch $U_\alpha$. 
They are subject to the following conditions: 
\begin{align}
\nabla_{\alpha \beta} \otimes 1+1 \otimes \nabla_{\beta \gamma} & =\sigma_{\alpha \beta \gamma}^* \nabla_{\alpha \gamma},
\\
B_\alpha+F(\nabla_{\alpha \beta}) & =B_\beta,
\end{align}
where $F(\nabla_{\alpha \beta})$ is the curvature 2-form associated to $\nabla_{\alpha \beta}$. 
These conditions are equivalent to
Eqs.\ 
\eqref{consistency cond 1}
and
\eqref{consistency cond 2},
respectively.}

As a warmup, 
we start by reviewing a connection on a line bundle. Let's consider a line bundle over a parameter space $X$. By taking an open covering $\{U_{\alpha}\}$ of $X$, a line bundle is characterized by a transition function $c_{\alpha\beta}^{(0)}:U_{\alpha\beta}\to U(1)$ which satisfies the cocycle condition 
\begin{eqnarray}\label{eq:cocycle_line}
    c_{\alpha\beta}^{(0)}c_{\beta\gamma}^{(0)}=c_{\alpha\gamma}^{(0)}
\end{eqnarray} 
on each triple intersection $U_{\alpha\beta\gamma}$. A connection of this line bundle is described by a set of $1$-forms $\{A_{\alpha}^{(1)}\}$ on each open set. They transform as 
\begin{eqnarray}\label{eq:consistency_line}
    A_{\alpha}^{(1)}=A_{\beta}^{(1)}+c_{\alpha\beta}^{(0)}dc_{\alpha\beta}^{(0)},
\end{eqnarray}
on each double intersection $U_{\alpha\beta}$. 

We note that for a given line bundle, there are infinitely many choices for a connection.
Even if 
a connection 
(and hence curvature) is
fixed, there still exist redundancies, i.e., gauge redundancies. 
We can redefine a connection and a transition function as follows:
\begin{eqnarray}\label{eq:gauge_line}
    c_{\alpha\beta}^{(0)}&\mapsto &c_{\alpha\beta}^{(0)}\xi_{\alpha}^{(0)}(\xi_{\beta}^{(0)})^{-1},\\
    A_{\alpha}^{(1)}&\mapsto&A_{\alpha}^{(1)}+(\xi_{\alpha}^{(0)})^{-1}d\xi_{\alpha}^{(0)}.
\end{eqnarray}
Here, $\xi_{\alpha}^{(0)}:U_{\alpha}\to U(1)$ is an arbitrary function on each open set. Under this transformation, we can check that the consistency conditions 
Eq.\ \eqref{eq:cocycle_line} and Eq.\ \eqref{eq:consistency_line} still hold. 
To simplify the description of this situation for a line bundle, we will introduce the notation of \v{C}ech differential. Let $f_{\alpha_1\alpha_2\cdots\alpha_k}$ be a set of $U(1)$-valued functions defined on each $k$-intersection $U_{\alpha_1\alpha_2\cdots\alpha_k}:=U_{\alpha_1}\cap U_{\alpha_2}\cap\cdots \cap U_{\alpha_k}$, we define another $U(1)$-valued function on a $k+1$-intersection $U_{\alpha_1\alpha_2\cdots\alpha_{k+1}}$ as
\begin{eqnarray}
    (\delta f)_{\alpha_1\alpha_2\cdots\alpha_{k+1}}:=f_{\alpha_2\alpha_3\cdots\alpha_{k+1}}(f_{\alpha_1\alpha_3\cdots\alpha_{k+1}})^{-1}\cdots (f_{\alpha_1\alpha_2\cdots\alpha_{k}})^{\pm},
\end{eqnarray}
where the sign of the last factor is $+$ if $k$ is even and $-$ if $k$ is odd. 
In other words, the operator $\delta$ for $U(1)$-valued functions simply multiplies the original function with alternating inversion. On the other hand, for $\mathbb{R}$-valued functions or differential forms, $\delta$ operates similarly by attaching a negative sign instead of performing an inversion. That is, for a set of differential forms $f_{\alpha_1\alpha_2\cdots\alpha_k}$ defined on each $k$-intersection $U_{\alpha_1\alpha_2\cdots\alpha_k}$, $\delta$ defines another differential form on a $k+1$-intersection $U_{\alpha_1\alpha_2\cdots\alpha_{k+1}}$ as
\begin{eqnarray}
    (\delta f)_{\alpha_1\alpha_2\cdots\alpha_{k+1}}:=f_{\alpha_2\alpha_3\cdots\alpha_{k+1}}-f_{\alpha_1\alpha_3\cdots\alpha_{k+1}}+\cdots \pm f_{\alpha_1\alpha_2\cdots\alpha_{k}}.
\end{eqnarray}
Here, the sign of the last factor is $+$ if $k$ is even and $-$ if $k$ is odd. By using this $\delta$, 
Eq.\ \eqref{eq:cocycle_line} and Eq.\ \eqref{eq:consistency_line} are recast into
\begin{eqnarray}
    (\delta c^{(0)})_{\alpha\beta\gamma}=1, 
    \quad
    (\delta A^{(0)})_{\alpha\beta}=d\log c_{\alpha\beta}^{(0)}.
\end{eqnarray}
Also, the gauge redundancy Eq.\ \eqref{eq:gauge_line} is 
\begin{eqnarray}
    c_{\alpha\beta}^{(0)}\mapsto c_{\alpha\beta}^{(0)}(\delta\xi^{(0)})_{\alpha\beta},
   \quad 
    A_{\alpha}^{(1)}\to A_{\alpha}^{(1)}+d\log \xi^{(0)}_{\alpha}.
\end{eqnarray}
By using a connection, we define a curvature as 
\begin{eqnarray}
    F^{(2)}:=dA_{\alpha}^{(1)}.
\end{eqnarray}
Since $F^{(2)}$ is defined in terms of $\{A_{\alpha}^{(1)}\}$, 
it seems that it must be glued non-trivially at intersections. 
However, according to Eq.\ \eqref{eq:consistency_line}, 
$F^{(2)}$ is glued identically as a differential form on intersections, so $F^{(2)}$ becomes a global $2$-form on $X$. 
Therefore, no indices with respect to patches are assigned. 
Note that the integral values of ${F^{(2)}}/{2\pi i}$ are quantized 
to integers on a closed orientable surface.
These properties are summarized as:
\begin{align}
  &
  \begin{tikzpicture}[line cap=round,line join=round,x=1.0cm,y=1.0cm, scale=0.25, baseline={([yshift=-.4ex]current bounding box.center)}, thick, shift={(0,0)}, scale=0.7]
  \def\tate{14} 
  \def\yoko{14} 
  \def\sen{7} 
  \def\gaisen{2} 
  \draw (0,0) -- (\yoko,0);
  \draw (0,5) -- (\yoko,5);
  \draw (0,10) -- (\yoko,10);
  \draw (0,0) -- (0,\tate);
  \draw (5,0) -- (5,\tate);
  \draw (10,0) -- (10,\tate);
  \draw[->] (4.5,7.5) -- (6,7.5);
  \draw[->] (9.5,2.5) -- (11,2.5);
  \draw[->] (7.5,4.5) -- (7.5,6);
  \node[anchor=center] at (2.5,7.5) {$A^{(1)}_{\alpha}$};
  \node[anchor=center] at (7.5,2.5) {$c^{(0)}_{\alpha\beta}$};
  \node[anchor=center] at (12.5,2.5) {$1$};
  \node[anchor=center] at (7.5,7.5) {$0$};
  \node[anchor=center] at (5,-2) {$\stackrel{\longrightarrow}{\delta}$};
  \node[anchor=center] at (10,-2) {$\stackrel{\longrightarrow}{\delta}$};
  \node[anchor=center] at (7,-5) {Connection};
  \node[anchor=east] at (0,5) {$d\log\, \uparrow$};
  \node[anchor=east] at (0,9) {$d\, \uparrow$};
\end{tikzpicture}
\qquad
\begin{tikzpicture}[line cap=round,line join=round,x=1.0cm,y=1.0cm, scale=0.25, baseline={([yshift=-.4ex]current bounding box.center)}, thick, shift={(0,0)}, scale=0.7]
  \def\tate{14} 
  \def\yoko{20} 
  \def\sen{7} 
  \def\gaisen{2} 
  \draw (0,0) -- (\yoko,0);
  \draw (0,5) -- (\yoko,5);
  \draw (0,10) -- (\yoko,10);
  \draw (0,0) -- (0,\tate);
  \draw (9,0) -- (9,\tate);
  \draw (19,0) -- (19,\tate);
  \draw[->] (4.5,4.5) -- (4.5,6);
  \draw[->] (8.5,2.5) -- (10,2.5);
  \node[anchor=center] at (4.5,7.5) {$d\log \xi^{(0)}_{\alpha}$};
  \node[anchor=center] at (4.5,2.5) {$\xi^{(0)}_{\alpha}$};
  \node[anchor=center] at (14,2.5) {$(\delta \xi^{(0)})_{\alpha\beta}$};
  \node[anchor=center] at (10,-5) {Redundancy};
\end{tikzpicture}
\qquad
\begin{tikzpicture}[line cap=round,line join=round,x=1.0cm,y=1.0cm, scale=0.25, baseline={([yshift=-.4ex]current bounding box.center)}, thick, shift={(0,0)}, scale=0.7]
  \def\tate{14} 
  \def\yoko{12} 
  \def\sen{7} 
  \def\gaisen{2} 
  \draw (0,0) -- (\yoko,0);
  \draw (0,5) -- (\yoko,5);
  \draw (0,10) -- (\yoko,10);
  \draw (0,0) -- (0,\tate);
  \draw (5,0) -- (5,\tate);
  \draw (10,0) -- (10,\tate);
  \draw[->] (2.5,9.5) -- (2.5,11);
  \node[anchor=center] at (2.5,12.5) {$F^{(2)}$};
  \node[anchor=center] at (2.5,7.5) {$A^{(1)}_{\alpha}$};
  \node[anchor=center] at (7.5,2.5) {$c^{(0)}_{\alpha\beta}$};
  \node[anchor=center] at (6,-5) {Curvature};
\end{tikzpicture}
\end{align}

As a natural generalization, we can consider a connection on a gerbe. 
As explained in Sec.\ \ref{sec:MPSgerbe}, 
a gerbe is topologically described by a $U(1)$-valued function $c_{\alpha\beta\gamma}^{(0)}:U_{\alpha\beta\gamma}\to U(1)$ such that
\begin{eqnarray}
    (\delta c^{(0)})_{\alpha\beta\gamma\delta}=1
\end{eqnarray}
on $U_{\alpha\beta\gamma\delta}$. 
A connection on this gerbe is given by
\begin{align}
  c^{(0)}_{\alpha\beta\gamma},
  \quad
  w^{(1)}_{\alpha\beta},
  \quad
  B^{(2)}_{\alpha}.
\end{align}
Here, $\{w_{\alpha\beta}^{(1)}\}$ is a set of $1$-forms on each intersection $U_{\alpha\beta}$, and $\{B^{(2)}_{\alpha}\}$ is a set of $2$-forms on each open set $U_{\alpha}$. 
We call $w^{(1)}_{\alpha\beta}$
and $B^{(2)}_{\alpha}$
the one-form and two-form connections, respectively.
These data are subject to the consistency conditions, 
\begin{align}
  &
\label{consistency cond 0}
  (\delta c^{(0)})_{\alpha\beta\gamma\delta}=1\\
  &
  (\delta w^{(1)})_{\alpha\beta\gamma}=d \log c^{(0)}_{\alpha\beta\gamma},
\label{consistency cond 1}
    \\
\label{consistency cond 2}
  &
   (\delta B^{(2)})_{\alpha\beta}=d w^{(1)}_{\alpha\beta}.
\end{align}
Similar to the case of line bundles, a connection on a gerbe has a gauge redundancy. This is given by
\begin{eqnarray}\label{eq:connection_gauge}
    c_{\alpha\beta\gamma}^{(0)}&\mapsto& c_{\alpha\beta\gamma}^{(0)}(\delta\xi^{(0)})_{\alpha\beta\gamma},\label{eq:consistency_gerbe_1}
    \\
    w_{\alpha\beta}^{(1)}&\mapsto& w_{\alpha\beta}^{(1)}+(\delta \xi^{(1)})_{\alpha\beta}+d\log\xi_{\alpha\beta}^{(0)},\label{eq:consistency_gerbe_2}\\
    B_{\alpha}^{(2)}&\mapsto& B_{\alpha}^{(2)}+d\xi_{\alpha}^{(1)}\label{eq:consistency_gerbe_3}.
\end{eqnarray}
Here, $\xi_{\alpha\beta}^{(0)}$ is an arbitrary $U(1)$-valued function on each intersection $U_{\alpha\beta}$ and $\xi_{\alpha}^{(1)}$ is an arbitrary $1$-form on each open set $U_{\alpha}$. 
We call 
the part of gauge transformations
relevant to
$\xi^{(0)}_{\alpha\beta}$
and 
$\xi^{(1)}_{\alpha}$
0-form and
1-form gauge transformations,
respectively.
Under this transformation, we can check that the consistency conditions Eqs.\ \eqref{eq:consistency_gerbe_1}-\eqref{eq:consistency_gerbe_3} remain hold true. 
By using a connection, we define a curvature as 
\begin{eqnarray}
    H^{(3)}:=dB_{\alpha}^{(2)}.
\end{eqnarray}
$H^{(3)}$ is defined by $\{B_{\alpha}^{(2)}\}$, so it seems that it must be glued non-trivially at intersections. However, according to Eq.\ \eqref{eq:consistency_gerbe_3}, $H^{(3)}$ glued identically as a differential form on intersections, so $H^{(3)}$ becomes a global $3$-form on $X$. Therefore, no indices with respect to patches are assigned. Note that the integral values of ${H^{(3)}}/{2\pi i}$ are quantized to integers on a closed orientable $3$-dimensional manifold.
These properties are summarized as follows:
\begin{align}
  &
  \begin{tikzpicture}[line cap=round,line join=round,x=1.0cm,y=1.0cm, scale=0.35, baseline={([yshift=-.4ex]current bounding box.center)}, thick, shift={(0,0)}, scale=0.55]
  \def\tate{19} 
  \def\yoko{19} 
  \def\sen{7} 
  \def\gaisen{2} 
  \draw (0,0) -- (\yoko,0);
  \draw (0,5) -- (\yoko,5);
  \draw (0,10) -- (\yoko,10);
  \draw (0,15) -- (\yoko,15);
  \draw (0,0) -- (0,\tate);
  \draw (5,0) -- (5,\tate);
  \draw (10,0) -- (10,\tate);
  \draw (15,0) -- (15,\tate);
  \draw[->] (4.5,12.5) -- (6,12.5);
  \draw[->] (7.5,9.5) -- (7.5,11);
  \draw[->] (9.5,7.5) -- (11,7.5);
  \draw[->] (12.5,4.5) -- (12.5,6);
  \draw[->] (14.5,2.5) -- (16,2.5);
  \node[anchor=center] at (2.5,12.5) {$B^{(2)}_{\alpha}$};
  \node[anchor=center] at (7.5,7.5) {$w^{(1)}_{\alpha\beta}$};
  \node[anchor=center] at (12.5,2.5) {$c^{(0)}_{\alpha\beta\gamma}$};
  \node[anchor=center] at (17.5,2.5) {$1$};
  \node[anchor=center] at (12.5,7.5) {$0$};
  \node[anchor=center] at (7.5,12.5) {$0$};
  \node[anchor=center] at (5,-2) {$\stackrel{\longrightarrow}{\delta}$};
  \node[anchor=center] at (10,-2) {$\stackrel{\longrightarrow}{\delta}$};
  \node[anchor=center] at (15,-2) {$\stackrel{\longrightarrow}{\delta}$};
  \node[anchor=center] at (9.5,-5) {Connection};
  \node[anchor=east] at (0,5) {$d\log \uparrow$};
  \node[anchor=east] at (0,10) {$d \uparrow$};
  \node[anchor=east] at (0,15) {$d \uparrow$};
\end{tikzpicture}
\qquad
\begin{tikzpicture}[line cap=round,line join=round,x=1.0cm,y=1.0cm, scale=0.35, baseline={([yshift=-.4ex]current bounding box.center)}, thick, shift={(0,0)}, scale=0.55]
  \def\tate{19} 
  \def\yoko{26} 
  \def\sen{7} 
  \def\gaisen{2} 
  \draw (0,0) -- (\yoko,0);
  \draw (0,5) -- (\yoko,5);
  \draw (0,10) -- (\yoko,10);
  \draw (0,15) -- (\yoko,15);
  \draw (0,0) -- (0,\tate);
  \draw (5,0) -- (5,\tate);
  \draw (15,0) -- (15,\tate);
  \draw (25,0) -- (25,\tate);
  \draw[->] (2.5,9.5) -- (2.5,10.9);
  \draw[->] (4.5,7.5) -- (5.9,7.5);
  \draw[->] (10,4.5) -- (10,5.7);
  \draw[->] (14.5,2.5) -- (16,2.5);
  \node[anchor=center] at (2.5,12.5) {$d\xi^{(1)}_{\alpha}$};
  \node[anchor=center] at (2.5,7.5) {$\xi^{(1)}_{\alpha}$};
  \node[anchor=center] at (10,8.6) {
  \begin{scriptsize}$(\delta \xi^{(1)})_{\alpha\beta}+$\end{scriptsize}};
  \node[anchor=center] at (10,6.7) {\begin{scriptsize}$d\log\xi^{(0)}_{\alpha\beta}$\end{scriptsize}};
  \node[anchor=center] at (10,2.5) {$\xi^{(0)}_{\alpha\beta}$};
  \node[anchor=center] at (20.5,2.5) {
  $(\delta \xi^{(0)})_{\alpha\beta\gamma}$
  };
  \node[anchor=center] at (13,-5) {Redundancy};
\end{tikzpicture}
\qquad
\begin{tikzpicture}[line cap=round,line join=round,x=1.0cm,y=1.0cm, scale=0.35, baseline={([yshift=-.4ex]current bounding box.center)}, thick, shift={(0,0)}, scale=0.55]
  \def\tate{19} 
  \def\yoko{16} 
  \def\sen{7} 
  \def\gaisen{2} 
  \draw (0,0) -- (\yoko,0);
  \draw (0,5) -- (\yoko,5);
  \draw (0,10) -- (\yoko,10);
  \draw (0,15) -- (\yoko,15);
  \draw (0,0) -- (0,\tate);
  \draw (5,0) -- (5,\tate);
  \draw (10,0) -- (10,\tate);
  \draw (15,0) -- (15,\tate);
  \draw [->](2.5,14.5) -- (2.5,16);
  \node[anchor=center] at (2.5,17.5) {$H^{(3)}$};
  \node[anchor=center] at (2.5,12.5) {$B^{(2)}_{\alpha}$};
  \node[anchor=center] at (7.5,7.5) {$w^{(1)}_{\alpha\beta}$};
  \node[anchor=center] at (12.5,2.5) {$c^{(0)}_{\alpha\beta\gamma}$};
  \node[anchor=center] at (8,-5) {Curvature};
\end{tikzpicture}
\end{align}

\subsection{Summary}
\label{summary}

The main purpose of this paper is to construct a gerbe connection from a family of normal MPS. 
This is analogous to the work by Berry
\cite{Berry1984}
who identified the Berry connection $a^{(1)}$ as
the wave function overlap
$a^{(1)}=\langle \psi|d\psi \rangle$.
By introducing a gerbe connection, we can easily compute the topological invariant. 
In this section, we summarize the results.

First, the 1-form connection on $U_{\alpha\beta}$ is given by 
\begin{align}
\label{sum: 1-form conn}
  w^{(1)}_{\alpha\beta}
  &=
  \begin{tikzpicture}[line cap=round,line join=round,x=1.0cm,y=1.0cm, scale=0.25, baseline={([yshift=-.4ex]current bounding box.center)}, thick, shift={(0,0)}, scale=0.7]
  \def\tate{2} 
  \def\yoko{5} 
  \def\sen{7} 
  \def\gaisen{2} 
  \draw (0,\tate) -- (\yoko,\tate);
  \draw (0,-\tate) -- (\yoko,-\tate);
  \draw (0,\tate) arc (90:270:\tate);
  \draw (\yoko,\tate) arc (90:-90:\tate);
  \filldraw[fill=white] (\yoko/2,-\tate) circle(3mm);
  \node[anchor=east] at (-\tate,0) 
  {$\Lambda_{\beta}^{L}$};
  \node[anchor=west] at (\yoko+\tate,0) 
  {$\Lambda_{\beta}^{R}$};
  \node[anchor=north] at (\yoko/2,-\tate) 
  {$d\log g_{\alpha\beta}$};
  \node[anchor=south] at (\yoko/2,\tate) 
  {${\color{white} d\log g_{\alpha\beta}}$};
\end{tikzpicture}\, .
\end{align}
Here, the white circle represents $d\log g_{\alpha\beta}$. The 2-form connection consists of two parts:
\begin{eqnarray}
\label{sum: 2-form conn a}
    B_{\alpha}^{(2)}=b_\alpha-b'_{\alpha}.
\end{eqnarray}
The first part $b_{\alpha}$ is defined as
\begin{eqnarray}
\label{sum: 2-form conn b}
    b_{\alpha}=\begin{tikzpicture}[line cap=round,line join=round,x=1.0cm,y=1.0cm, scale=0.25, baseline={([yshift=-.4ex]current bounding box.center)}, thick, shift={(0,0)}, scale=0.7]
  \def\tate{2} 
  \def\yoko{10} 
  \def\sen{7} 
  \def\gaisen{2} 
  \def\siten{1}
  \pgfmathsetmacro\hako{\yoko-\siten}
  \draw (0,\tate) -- (\siten,\tate);
  \draw (\hako,\tate) -- (\yoko,\tate);
  \draw (0,-\tate) -- (\siten,-\tate);
  \draw (\hako,-\tate) -- (\yoko,-\tate);
  \draw (0,\tate) arc (90:270:\tate);
  \draw (\yoko,\tate) arc (90:-90:\tate);
  \filldraw[fill=black] (-\tate,0) circle(3mm);
  \node[anchor=east] at (-\tate,0) 
   {$d\Lambda_{\alpha}^{L}$};
  \draw (\siten,-\tate-1) -- (\siten,\tate+1);
  \node at (\hako/2+\siten/2,0) 
  {$\cfrac{1}{1-T'_{\alpha}}$
  };
  \draw (\hako,-\tate-1) -- (\hako,\tate+1);
  \draw (\siten,-\tate-1) -- (\hako,-\tate-1);
  \draw (\siten,\tate+1) -- (\hako,\tate+1);
  \draw (\yoko,-\tate) -- (\yoko,\tate);
  \filldraw[fill=black] (\yoko,-\tate) circle(3mm);
  \node[anchor=west] at (\yoko+\tate/2,-\tate) 
  {$dA_{\alpha}$};
\end{tikzpicture}.
\end{eqnarray}
Here, each black dot represents the exterior derivative and $T^{\prime}_{\alpha}$ represents the reduced transfer matrix defined
in Eq.\ \eqref{eq:reduced_transfer}. 
The second part $b_{\alpha}^{\prime}$ 
is defined by
\begin{eqnarray}
    b'_\alpha:=\sum_{\alpha_0} \rho_{\alpha_0} (x_{\alpha_0 \alpha}+ y^0_{\alpha_0 \alpha}).
\end{eqnarray}
Here, $\rho_{\alpha}$ is a partition of unity of the parameter space $X$ and 
$x_{\alpha\beta}, y_{\alpha\beta}^{0}$ are defined in 
Eqs.\ \eqref{eq:x} and \eqref{eq:y0}.

We will describe the details of these definitions and, also, gauge invariance 
in Sec.\ \ref{Construction of an MPS gerbe connection}. 
We note that the regular Berry connection 
undergoes the gauge transformation 
$a^{(1)}
\to a^{(1)} + d \theta
$
under the gauge transformation
$|\psi\rangle \to e^{i\theta} |\psi\rangle$.
In MPS, the gauge transformation is implemented as 
$A^s \to e^{i \theta/L} A^s$
where $L$ is the total length of the chain,
which is the $U(1)$ part 
in Eq.\ \eqref{MPS gauge}.
Similarly, 
the gerbe gauge transformations
$\xi^{(0)}$
and $\xi^{(1)}$
are expected to be related to 
the MPS gauge transformation
\eqref{MPS gauge}.
First, 
we expect that 
the 1-form gauge transformation 
is associated with the 
MPS gauge transformation, 
$
A^s_{\alpha} \to 
g_{\alpha}
A^s_{\alpha}
g_{\alpha}^{\dag}
$.
Also, 
$c_{\alpha\beta\gamma}^{(0)}$
(the Dixmier-Douady class)
is given
in terms of a $U(\mathsf{D})$ lift of $g_{\alpha\beta}$,
$
c^{(0)}_{\alpha\beta\gamma}
=
\mathrm{tr}\left[
  \hat{g}_{\alpha\beta}\hat{g}_{\beta\gamma}\hat{g}_{\gamma\alpha}
\right]
$
(for the case of constant-rank MPS gerbe).
Thus, the ambiguity of the lift is the 0-form gauge transformation,
$
c^{(0)}_{\alpha\beta\gamma}
\longrightarrow
c^{(0)}_{\alpha\beta\gamma}
e^{i \phi_{\alpha\beta}}e^{ i \phi_{\beta\gamma}}
e^{i \phi_{\gamma\alpha}}
$.

At first glance, 
these definitions may only work when the matrix rank of the MPS matrices is constant. 
However, we will see that these definitions can be applied to general MPS gerbes. 
In Sec.\ \ref{Examples}, using this gerbe connection, we will calculate the higher Berry curvature for two models and provide examples where the invariant becomes non-trivial.

\section{Construction of an MPS gerbe connection}
\label{Construction of an MPS gerbe connection}

In this section, we define a gerbe connection by using MPS representations. In Sec.\ \ref{sec:constant}, as a warm-up, 
we first propose a gerbe connection for a constant-rank MPS gerbe and confirm the consistency condition. 
In this case, however, it is known that the higher Berry curvature is always trivial \cite{OR23,qi2023charting}.
By slightly modifying the construction for the constant rank case, we can construct a gerbe connection for a general MPS gerbe, including non-constant rank gerbes.
We will explain this point in Sec.\ 
\ref{General MPS gerbe}. 
In Sec.\ \ref{sec:MPSgerbe}, we discuss the behaviors of the gerbe connection under gauge transformations of MPS representation. Consequently, the change is absorbed into the gauge redundancy 
Eq.\ \eqref{eq:connection_gauge} of a gerbe connection.

\subsection{Constant MPS gerbe} 
\label{sec:constant}

\subsubsection{The Dixmier-Douady class $c_{\alpha\beta\gamma}^{(0)}$}

As explained in 
Sec.\ \ref{sec:MPSgerbe}, 
we can extract the Dixmier-Douady class 
as a product of two fixed points:
\begin{eqnarray}
    \Lambda_{\alpha\beta}^{R}\Lambda_{\beta\gamma}^{R}=c_{\alpha\beta\gamma}^{(0)}\Lambda_{\alpha\gamma}^{R}.
\end{eqnarray}
By multiplying the left fixed point $\Lambda_{\alpha\gamma}^{L}$ and taking a trace, we obtain  
\begin{eqnarray}
    c_{\alpha\beta\gamma}^{(0)}=\tr{\Lambda_{\alpha\gamma}^{L}\Lambda_{\alpha\beta}^{R}\Lambda_{\beta\gamma}^{R}}.
\end{eqnarray}
Here, we used the normalization condition 
Eq.\ \eqref{normalization mixed fixed pt}. 
Remark that the right-hand side can be regarded 
as an overlap of three MPS. 
For this reason, the right-hand side is referred to as the triple inner product
\cite{OR23}.

\subsubsection{One-form connection $w_{\alpha\beta}^{(1)}$}

Let us now define a $1$-form connection $w_{\alpha\beta}^{(1)}$: this is a $1$-form on 2-intersection $U_{\alpha\beta}$. 
For now,
we consider a constant-rank MPS gerbe. 
Therefore, we have a transition function $g_{\alpha\beta}$ on each 2-intersection $U_{\alpha\beta}$. 
By using the transition function, we define a $1$-form connection $w_{\alpha\beta}^{(1)}$ as
\begin{align}
\label{1-form conn}
  w^{(1)}_{\alpha\beta}
  &
  =
  \Lambda^L_{\beta} \cdot 
  (1_\mathsf{D}\otimes d\log g_{\alpha\beta})
  \cdot \Lambda^R_{\beta}
  \nonumber \\
  &=
  \sum_{i,j,k,l}
  (\Lambda^L_{\beta})_{(i,j)}
  (1_\mathsf{D}\otimes d\log g_{\alpha\beta})_{(i,j)(k,l)}
  \cdot (\Lambda^R_{\beta})_{(k,l)}
  \nonumber \\
  &=
  \begin{tikzpicture}[line cap=round,line join=round,x=1.0cm,y=1.0cm, scale=0.25, baseline={([yshift=-.4ex]current bounding box.center)}, thick, shift={(0,0)}, scale=0.7]
  \def\tate{2} 
  \def\yoko{5} 
  \def\sen{7} 
  \def\gaisen{2} 
  \draw (0,\tate) -- (\yoko,\tate);
  \draw (0,-\tate) -- (\yoko,-\tate);
  \draw (0,\tate) arc (90:270:\tate);
  \draw (\yoko,\tate) arc (90:-90:\tate);
  \filldraw[fill=white] (\yoko/2,-\tate) circle(3mm);
  \node[anchor=east] at (-\tate,0) 
  {$\Lambda_{\beta}^{L}$};
  \node[anchor=west] at (\yoko+\tate,0) 
  {$\Lambda_{\beta}^{R}$};
  \node[anchor=north] at (\yoko/2,-\tate) 
  {$d\log g_{\alpha\beta}$};
\end{tikzpicture}\, .
\end{align}
Here, in the first line,
$\Lambda^R_{\beta}$
and
$\Lambda^L_{\beta}$
are considered
as 
states in the doubled 
virtual Hilbert space
("bra" and "ket", respectively),
and 
$
(1_\mathsf{D}\otimes d\log g_{\alpha\beta})
$ 
is an operator acting on 
the doubled Hilbert space.
We will use similar notations henceforth.
If we use the right canonical condition\footnote{In the following, we take the right canonical gauge.}, this definition is equivalent to
\begin{align}
  w^{(1)}_{\alpha\beta}=
  \begin{tikzpicture}[line cap=round,line join=round,x=1.0cm,y=1.0cm, scale=0.25, baseline={([yshift=-.4ex]current bounding box.center)}, thick, shift={(0,0)}, scale=0.7]
  \def\tate{2} 
  \def\yoko{4} 
  \def\sen{7} 
  \def\gaisen{2} 
  \draw (0,\tate) -- (\yoko,\tate);
  \draw (0,-\tate) -- (\yoko,-\tate);
  \draw (0,\tate) arc (90:270:\tate);
  \draw (\yoko,\tate) arc (90:-90:\tate);
  \filldraw[fill=black] (\yoko+\tate,0) circle(3mm);
  \node[anchor=east] at (-\tate,0) 
  {$\Lambda_{\alpha\beta}^{L}$};
  \node[anchor=west] at (\yoko+\tate,0) 
  {$d\Lambda_{\alpha\beta}^{R}$};
\end{tikzpicture}.
\end{align}
Then, we can show that the consistency condition
\eqref{consistency cond 1}:
\begin{align}\label{eq:1st_consistency}
    (\delta w^{(1)})_{\alpha\beta\gamma}=d\log c^{(0)}_{\alpha\beta\gamma}.
\end{align}
\begin{proof}
Let's consider the quantity
$
    \Lambda_\gamma^{L}(g_{\alpha\beta}g_{\beta\gamma})^{\dagger}d(g_{\alpha\beta}g_{\beta\gamma})\Lambda_\gamma^{R}
$
and
evaluate it in two different ways.
First, by using the Leibniz rule,
\begin{eqnarray}
    \Lambda_\gamma^{L}(g_{\alpha\beta}g_{\beta\gamma})^{\dagger}d(g_{\alpha\beta}g_{\beta\gamma})\Lambda_\gamma^{R}=\Lambda_\gamma^{L}(g_{\beta\gamma}^{\dagger}d\log g_{\alpha\beta}g_{\beta\gamma}+d\log g_{\beta\gamma})\Lambda_\gamma^{R}.
\end{eqnarray}
On the other hand, by using the definition of $c_{\alpha\beta\gamma}$,
\begin{eqnarray}\Lambda_\gamma^{L}(g_{\alpha\beta}g_{\beta\gamma})^{\dagger}d(g_{\alpha\beta}g_{\beta\gamma})\Lambda_\gamma^{R}=\Lambda_\gamma^{L}c_{\alpha\beta\gamma}^{\ast}g_{\alpha\gamma}^{\dagger}(dc_{\alpha\beta\gamma}g_{\alpha\gamma}+c_{\alpha\beta\gamma}dg_{\alpha\gamma})\Lambda_\gamma^{R}=\Lambda_\gamma^{L}(d\log c_{\alpha\beta\gamma}+d\log g_{\alpha\gamma})\Lambda_\gamma^{R}.
\end{eqnarray}
By taking the trace of both expressions, we obtain Eq.\ (\ref{eq:1st_consistency}). 
\end{proof}

In the above definition of the 1-form connection, 
we choose a particular gauge of the MPS representation.
While the MPS gauge transformation changes our 1-form connection, 
%
this change can be compensated by 
the 1-form gauge transformation of
the MPS gerbe connection 
in Eq.\ \eqref{eq:consistency_gerbe_2}.
We will discuss this point
in Sec.\ \ref{Gauge redundancy}.

\subsubsection{Two-form connection $B_{\alpha}^{(2)}$}

\paragraph{Fixed-point MPS}

Next, we discuss the two-form connection $B^{(2)}_{\alpha}$.
Here, we first focus on the case of fixed-point MPS,
for which the expression of the two-form connection simplifies.  
Here, fixed-point MPS are MPS whose transfer matrix
for which the spectrum of Lyapunov exponents consists of one maximal one and all the others are zero. 
Hence, for a fixed point MPS, 
the spectral decomposition of
the transfer matrix 
consists of a single term,
%
\begin{align}
\label{fixed pt}
  \begin{tikzpicture}[line cap=round,line join=round,x=1.0cm,y=1.0cm, scale=0.25, baseline={([yshift=-.4ex]current bounding box.center)}, thick, shift={(0,0)}, scale=0.7]
  \def\tate{2} 
  \def\yoko{1} 
  \def\sen{7} 
  \def\gaisen{2} 
  \draw (0,\tate) -- (\yoko+\yoko,\tate);
  \draw (0,-\tate) -- (\yoko + \yoko,-\tate);
  \draw (\yoko,-\tate) -- (\yoko,\tate);
\end{tikzpicture}\, 
=
\begin{tikzpicture}[line cap=round,line join=round,x=1.0cm,y=1.0cm, scale=0.25, baseline={([yshift=-.4ex]current bounding box.center)}, thick, shift={(0,0)}, scale=0.7]
  \def\tate{2} 
  \def\yoko{10} 
  \def\sen{7} 
  \def\gaisen{2} 
  \draw (5,\tate) arc (90:270:\tate);
  \draw (0,\tate) arc (90:-90:\tate);
\end{tikzpicture}\, .
\end{align}
We note that fixed-point MPS can be obtained by 
blocking or renormalizing $n$ sites into a single site.
Namely, 
since $\left\{A^s\right\}$ is normal and in the canonical form, the maximal eigenvalue of the transfer matrix is 1 and
the norms of the others are less than 1. 
Thus, for large enough $n$, the spectrum of $T^n$ is a subset of $[0, \epsilon) \cup\{1\}$ for some small $\epsilon$, i.e., $T^n$ is almost a projection. In particular, the eigenspace corresponding to the eigenvalue 1 is 1-dimensional and spanned by $1_\mathsf{D}$. 

For fixed-point MPS, let's consider the following quantity  $b^{0}_{\alpha}$: 
\begin{align}
  b^{0}_{\alpha}
  &=
  \sum_{i,j,k,l}
  \sum_s
  (d\Lambda^L_{\alpha})_{(i,j)}
  (A^{s*}_{\alpha} \otimes dA^s_{\alpha})_{(i,j),(k,l)}
  (1_\mathsf{D})_{(k,l)}
 \nonumber \\
   &=
  \begin{tikzpicture}[line cap=round,line join=round,x=1.0cm,y=1.0cm, scale=0.25, baseline={([yshift=-.4ex]current bounding box.center)}, thick, shift={(0,0)}, scale=0.7]
  \def\tate{2} 
  \def\yoko{4} 
  \def\sen{7} 
  \def\gaisen{2} 
  \draw (0,\tate) -- (\yoko,\tate);
  \draw (0,-\tate) -- (\yoko,-\tate);
  \draw (0,\tate) arc (90:270:\tate);
  \draw (\yoko,\tate) arc (90:-90:\tate);
  \filldraw[fill=black] (-\tate,0) circle(3mm);
  \node[anchor=east] at (-\tate,0) 
  {$d\Lambda_{\alpha}^{L}$};
  \draw (2,-\tate) -- (2,\tate);
  \draw (2,-\tate) -- (2,\tate);
  \filldraw[fill=black] (2,-\tate) circle(3mm);
    \node[anchor=east] at (4,-\tate-1.5) 
  {$dA_{\alpha}$};
\end{tikzpicture}\, .
  \label{2-form conn}
\end{align}
Note that $b^0_{\alpha}$ can be rewritten in different ways using the identities 
\begin{gather}
\label{deriv eig eq}
  \begin{tikzpicture}[line cap=round,line join=round,x=1.0cm,y=1.0cm, scale=0.25, baseline={([yshift=-.4ex]current bounding box.center)}, thick, shift={(0,0)}, scale=0.7]
  \def\tate{2} 
  \def\yoko{1} 
  \def\sen{7} 
  \def\gaisen{2} 
  \draw (0,\tate) -- (\yoko,\tate);
  \draw (0,-\tate) -- (\yoko,-\tate);
  \draw (\yoko,\tate) arc (90:-90:\tate);
  \draw (\yoko,-\tate) -- (\yoko,\tate);
  \node[anchor=west] at (3,0) 
  {$+$};
 \draw (7,\tate) arc (90:-90:\tate);
 \draw (7,-\tate) -- (7,\tate);
    \filldraw[fill=black] (1,\tate) circle(3mm);
    \filldraw[fill=black] (7,-\tate) circle(3mm);
          \draw (6,\tate) -- (7,\tate);
  \draw (6,-\tate) -- (7,-\tate);
      \node[anchor=west] at (9,0) 
  {$=0$};
\end{tikzpicture}\, ,
\quad 
  \begin{tikzpicture}[line cap=round,line join=round,x=1.0cm,y=1.0cm, scale=0.3, baseline={([yshift=-.4ex]current bounding box.center)}, thick, shift={(0,0)}, scale=0.7]
  \def\tate{2} 
  \def\yoko{1} 
  \def\sen{7} 
  \def\gaisen{2} 
  \draw (-5,\tate) arc (90:270:\tate);
  \filldraw[fill=black] (-7,0) circle(3mm);
  \draw (-5,-\tate) -- (-5,\tate);
  \draw (-5,\tate) -- (-4,\tate);
  \draw (-5,-\tate) -- (-4,-\tate);
  \node[anchor=west] at (-4,0) {$+$};
  \draw (1,\tate) arc (90:270:\tate);
  \draw (1,-\tate) -- (1,\tate);
  \draw (1,\tate) -- (2,\tate);
  \draw (1,-\tate) -- (2,-\tate);
  \filldraw[fill=black] (1,\tate) circle(3mm);
  \node[anchor=west] at (2,0) {$+$};
  \draw (7,\tate) arc (90:270:\tate);
  \draw (7,-\tate) -- (7,\tate);
  \draw (7,\tate) -- (8,\tate);
  \draw (7,-\tate) -- (8,-\tate);
  \filldraw[fill=black] (7,-\tate) circle(3mm);
   \node[anchor=west] at (8,0) {$=$};
  %
  \draw (13,\tate) arc (90:270:\tate);
    \filldraw[fill=black] (11,0) circle(3mm);
\end{tikzpicture}\, ,
\end{gather}
obtained by taking the derivative of the eigen equation
\eqref{eig eq}.
As we will see below, $b^{0}_{\alpha}$ becomes the $2$-form connection $B^{(2)}_{\alpha}$ for simple models. 
However, in the general case, we need to introduce some corrections to $b^{0}_{\alpha}$. 
Let's see how $b^{(2)}_{\alpha}$ behaves under 
$\delta$.
After a simple calculation, we obtain  
\begin{align}
(\delta b^{0})_{\alpha \beta}=dw_{\alpha\beta}+x_{\alpha\beta}^{0}+y_{\alpha\beta}^{0}-y_{\alpha\beta}^{1}-z_{\alpha\beta}^{1},
\end{align}
where 
\begin{eqnarray}
    x_{\alpha\beta}^{0}&=&\Lambda^L_{\beta}\cdot (d\log g_{\alpha\beta}^{T}\otimes1_{\mathsf{D}_{\beta}}-1_{\mathsf{D}_{\beta}}\otimes d\log g_{\alpha\beta})\cdot T[A_{\beta};dA_{\beta}]\cdot 1_{\mathsf{D}_{\beta}}
    \nonumber \\
    &=&
  \begin{tikzpicture}[line cap=round,line join=round,x=1.0cm,y=1.0cm, scale=0.25, baseline={([yshift=-.4ex]current bounding box.center)}, thick, shift={(0,0)}, scale=0.7]
  \def\tate{2} 
  \def\yoko{4} 
  \def\sen{7} 
  \def\gaisen{2} 
  \draw (0,\tate) -- (\yoko,\tate);
  \draw (0,-\tate) -- (\yoko,-\tate);
  \draw (0,\tate) arc (90:270:\tate);
  \draw (\yoko,\tate) arc (90:-90:\tate);
  \filldraw[fill=white] (1,\tate) circle(3mm);
  \draw (2,-\tate) -- (2,\tate);
  \filldraw[fill=black] (2,-\tate) circle(3mm);
\end{tikzpicture}
-
  \begin{tikzpicture}[line cap=round,line join=round,x=1.0cm,y=1.0cm, scale=0.25, baseline={([yshift=-.4ex]current bounding box.center)}, thick, shift={(0,0)}, scale=0.7]
  \def\tate{2} 
  \def\yoko{4} 
  \def\sen{7} 
  \def\gaisen{2} 
  \draw (0,\tate) -- (\yoko,\tate);
  \draw (0,-\tate) -- (\yoko,-\tate);
  \draw (0,\tate) arc (90:270:\tate);
  \draw (\yoko,\tate) arc (90:-90:\tate);
  \filldraw[fill=white] (1,-\tate) circle(3mm);
  \draw (2,-\tate) -- (2,\tate);
  \filldraw[fill=black] (2,-\tate) circle(3mm);
\end{tikzpicture}\, ,
\nonumber \\
    y^{0}_{\alpha\beta}&=&\Lambda^L_{\beta}\cdot (d\log g_{\alpha\beta}^{T}\otimes1_{\mathsf{D}_{\beta}}-1_{\mathsf{D}_{\beta}}\otimes d\log g_{\alpha\beta})\cdot (1_{\mathsf{D}_{\beta}}\otimes d\log g_{\alpha\beta}) \cdot 1_{\mathsf{D}_\beta}
    \nonumber \\
    &=&
  \begin{tikzpicture}[line cap=round,line join=round,x=1.0cm,y=1.0cm, scale=0.25, baseline={([yshift=-.4ex]current bounding box.center)}, thick, shift={(0,0)}, scale=0.7]
  \def\tate{2} 
  \def\yoko{4} 
  \def\sen{7} 
  \def\gaisen{2} 
  \draw (0,\tate) -- (\yoko,\tate);
  \draw (0,-\tate) -- (\yoko,-\tate);
  \draw (0,\tate) arc (90:270:\tate);
  \draw (\yoko,\tate) arc (90:-90:\tate);
  \filldraw[fill=white] (1,\tate) circle(3mm);
  \filldraw[fill=white] (3,-\tate) circle(3mm);
\end{tikzpicture}
-
  \begin{tikzpicture}[line cap=round,line join=round,x=1.0cm,y=1.0cm, scale=0.25, baseline={([yshift=-.4ex]current bounding box.center)}, thick, shift={(0,0)}, scale=0.7]
  \def\tate{2} 
  \def\yoko{4} 
  \def\sen{7} 
  \def\gaisen{2} 
  \draw (0,\tate) -- (\yoko,\tate);
  \draw (0,-\tate) -- (\yoko,-\tate);
  \draw (0,\tate) arc (90:270:\tate);
  \draw (\yoko,\tate) arc (90:-90:\tate);
  \filldraw[fill=white] (1,-\tate) circle(3mm);
  \filldraw[fill=white] (3,-\tate) circle(3mm);
\end{tikzpicture}\, ,
\nonumber \\
    y_{\alpha\beta}^{1}&=&\Lambda^L_{\beta}\cdot(d\log g_{\alpha\beta}^{T}\otimes1_{\mathsf{D}_{\beta}}-1_{\mathsf{D}_{\beta}}\otimes d\log g_{\alpha\beta})\cdot T_{A_{\beta}} \cdot (1_{\mathsf{D}_\beta}\otimes d\log g_{\alpha\beta})\cdot 1_{\mathsf{D}_\beta}
    \nonumber \\
    &=&
  \begin{tikzpicture}[line cap=round,line join=round,x=1.0cm,y=1.0cm, scale=0.25, baseline={([yshift=-.4ex]current bounding box.center)}, thick, shift={(0,0)}, scale=0.7]
  \def\tate{2} 
  \def\yoko{4} 
  \def\sen{7} 
  \def\gaisen{2} 
  \draw (0,\tate) -- (\yoko,\tate);
  \draw (0,-\tate) -- (\yoko,-\tate);
  \draw (0,\tate) arc (90:270:\tate);
  \draw (\yoko,\tate) arc (90:-90:\tate);
  \filldraw[fill=white] (1,\tate) circle(3mm);
  \draw (2,-\tate) -- (2,\tate);
  \filldraw[fill=white] (3,-\tate) circle(3mm);
\end{tikzpicture}
-
  \begin{tikzpicture}[line cap=round,line join=round,x=1.0cm,y=1.0cm, scale=0.25, baseline={([yshift=-.4ex]current bounding box.center)}, thick, shift={(0,0)}, scale=0.7]
  \def\tate{2} 
  \def\yoko{4} 
  \def\sen{7} 
  \def\gaisen{2} 
  \draw (0,\tate) -- (\yoko,\tate);
  \draw (0,-\tate) -- (\yoko,-\tate);
  \draw (0,\tate) arc (90:270:\tate);
  \draw (\yoko,\tate) arc (90:-90:\tate);
  \filldraw[fill=white] (1,-\tate) circle(3mm);
  \draw (2,-\tate) -- (2,\tate);
  \filldraw[fill=white] (3,-\tate) circle(3mm);
\end{tikzpicture}\, ,
\nonumber \\
    z^{1}_{\alpha\beta}&=&d\Lambda^L_{\beta}\cdot T_{A_{\beta}} \cdot (1_{\mathsf{D}_\beta}\otimes d\log g_{\alpha\beta})\cdot 1_{\mathsf{D}_\beta}
    \nonumber \\
    &=&
  \begin{tikzpicture}[line cap=round,line join=round,x=1.0cm,y=1.0cm, scale=0.25, baseline={([yshift=-.4ex]current bounding box.center)}, thick, shift={(0,0)}, scale=0.7]
  \def\tate{2} 
  \def\yoko{4} 
  \def\sen{7} 
  \def\gaisen{2} 
  \draw (0,\tate) -- (\yoko,\tate);
  \draw (0,-\tate) -- (\yoko,-\tate);
  \draw (0,\tate) arc (90:270:\tate);
  \draw (\yoko,\tate) arc (90:-90:\tate);
  \filldraw[fill=black] (-2,0) circle(3mm);
  \draw (2,-\tate) -- (2,\tate);
  \filldraw[fill=white] (3,-\tate) circle(3mm);
\end{tikzpicture}\, .\label{eq:y0}
\end{eqnarray}
Here, 
$T[A;B]:= \sum_s A^{s*}\otimes B^s$.

By using the fixed point condition, we can show that $y^{1}_{\alpha\beta}=z^{1}_{\alpha\beta}=0$.
Thus the above calculation is  summarized as
$
(\delta b^{0})_{\alpha \beta}=
d w_{\alpha \beta}+x_{\alpha \beta}^{0}+y_{\alpha \beta}^{0}.
$
Unfortunately, this is not quite the same as Eq.\ \eqref{consistency cond 2}. 
We can however "correct" or modify the definition of
$b^{0}_{\alpha}$ properly as follows. 
From 
$(\delta b^0)_{\alpha \beta}=
d w_{\alpha \beta}+x^{0}_{\alpha \beta}+y^{0}_{\alpha\beta}$
we can easily check that $(\delta x+\delta y)_{\alpha \beta \gamma}=0$. By using the generalized Mayer-Vietoris theorem, there is a 2-form $b'^{0}_\alpha$ on $U_\alpha$ such that 
$\delta(b'^{0})_{\alpha \beta}=x^{0}_{\alpha \beta}+y^{0}_{\alpha\beta}$
\footnote{
See Prop.\ 8 of Bott-Tu.}.
We can explicitly construct ${b'}^{0}_\alpha$ by using a partition of unity. Let $\left\{\rho_\alpha\right\}$ be a partition of unity associated to the open covering $\left\{U_\alpha\right\}^7$. 
Then we define $b'^{0}_\alpha$ by
\begin{eqnarray}
    b'^{0}_\alpha:=\sum_{\alpha_0} \rho_{\alpha_0} (x^0_{\alpha_0 \alpha}+ y^0_{\alpha_0 \alpha}).
\end{eqnarray}
Therefore, by defining a modified 2-from connection as
\begin{align}
B^{(2)}_\alpha=b^0_\alpha-b'^{0}_\alpha,
\end{align}
this satisfies the consistency condition Eq.\ \eqref{consistency cond 2},
$(\delta B^{(2)})_{\alpha\beta}=d w_{\alpha\beta}^{(1)}$.
Consequently, 
we constructed
a gerbe connection for fixed point MPS
\footnote{This gerbe connection and curvature depend on the choice of a partition of unity. Therefore, they are unphysical as differential forms. 
Similar dependence appears in Ref.\ \cite{KS20-1}.}
\footnote{Although a general normal MPS is not a fixed point MPS, 
by renormalizing $n$ sites into one site, eigenvalues other than one are exponentially suppressed. 
Therefore, practically, 
by taking $n$ to be sufficiently large, it is possible to calculate 
physical observables, 
including quantized topological invariants in particular.}.

The higher Berry curvature is obtained by applying the exterior derivative to $B_{\alpha}^{(2)}$,
$H^{(3)}=dB^{(2)}_{\alpha}$,
and the integration value of 
${H^{(3)}}/{2\pi i}$ 
over a closed orientable manifold is quantized to integers.
In particular, for fixed-point MPSs and when $b'_\alpha=0$ for all patches, the higher Berry curvature form is recast into
\begin{eqnarray}
    H^{(3)}=\begin{tikzpicture}[line cap=round,line join=round,x=1.0cm,y=1.0cm, scale=0.25, baseline={([yshift=-.4ex]current bounding box.center)}, thick, shift={(0,0)}, scale=0.7]
  \def\tate{2} 
  \def\yoko{4} 
  \def\sen{7} 
  \def\gaisen{2} 
  \draw (0,\tate) -- (\yoko,\tate);
  \draw (0,-\tate) -- (\yoko,-\tate);
  \draw (0,\tate) arc (90:270:\tate);
  \draw (\yoko,\tate) arc (90:-90:\tate);
  \filldraw[fill=black] (-\tate,0) circle(3mm);
  \node[anchor=east] at (-\tate,0) 
  {$d\Lambda_{\alpha}^{L}$};
  \draw (2,-\tate) -- (2,\tate);
  \draw (2,-\tate) -- (2,\tate);
  \filldraw[fill=black] (2,-\tate) circle(3mm);
    \node[anchor=north] at (2,-\tate) 
  {$dA_{\alpha}$};
  \filldraw[fill=black] (2,\tate) circle(3mm);
    \node[anchor=south] at (2,\tate) 
  {$dA_{\alpha}$};
\end{tikzpicture}\, .
\end{eqnarray}

\paragraph{Non fixed-point MPS}

To generalize this construction for non-fixed point MPS, let's consider the following quantity as a generalization of $b^0_{\alpha}$:
\begin{align}
\label{eq:def_of_b}
    b_\alpha:&=\sum^{\infty}_{k=0}d\Lambda^L_{\alpha}\cdot (T_{A_{\alpha}})^k \cdot T[A_{\alpha};dA_{\alpha}]\cdot 1_{\mathsf{D}_{\alpha}}
    \nonumber \\
  &=
  \sum_{k=0}^{\infty}
  \begin{tikzpicture}[line cap=round,line join=round,x=1.0cm,y=1.0cm, scale=0.25, baseline={([yshift=-.4ex]current bounding box.center)}, thick, shift={(0,0)}, scale=0.7]
  \def\tate{2} 
  \def\yoko{8} 
  \def\sen{7} 
  \def\gaisen{2} 
  \draw (0,\tate) -- (\yoko,\tate);
  \draw (0,-\tate) -- (\yoko,-\tate);
  \draw (0,\tate) arc (90:270:\tate);
  \draw (\yoko,\tate) arc (90:-90:\tate);
  \filldraw[fill=black] (-\tate,0) circle(3mm);
  \node[anchor=east] at (-\tate,0) 
  {$d\Lambda_{\alpha}^{L}$};
  \draw (0,-\tate) -- (0,\tate);
  \draw (1,-\tate) -- (1,\tate);
  \node[anchor=east] at (6,0) 
  {$\cdots$};
  \draw (7,-\tate) -- (7,\tate);
  \draw (8,-\tate) -- (8,\tate);
  \filldraw[fill=black] (8,-\tate) circle(3mm);
    \node[anchor=north] at (8,-\tate) 
  {$dA_{\alpha}$};
    \node[anchor=south] at (10,\tate) 
  {${\color{white} dA_{\alpha}}$};
\end{tikzpicture}\, .
\
\end{align}
Namely, 
we insert in \eqref{2-form conn}
products of transfer matrices $T_{A_{\alpha}}$
on the left 
of $T[A_{\alpha};dA_{\alpha}]$.
We note that inserting 
$T_{A_{\alpha}}$ on the right 
of $T[A_{\alpha};dA_{\alpha}]$
has no effect since $1_{\mathsf{D}_{\alpha}}$
is the right fixed point.
One can readily check that using the fixed point condition \eqref{fixed pt}
$b_{\alpha}$ reduces to $b^0_{\alpha}$.

To see how $b_{\alpha}$
transforms under
\v{C}ech differential $\delta$,
let us first consider 
\begin{align}
    b_{\alpha}^{k}
  =
  d\Lambda_{\alpha}^{L}\cdot (T_{A_{\alpha}})^{k}\cdot T[A_{\alpha};dA_{\alpha}]\cdot 1_{\mathsf{D}_{\alpha}},
\end{align}
and note that
\begin{eqnarray}
    (\delta b^{k})_{\alpha\beta}=x_{\alpha\beta}^{k}+y_{\alpha\beta}^{k}+z_{\alpha\beta}^{k}-y_{\alpha\beta}^{k+1}-z_{\alpha\beta}^{k+1},
\end{eqnarray}
where
\begin{eqnarray}
    x_{\alpha\beta}^{k}&=&\Lambda^L_{\beta}\cdot (d\log g_{\alpha\beta}^{T}\otimes 1_{\mathsf{D}_\beta}-1_{\mathsf{D}_\beta}\otimes d\log g_{\alpha\beta})\cdot (T_{A_{\beta}})^{k}\cdot T[A_\beta;dA_\beta]\cdot 1_{\mathsf{D}_\beta}
    \nonumber \\
    &=&
    \begin{tikzpicture}[line cap=round,line join=round,x=1.0cm,y=1.0cm, scale=0.25, baseline={([yshift=-.4ex]current bounding box.center)}, thick, shift={(0,0)}, scale=0.7]
  \def\tate{2} 
  \def\yoko{8} 
  \def\sen{7} 
  \def\gaisen{2} 
  \draw (0,\tate) -- (\yoko,\tate);
  \draw (0,-\tate) -- (\yoko,-\tate);
  \draw (0,\tate) arc (90:270:\tate);
  \draw (\yoko,\tate) arc (90:-90:\tate);
  \filldraw[fill=white] (0,\tate) circle(3mm);
  \node[anchor=south] at (0,\tate) 
  {$d\log g_{\alpha\beta}^{T}$};
  \draw (1,-\tate) -- (1,\tate);
  \node[anchor=east] at (6,0) 
  {$\cdots$};
  \draw (7,-\tate) -- (7,\tate);
  \draw (8,-\tate) -- (8,\tate);
  \filldraw[fill=black] (8,-\tate) circle(3mm);
    \node[anchor=east] at (10,-\tate-1.5) 
  {$dA_{\alpha}$};
\end{tikzpicture}
-
\hspace{-4mm}
\begin{tikzpicture}[line cap=round,line join=round,x=1.0cm,y=1.0cm, scale=0.25, baseline={([yshift=-.4ex]current bounding box.center)}, thick, shift={(0,0)}, scale=0.7]
  \def\tate{2} 
  \def\yoko{8} 
  \def\sen{7} 
  \def\gaisen{2} 
  \draw (0,\tate) -- (\yoko,\tate);
  \draw (0,-\tate) -- (\yoko,-\tate);
  \draw (0,\tate) arc (90:270:\tate);
  \draw (\yoko,\tate) arc (90:-90:\tate);
  \filldraw[fill=white] (0,-\tate) circle(3mm);
  \node[anchor=north] at (0,-\tate) 
  {$d\log g_{\alpha\beta}$};
  \node[anchor=south] at (0,\tate) 
  {${\color{white} d\log g_{\alpha\beta}}$};
  \draw (1,-\tate) -- (1,\tate);
  \node[anchor=east] at (6,0) 
  {$\cdots$};
  \draw (7,-\tate) -- (7,\tate);
  \draw (8,-\tate) -- (8,\tate);
  \filldraw[fill=black] (8,-\tate) circle(3mm);
    \node[anchor=north] at (10,-\tate) 
  {$dA_{\alpha}$};
\end{tikzpicture},\\
    y^{k}_{\alpha\beta}&=&\Lambda^L_{\beta}\cdot (d\log g_{\alpha\beta}^{T}\otimes 1_{\mathsf{D}_\beta}-1_{\mathsf{D}_\beta}\otimes d\log g_{\alpha\beta})\cdot (T_{A_{\beta}})^{k}\cdot (1_{\mathsf{D}_\beta}\otimes d\log g_{\alpha\beta})\cdot 1_{\mathsf{D}_\beta}
    \nonumber \\
    &=&
    \begin{tikzpicture}[line cap=round,line join=round,x=1.0cm,y=1.0cm, scale=0.25, baseline={([yshift=-.4ex]current bounding box.center)}, thick, shift={(0,0)}, scale=0.7]
  \def\tate{2} 
  \def\yoko{8} 
  \def\sen{7} 
  \def\gaisen{2} 
  \draw (0,\tate) -- (\yoko,\tate);
  \draw (0,-\tate) -- (\yoko,-\tate);
  \draw (0,\tate) arc (90:270:\tate);
  \draw (\yoko,\tate) arc (90:-90:\tate);
  \filldraw[fill=white] (0,\tate) circle(3mm);
  \node[anchor=south] at (0,\tate) 
  {$d\log g_{\alpha\beta}^{T}$};
  \draw (1,-\tate) -- (1,\tate);
  \node[anchor=east] at (6,0) 
  {$\cdots$};
  \draw (7,-\tate) -- (7,\tate);
  \filldraw[fill=white] (8,-\tate) circle(3mm);
    \node[anchor=east] at (10,-\tate-1.5) 
  {$d\log g_{\alpha\beta}$};
\end{tikzpicture}
-
\hspace{-4mm}
\begin{tikzpicture}[line cap=round,line join=round,x=1.0cm,y=1.0cm, scale=0.25, baseline={([yshift=-.4ex]current bounding box.center)}, thick, shift={(0,0)}, scale=0.7]
  \def\tate{2} 
  \def\yoko{8} 
  \def\sen{7} 
  \def\gaisen{2} 
  \draw (0,\tate) -- (\yoko,\tate);
  \draw (0,-\tate) -- (\yoko,-\tate);
  \draw (0,\tate) arc (90:270:\tate);
  \draw (\yoko,\tate) arc (90:-90:\tate);
  \filldraw[fill=white] (0,-\tate) circle(3mm);
  \node[anchor=north] at (0,-\tate) 
  {$d\log g_{\alpha\beta}$};
  \node[anchor=south] at (0,\tate) 
  {${\color{white} d\log g_{\alpha\beta}}$};
  \draw (1,-\tate) -- (1,\tate);
  \node[anchor=east] at (6,0) 
  {$\cdots$};
  \draw (7,-\tate) -- (7,\tate);
  \filldraw[fill=white] (8,-\tate) circle(3mm);
    \node[anchor=north] at (10,-\tate) 
  {$d\log g_{\alpha\beta}$};
\end{tikzpicture},
\\
    z^{k}_{\alpha\beta}&=&d\Lambda^L_{\beta}\cdot  (T_{A_{\beta}})^{k} \cdot (1_{\mathsf{D}_\beta}\otimes d\log g_{\alpha\beta})\cdot 1_{\mathsf{D}_\beta}
   \nonumber \\
    &=&
    \begin{tikzpicture}[line cap=round,line join=round,x=1.0cm,y=1.0cm, scale=0.25, baseline={([yshift=-.4ex]current bounding box.center)}, thick, shift={(0,0)}, scale=0.7]
  \def\tate{2} 
  \def\yoko{8} 
  \def\sen{7} 
  \def\gaisen{2} 
  \draw (0,\tate) -- (\yoko,\tate);
  \draw (0,-\tate) -- (\yoko,-\tate);
  \draw (0,\tate) arc (90:270:\tate);
  \draw (\yoko,\tate) arc (90:-90:\tate);
  \filldraw[fill=black] (-\tate,0) circle(3mm);
  \node[anchor=east] at (-\tate,0) 
  {$d\Lambda_{\beta}^{L}$};
  \node[anchor=south] at (0,\tate) 
  {${\color{white} d\log g_{\alpha\beta}}$};
  \draw (1,-\tate) -- (1,\tate);
  \node[anchor=east] at (6,0) 
  {$\cdots$};
  \draw (7,-\tate) -- (7,\tate);
  \filldraw[fill=white] (8,-\tate) circle(3mm);
    \node[anchor=north] at (10,-\tate) 
  {$d\log g_{\alpha\beta}$};
\end{tikzpicture}.
\end{eqnarray}
Note that $z_{\alpha\beta}^{0}=dw_{\alpha\beta}^{(1)}$. 
Therefore, under 
$\delta$,
$
    b^{< n}_{\alpha}:=\sum_{k=0}^{n-1} b^{k}_{\alpha}
$
transforms as
\begin{eqnarray}\label{eq:gauge_behave}
    (\delta b^{<n})_{\alpha\beta}=dw_{\alpha\beta}^{0}+y^{0}_{\alpha\beta}+x_{\alpha\beta}^{<n}-y^{n}_{\alpha\beta}-z^{n}_{\alpha\beta},
\end{eqnarray}
where $x_{\alpha\beta}^{<n}:=\sum_{k=0}^{n-1}x_{\alpha\beta}^{k}$. We can show that 
$ \lim_{n\to\infty}y^{n}_{\alpha\beta}=\lim_{n\to\infty}z^{n}_{\alpha\beta}=0.
$
Therefore, by taking the limit of Eq.\ (\ref{eq:gauge_behave}),
$ b_{\alpha}:=\lim_{n\to\infty}b^{<n}_{\alpha}=\sum_{k=0}^{\infty}b_{\alpha}^{k}
$
satisfies 
\begin{eqnarray}
    (\delta b)_{\alpha\beta}=dw_{\alpha\beta}^{(1)}+y^{0}_{\alpha\beta}+x_{\alpha\beta},
\end{eqnarray}
where $x_{\alpha\beta}:=\lim_{n\to\infty}x^{<n}_{\alpha\beta}$. Similarly to the fixed point case, $y^{0}_{\alpha\beta}+x_{\alpha\beta}$ satisfies $(\delta y^{0}+\delta x)_{\alpha\beta\gamma}=0$. 
Thus, we introduce
\begin{eqnarray}
    b'_\alpha:=\sum_{\alpha_0} \rho_{\alpha_0} (x_{\alpha_0 \alpha}+ y^0_{\alpha_0 \alpha}),
\end{eqnarray}
and define the $2$-form connection as 
\begin{eqnarray}
    B_{\alpha}^{(2)}:=b_{\alpha}-b'_{\alpha}.
\end{eqnarray}
Then, this satisfies the consistency condition Eq.\ \eqref{consistency cond 2},
$
    (\delta B^{(2)})_{\alpha\beta}=d w_{\alpha\beta}^{(1)}
$.
Consequently, the data $(\{B^{(2)}_{\alpha}\},\{w_{\alpha\beta}^{(1)}\},\{c^{(0)}_{\alpha\beta\gamma}\})$ satisfies the consistency conditions,
Eqs.\ 
\eqref{consistency cond 0},
\eqref{consistency cond 1},
and
\eqref{consistency cond 2},
i.e., this is a gerbe connection on the MPS gerbe.

We can generalize this construction to a general MPS gerbe, including a non-constant rank case. 
Before moving on to that situation, we comment on the convergence of the $2$-form connection. 
To clarify the convergence of the expression \eqref{eq:def_of_b}, 
we pull out from the transfer matrix $T_{\alpha}$ 
the projector $P_\alpha$ on the space with unit eigenvalue, 
\begin{align}\label{eq:reduced_transfer}
T_\alpha=T'_{\alpha}+P_{\alpha}, \quad T^m_{\alpha}=(T'_{\alpha}+P_{\alpha})^m=T'^m_{\alpha}+P_{\alpha}
\end{align}
since $T'_{\alpha}$ and $P_{\alpha}$ are orthogonal and $P^m_{\alpha}=P_{\alpha}$.
We will refer to $T'_{\alpha}$ as a reduced transfer matrix.
By using the left and right fixed points, $P_{\alpha}$ can be written as  $P_\alpha=\Lambda^R_\alpha\otimes\Lambda_\alpha^L$. 
%
Then,
\begin{align}
\sum_{m } T^m_{\alpha}=(1-T'_\alpha)^{-1}+\sum_{m } P_{\alpha}
\end{align}
since $T'_{\alpha}$ and $P_{\alpha}$ are orthogonal and $P^m_{\alpha}=P_{\alpha}$. 
Diagrammatically, this relation 
is represented as 
%
\begin{align}
\sum_{m}\,
\begin{tikzpicture}[line cap=round,line join=round,x=1.0cm,y=1.0cm, scale=0.25, baseline={([yshift=-.4ex]current bounding box.center)}, thick, shift={(0,0)}, scale=0.7]
  \def\tate{2} 
  \def\yoko{10} 
  \def\sen{7} 
  \def\gaisen{2} 
  \draw (0,\tate) -- (\yoko,\tate);
  \draw (0,-\tate) -- (\yoko,-\tate);
  \draw (2,-\tate) -- (2,\tate);
  \draw (3,-\tate) -- (3,\tate);
  \node[anchor=east] at (7,0) 
  {$\cdots$};
  \draw (8,-\tate) -- (8,\tate);
\end{tikzpicture}
\,\,
=
\,\,
\begin{tikzpicture}[line cap=round,line join=round,x=1.0cm,y=1.0cm, scale=0.25, baseline={([yshift=-.4ex]current bounding box.center)}, thick, shift={(0,0)}, scale=0.7]
  \def\tate{2} 
  \def\yoko{11} 
  \def\sen{7} 
  \def\gaisen{2} 
  \def\siten{2}
  \pgfmathsetmacro\hako{\yoko-\siten}
  \draw (0,\tate) -- (\siten,\tate);
  \draw (\hako,\tate) -- (\yoko,\tate);
  \draw (0,-\tate) -- (\siten,-\tate);
  \draw (\hako,-\tate) -- (\yoko,-\tate);
  \draw (\siten,-\tate-1) -- (\siten,\tate+1);
  \node at (\hako/2+\siten/2,0) 
  {$\cfrac{1}{1-T'_{\alpha}}$
  };
  \draw (\hako,-\tate-1) -- (\hako,\tate+1);
  \draw (\siten,-\tate-1) -- (\hako,-\tate-1);
  \draw (\siten,\tate+1) -- (\hako,\tate+1);
\end{tikzpicture}
\, 
\,
+
\,\,
\sum_m\,  
\begin{tikzpicture}[line cap=round,line join=round,x=1.0cm,y=1.0cm, scale=0.25, baseline={([yshift=-.4ex]current bounding box.center)}, thick, shift={(0,0)}, scale=0.7]
  \def\tate{2} 
  \def\yoko{10} 
  \def\sen{7} 
  \def\gaisen{2} 
  \draw (5,\tate) arc (90:270:\tate);
  \draw (0,\tate) arc (90:-90:\tate);
\end{tikzpicture}\, .
\end{align}
(See, e.g., Ref.\ \cite{Vanderstraeten_2019}
for similar calculations.)
By using this decomposition, 
Eq.\ \eqref{eq:def_of_b} is recast into
\begin{align}
    b_\alpha&=
    d\Lambda^L_{\alpha}\cdot \frac{1}{1-T'_{\alpha}} \cdot T[A_{\alpha};dA_{\alpha}]\cdot 1_{\mathsf{D}_{\alpha}}+\sum_m d\Lambda_{\alpha}^L\cdot \Lambda^R_\alpha\otimes\Lambda_\alpha^L \cdot T[A_{\alpha};dA_{\alpha}]\cdot 1_{\mathsf{D}_{\alpha}}
    \nonumber 
    \\
    &=
    d\Lambda^L_{\alpha}\cdot \frac{1}{1-T'_{\alpha}} \cdot T[A_{\alpha};dA_{\alpha}]\cdot 1_{\mathsf{D}_{\alpha}}
    \nonumber \\
&
=\begin{tikzpicture}[line cap=round,line join=round,x=1.0cm,y=1.0cm, scale=0.25, baseline={([yshift=-.4ex]current bounding box.center)}, thick, shift={(0,0)}, scale=0.7]
  \def\tate{2} 
  \def\yoko{10} 
  \def\sen{7} 
  \def\gaisen{2} 
  \def\siten{1}
  \pgfmathsetmacro\hako{\yoko-\siten}
  \draw (0,\tate) -- (\siten,\tate);
  \draw (\hako,\tate) -- (\yoko,\tate);
  \draw (0,-\tate) -- (\siten,-\tate);
  \draw (\hako,-\tate) -- (\yoko,-\tate);
  \draw (0,\tate) arc (90:270:\tate);
  \draw (\yoko,\tate) arc (90:-90:\tate);
  \filldraw[fill=black] (-\tate,0) circle(3mm);
  \node[anchor=east] at (-\tate,0) 
   {$d\Lambda_{\alpha}^{L}$};
  \draw (\siten,-\tate-1) -- (\siten,\tate+1);
  \node at (\hako/2+\siten/2,0) 
  {$\cfrac{1}{1-T'_{\alpha}}$
  };
  \draw (\hako,-\tate-1) -- (\hako,\tate+1);
  \draw (\siten,-\tate-1) -- (\hako,-\tate-1);
  \draw (\siten,\tate+1) -- (\hako,\tate+1);
  \draw (\yoko,-\tate) -- (\yoko,\tate);
  \filldraw[fill=black] (\yoko,-\tate) circle(3mm);
  \node[anchor=west] at (\yoko+\tate/2,-\tate) 
  {$dA_{\alpha}$};
\end{tikzpicture}\, .
\end{align}
Here, we noted the relation
\begin{align}
d\Lambda_{\alpha}^L\cdot \Lambda^R_\alpha=d(\Lambda_{\alpha}^L\cdot \Lambda^R_\alpha)= 
\begin{tikzpicture}[line cap=round,line join=round,x=1.0cm,y=1.0cm, scale=0.25, baseline={([yshift=-.4ex]current bounding box.center)}, thick, shift={(0,0)}, scale=0.7]
  \def\tate{2} 
  \def\yoko{10} 
  \def\sen{7} 
  \def\gaisen{2} 
  \draw (0,\tate) arc (90:270:\tate);
  \draw (0,\tate) arc (90:-90:\tate);
 \filldraw[fill=black] (-\tate,0) circle(3mm);
    \node[anchor=east] at (-3,0) 
  {$d\Lambda^L_{\alpha}$};
\end{tikzpicture}\, 
=
0.
\end{align}
Similarly, $x_{\alpha\beta}$ can be written as 
\begin{eqnarray}\label{eq:x}
    x_{\alpha\beta}&=&\Lambda^L_{\beta}\cdot (d\log g_{\alpha\beta}^{T}\otimes 1_{\mathsf{D}_\beta}-1_{\mathsf{D}_\beta}\otimes d\log g_{\alpha\beta})\cdot \frac{1}{1-T'_{\beta}} \cdot T[A_{\beta};dA_{\beta}]\cdot 1_{\mathsf{D}_{\beta}}
    \nonumber \\
    &=&\begin{tikzpicture}[line cap=round,line join=round,x=1.0cm,y=1.0cm, scale=0.25, baseline={([yshift=-.4ex]current bounding box.center)}, thick, shift={(0,0)}, scale=0.7]
  \def\tate{2} 
  \def\yoko{10.5} 
  \def\sen{7} 
  \def\gaisen{2} 
  \def\siten{1.5}
  \pgfmathsetmacro\hako{\yoko-\siten}
  \draw (0,\tate) -- (\siten,\tate);
  \draw (\hako,\tate) -- (\yoko,\tate);
  \draw (0,-\tate) -- (\siten,-\tate);
  \draw (\hako,-\tate) -- (\yoko,-\tate);
  \draw (0,\tate) arc (90:270:\tate);
  \draw (\yoko,\tate) arc (90:-90:\tate);
  \filldraw[fill=white] (0.5,\tate) circle(3mm);
  \draw (\siten,-\tate-1) -- (\siten,\tate+1);
  \node at (\hako/2+\siten/2,0) 
  {$\cfrac{1}{1-T'_{\beta}}$
  };
  \draw (\hako,-\tate-1) -- (\hako,\tate+1);
  \draw (\siten,-\tate-1) -- (\hako,-\tate-1);
  \draw (\siten,\tate+1) -- (\hako,\tate+1);
  \draw (\yoko,-\tate) -- (\yoko,\tate);
  \filldraw[fill=black] (\yoko,-\tate) circle(3mm);
  \node[anchor=west] at (\yoko+\tate/2,-\tate) 
  {$dA_{\beta}$};
\end{tikzpicture}
\hspace{-5mm}
\,\,
-
\,\,
\begin{tikzpicture}[line cap=round,line join=round,x=1.0cm,y=1.0cm, scale=0.25, baseline={([yshift=-.4ex]current bounding box.center)}, thick, shift={(0,0)}, scale=0.7]
  \def\tate{2} 
  \def\yoko{10.5} 
  \def\sen{7} 
  \def\gaisen{2} 
  \def\siten{1.5}
  \pgfmathsetmacro\hako{\yoko-\siten}
  \draw (0,\tate) -- (\siten,\tate);
  \draw (\hako,\tate) -- (\yoko,\tate);
  \draw (0,-\tate) -- (\siten,-\tate);
  \draw (\hako,-\tate) -- (\yoko,-\tate);
  \draw (0,\tate) arc (90:270:\tate);
  \draw (\yoko,\tate) arc (90:-90:\tate);
  \filldraw[fill=white] (0.5,-\tate) circle(3mm);
  \draw (\siten,-\tate-1) -- (\siten,\tate+1);
  \node at (\hako/2+\siten/2,0) 
  {$\cfrac{1}{1-T'_{\beta}}$
  };
  \draw (\hako,-\tate-1) -- (\hako,\tate+1);
  \draw (\siten,-\tate-1) -- (\hako,-\tate-1);
  \draw (\siten,\tate+1) -- (\hako,\tate+1);
  \draw (\yoko,-\tate) -- (\yoko,\tate);
  \filldraw[fill=black] (\yoko,-\tate) circle(3mm);
  \node[anchor=west] at (\yoko+\tate/2,-\tate) 
  {$dA_{\beta}$};
\end{tikzpicture}
\end{eqnarray}
Thus $B^{(2)}_\alpha$ is a well-defined quantity.

In the above definition of 
the 2-form connection, 
we choose a particular gauge 
of the MPS representation.
While the MPS gauge transformation changes our 2-form connection, 
%
this change can be compensated by 
the 1-form gauge transformation of
the MPS gerbe connection in 
Eq.\ \eqref{eq:consistency_gerbe_3}.
We will discuss this point
in Sec.\ \ref{Gauge redundancy}.

So far, we discussed a constant-rank MPS gerbe. However, for such an MPS gerbe, the integral of the $3$-form curvature cannot be nontrivial
\cite{OR23,OTS23,XMAJMDAM21}. 
In fact, $(B^{(2)}=0,w_{\alpha\beta}^{(1)}=\tr{d\log g_{\alpha\beta}},c_{\alpha\beta\gamma}^{(0)})$ 
satisfies the consistency condition for a constant-rank 
MPS gerbe and $\int {H^{(3)}}/{2\pi i}=0$. Since the integral value does not depend on a choice of a higher connection, constant-rank MPS gerbe has a topologically trivial higher Berry curvature\footnote{Note that the torsion part of the higher Berry class can be nontrivial. See Ref.\ \cite{OTS23} for nontrivial examples.}. In the next section, we consider the case where the bond dimension is not constant. In this case, the integral value of the higher Berry curvature can be nontrivial.

\subsection{General MPS gerbe}
\label{General MPS gerbe}


In a general situation, a family of invertible states are described 
by essentially normal MPS \cite{MGSC18,OR23}. 
Here, MPS matrices 
$\{\tilde{A}^s_{\alpha}\}$
are essentially normal 
if there exists a normal MPS $\{A^{s}_{\alpha}\}$ 
such that two MPS representations 
for a one-dimensional chain of length $L$,
$\ket{\{A^s_{\alpha}\}}_{L}$ and 
$\ket{\{\tilde{A}^{s}_{\alpha}\}}_L$,
are physically the same state 
for any $L$. A typical example is 
\begin{eqnarray}
    \tilde{A}^s_\alpha=\left(\begin{array}{cc}
       A^s_\alpha  & 0 \\
       Y^s_\alpha  & 0
    \end{array}
    \right)
\end{eqnarray}
where $\{A^s_{\alpha}\}$ is normal MPS representation. 

We will refer to $\{A^s_{\alpha}\}$ as a normal part of 
$\{\tilde{A}^s_{\alpha}\}$\footnote{We will assume the right canonical condition on the normal part.}. An essentially normal MPS has a nontrivial invariant subspace in the virtual Hilbert space, i.e., there exists a projection operator $p_\alpha$ such that $p_\alpha \tilde{A}^s_{\alpha}p_\alpha=\tilde{A}^s_{\alpha}p_\alpha$. 
By replacing 
$\tilde{A}^s_{\alpha}$ with $p_\alpha \tilde{A}^s_{\alpha}p_\alpha+p_\alpha^{\perp} \tilde{A}^s_{\alpha}p_\alpha^{\perp}$, 
we can set $Y^s_{\alpha}=0$ without loss of generality. 
Here, $p_\alpha^{\perp}:=1_{\mathsf{D}_{\alpha}}-p_\alpha$. 
Thus, we will assume that 
\begin{eqnarray}\label{eq:ess_normal_0}
    \tilde{A}^s_\alpha=\left(\begin{array}{cc}
       A^s_\alpha  & 0 \\
       0  & 0
    \end{array}
    \right)
\end{eqnarray}
by taking a suitable basis. The left and right fixed points of the transfer matrix are unique, and in the basis in Eq.\ \eqref{eq:ess_normal_0}, it is given by
\begin{align}
    \tilde{\Lambda}^R_\alpha:=\left(\begin{array}{cc}
       1_{\mathsf{D}_{\alpha}}  & 0 \\
       0  & 0
    \end{array}
    \right), 
    \quad
    \tilde{\Lambda}^L_\alpha:=\left(\begin{array}{cc}
       \Lambda^L_{\alpha}  & 0 \\
       0  & 0
    \end{array}
    \right).
\end{align}
Here, $\mathsf{D}_{\alpha}$ is a bond dimension of the normal part, and $\Lambda_{\alpha}^{L}$ is the left fixed point of the normal part.

Now, using a similar discussion as in Sec.\ \ref{sec:constant}, let's construct a gerbe connection on an MPS gerbe. We consider a family of essentially normal MPS parametrized by $X$, and take an open covering $\{U_{\alpha}\}$ of $X$. At each point $x\in U_{\alpha\beta}$, we have two physically equivalent MPS matrices $\{\tilde{A}^s_{\alpha}(x)\}$ and $\{\tilde{A}^s_{\beta}(x)\}$. 
However, since the sizes of these matrices are different in general, 
we cannot define a transition function $g_{\alpha\beta}$ as it stands. 
Instead of considering the transition function, let's consider the right and left fixed point of the mixed transfer matrix:
\begin{eqnarray}
    T_{\alpha\beta}\cdot \tilde{\Lambda}_{\alpha\beta}^{R}= \tilde{\Lambda}_{\alpha\beta}^{R},
    \quad \tilde{\Lambda}_{\alpha\beta}^{R} \cdot T_{\alpha\beta}= \tilde{\Lambda}_{\alpha\beta}^{L}. 
\end{eqnarray}
The existence and uniqueness of the right and left fixed points 
of essentially normal MPS is guaranteed
\cite{OR23}. 
Explicitly, by using the basis in Eq.\ \eqref{eq:ess_normal_0}, $\tilde{\Lambda}_{\alpha\beta}^{R}$ and $\tilde{\Lambda}_{\alpha\beta}^{L}$ 
are given by
\begin{align}
    \tilde{\Lambda}_{\alpha\beta}^{R}
    &=\left(\begin{array}{cc}
       \Lambda^{R}_{\alpha\beta}  & 0 \\
       0  & 0
    \end{array}
    \right)
    =\left(\begin{array}{cc}
       g_{\alpha\beta}  & 0 \\
       0  & 0
    \end{array}
    \right),
    \nonumber \\
    \tilde{\Lambda}_{\alpha\beta}^{L}
    &=\left(\begin{array}{cc}
       \Lambda^{L}_{\alpha\beta}  & 0 \\
       0  & 0
    \end{array}
    \right)
    =\left(\begin{array}{cc}
       \Lambda_{\beta}^{L}g_{\beta\alpha}  & 0 \\
       0  & 0
    \end{array}
    \right).
\end{align}
Here, $g_{\alpha\beta}$ is the transition function of the normal part. 
Also, the $\Lambda$s without tilde are the fixed points of the transfer matrices of the normal part. 
Since $\tilde{\Lambda}^{R}_{\alpha\beta}$ behaves like a transition function, 
we will denote $\tilde{g}_{\alpha\beta}:=\tilde{\Lambda}^{R}_{\alpha\beta}$\footnote{In \cite{acuaviva2022minimal},
by extending the gauge transformation group to the closure of gauge orbits, they found that these can be connected by gauge transformations. 
It is noteworthy that $\tilde{\Lambda}_{\alpha\beta}^{R}$ is similar to the extended transition function in their formalism. We will shortly mention this similarity in Sec.\ \ref{sec:minimal}.}. 
By putting a tilde on all symbols in Sec.\ \ref{sec:constant}, 
we obtain a gerbe connection. 
Remark that since $\tilde{g}_{\alpha\beta}$ is not invertible, we cannot define the logarithmic differentiation in the usual sense. 
However, our definition is $d\log \tilde{g}_{\alpha\beta}:=\tilde{g}_{\alpha\beta}^{\dagger}d\tilde{g}_{\alpha\beta}$, and this is a well-defined quantity. We will demonstrate an explicit computation of the invariant in Sec.\ \ref{Examples}.

\subsection{Gauge redundancy}
\label{Gauge redundancy}

As we mentioned in Sec.\ \ref{sec:infinite}, an MPS representation has gauge redundancies. 
Thus, the MPS gerbe connections transform under the gauge transformation \eqref{MPS gauge} of the MPS matrices. 
Furthermore, we implicitly fixed the phase of the fixed point matrices $\Lambda^{L}_{\alpha\beta}$s. 
Thus, the MPS gerbe connections also transform under a redefinition of the phase of fixed point matrices. However, in this section, we show that the changes in the connection under these redefinitions can be absorbed as a gauge redundancy of 
gerbe connections \eqref{eq:connection_gauge}. 

First, let's consider the redefinition
\begin{eqnarray}
    \Lambda_{\alpha\beta}^{R}\mapsto \xi_{\alpha\beta}^{(0)}\Lambda_{\alpha\beta}^{R}, 
    \quad
    \Lambda_{\alpha\beta}^{L}\mapsto (\xi_{\alpha\beta}^{(0)})^{-1}\Lambda_{\alpha\beta}^{L}, 
\end{eqnarray}
for an arbitrary $U(1)$-valued function $\xi_{\alpha\beta}^{(0)}$. Note that, due to the normalization condition Eq.\ \eqref{normalization mixed fixed pt}, the phases of $\Lambda_{\alpha\beta}^{R}$ and $\Lambda_{\alpha\beta}^{L}$ rotate in opposite directions. 
Under this, the triple inner product transforms as 
\begin{eqnarray}
    c_{\alpha\beta\gamma}^{(0)}\mapsto c_{\alpha\beta\gamma}^{(0)}(\delta\xi)_{\alpha\beta\gamma},
\end{eqnarray}
and the $1$-form connection transforms as 
\begin{eqnarray}
    w_{\alpha\beta}^{(1)}\mapsto w_{\alpha\beta}^{(1)}+d\log\xi_{\alpha\beta}^{(0)}.
\end{eqnarray}
Finally, $B_{\alpha}^{(2)}$ is invariant under this transformation. This is a part of the gauge redundancy of the connection explained 
in Eqs.\ \eqref{eq:consistency_gerbe_1}-\eqref{eq:consistency_gerbe_3}.

Next, let's consider the redefinition
\begin{eqnarray}\label{eq:MPS gauge gauge}
    A^s_{\alpha}\mapsto g_{\alpha}A^s_{\alpha}g_{\alpha}^{\dagger}
\end{eqnarray}
by an arbitrary unitary matrix $g_{\alpha}$. 
Under this transformation, the $1$-form connection transforms as
\begin{eqnarray}
    w_{\alpha\beta}^{(1)}\mapsto w_{\alpha\beta}^{(1)}+(\delta\xi^{(1)})_{\alpha\beta},
\end{eqnarray}
where
\begin{eqnarray}
    \xi_{\alpha}^{(1)}=\Lambda_{\alpha}^{L}\cdot (1_{\mathsf{D}_\alpha}\otimes d\log g_{\alpha})\cdot 1_{\mathsf{D}_\alpha}.
\end{eqnarray}
On the other hand, the $2$-form connection transforms as 
\begin{eqnarray}
    B^{(2)}_{\alpha}\mapsto B^{(2)}_{\alpha}+d\xi_{\alpha}^{(1)}.
\end{eqnarray}
This can be easily checked by using the property $(\delta B)_{\alpha\beta}=dw_{\alpha\beta}^{(1)}$, since the gauge transformation Eq.\ \eqref{eq:MPS gauge gauge} is similar to the patch transformation. 
The most nontrivial part is the transformation of the connections under the gauge transformation
\begin{eqnarray}
    A^s_{\alpha}\mapsto e^{i\theta_{\alpha}}A^s_{\alpha}.
\end{eqnarray}
Under this redefinition, $w_{\alpha\beta}^{(1)}$ is obviously invariant 
since it does not contain the MPS matrices itself. 
On the other hand, $b_{\alpha}$ transforms as 
\begin{eqnarray}
    b_\alpha=\begin{tikzpicture}[line cap=round,line join=round,x=1.0cm,y=1.0cm, scale=0.25, baseline={([yshift=-.4ex]current bounding box.center)}, thick, shift={(0,0)}, scale=0.7]
  \def\tate{2} 
  \def\yoko{10} 
  \def\sen{7} 
  \def\gaisen{2} 
  \def\siten{1}
  \pgfmathsetmacro\hako{\yoko-\siten}
  \draw (0,\tate) -- (\siten,\tate);
  \draw (\hako,\tate) -- (\yoko,\tate);
  \draw (0,-\tate) -- (\siten,-\tate);
  \draw (\hako,-\tate) -- (\yoko,-\tate);
  \draw (0,\tate) arc (90:270:\tate);
  \draw (\yoko,\tate) arc (90:-90:\tate);
  \filldraw[fill=black] (-\tate,0) circle(3mm);
  \node[anchor=east] at (-\tate,0) 
   {$d\Lambda_{\alpha}^{L}$};
  \draw (\siten,-\tate-1) -- (\siten,\tate+1);
  \node at (\hako/2+\siten/2,0) 
  {$\cfrac{1}{1-T'_{\alpha}}$
  };
  \draw (\hako,-\tate-1) -- (\hako,\tate+1);
  \draw (\siten,-\tate-1) -- (\hako,-\tate-1);
  \draw (\siten,\tate+1) -- (\hako,\tate+1);
  \draw (\yoko,-\tate) -- (\yoko,\tate);
  \filldraw[fill=black] (\yoko,-\tate) circle(3mm);
  \node[anchor=west] at (\yoko+\tate/2,-\tate) 
  {$dA_{\alpha}$};
\end{tikzpicture}
    &\mapsto& b_{\alpha}+ de^{i\theta_\alpha}
    \begin{tikzpicture}[line cap=round,line join=round,x=1.0cm,y=1.0cm, scale=0.25, baseline={([yshift=-.4ex]current bounding box.center)}, thick, shift={(0,0)}, scale=0.7]
  \def\tate{2} 
  \def\yoko{10} 
  \def\sen{7} 
  \def\gaisen{2} 
  \def\siten{1}
  \pgfmathsetmacro\hako{\yoko-\siten}
  \draw (0,\tate) -- (\siten,\tate);
  \draw (\hako,\tate) -- (\yoko,\tate);
  \draw (0,-\tate) -- (\siten,-\tate);
  \draw (\hako,-\tate) -- (\yoko,-\tate);
  \draw (0,\tate) arc (90:270:\tate);
  \draw (\yoko,\tate) arc (90:-90:\tate);
  \filldraw[fill=black] (-\tate,0) circle(3mm);
  \node[anchor=east] at (-\tate,0) 
   {$d\Lambda_{\alpha}^{L}$};
  \draw (\siten,-\tate-1) -- (\siten,\tate+1);
  \node at (\hako/2+\siten/2,0) 
  {$\cfrac{1}{1-T'_{\alpha}}$
  };
  \draw (\hako,-\tate-1) -- (\hako,\tate+1);
  \draw (\siten,-\tate-1) -- (\hako,-\tate-1);
  \draw (\siten,\tate+1) -- (\hako,\tate+1);
  \draw (\yoko,-\tate) -- (\yoko,\tate);
\end{tikzpicture}
\nonumber \\
&=&b_{\alpha}+ de^{i\theta_\alpha}\begin{tikzpicture}[line cap=round,line join=round,x=1.0cm,y=1.0cm, scale=0.25, baseline={([yshift=-.4ex]current bounding box.center)}, thick, shift={(0,0)}, scale=0.7]
  \def\tate{2} 
  \def\yoko{4} 
  \def\sen{7} 
  \def\gaisen{2} 
  \draw (0,\tate) -- (\yoko,\tate);
  \draw (0,-\tate) -- (\yoko,-\tate);
  \draw (0,\tate) arc (90:270:\tate);
  \draw (\yoko,\tate) arc (90:-90:\tate);
  \filldraw[fill=black] (-\tate,0) circle(3mm);
  \node[anchor=east] at (-\tate,0) 
  {$d\Lambda_{\alpha}^{L}$};
  \node[anchor=west] at (\yoko+\tate,0) 
  {$\Lambda_{\alpha}^{R}$};
\end{tikzpicture}
\nonumber \\
&=&b_\alpha.
\end{eqnarray}
Similarly, we can show that $b'_{\alpha}$ is also invariant. Therefore, $2$-form connection $B_{\alpha}^{(2)}=b_\alpha-b'_{\alpha}$ is invariant under this transformation. Therefore, the redundancy associated with the MPS representation can be absorbed as gauge transformations of the connection.

\section{Examples}
\label{Examples}

In this section, we compute the higher Berry phase for two models. In Sec.\ \ref{sec:example1}, we compute the higher Berry curvature for the model parametrized by $X=S^3$. Since $\cohoZ{3}{S^3}\simeq\mathbb{Z}$, the higher Berry curvature can be nontrivial.
This model is proposed in \cite{XMAJMDAM21} and the non-triviality as a family is confirmed from several perspectives
\cite{XMAJMDAM21,shiozaki2023higher,QSWSPBH23}. 
The second model is obtained by deforming the Su–Schrieffer–Heeger type model parametrized by $X=S^2\times S^1$. Since $\cohoZ{3}{S^2\times S^1}\simeq\mathbb{Z}$, the higher Berry curvature can be nontrivial.  For this case, the higher Berry curvature can be regarded as a topological invariant of the Thouless pump phenomena in many-body systems.

\subsection{Example 1: $X=S^3$}
\label{sec:example1}

Let us demonstrate our formula by using a specific example introduced and discussed in
Ref.\ \cite{Wen_2023}.
It is defined by the  
following family of uniquely gapped Hamiltonian parametrized over the 3-dimensional sphere 
$S^3=\{\vec{w}=(w_1, w_2, w_3, w_4) 
| \sum_{\mu=1}^4 w_\mu^2=1\}:$
\begin{align}
\label{Wen model}
H(\vec{w})=\sum_{p \in \mathbb{Z}} H_p(\vec{w})+\sum_{p \in 2 \mathbb{Z}+1} H_{p, p+1}^{\text {odd, } \mathrm{N}}(\vec{w})+\sum_{p \in 2 \mathbb{Z}} H_{p, p+1}^{\text {even,S }}(\vec{w}),
\end{align}
where each term is defined by
\begin{align}
\begin{aligned}
H_p(\vec{w}) & =(-1)^p
\left(w_1 \sigma^1_p+w_2 
\sigma^2_p+w_3 \sigma^3_p\right), \\
H_{p, p+1}^{\text {odd,N }}(\vec{w}) & =g^{\mathrm{N}}(\vec{w}) \sum_{i=1}^3 \sigma_p^i \sigma_{p+1}^i, 
\quad 
H_{p, p+1}^{\text {even,S }}(\vec{w})  =g^{\mathrm{S}}(\vec{w}) \sum_{i=1}^3 \sigma_p^i \sigma_{p+1}^i.
\end{aligned}
\end{align}
Here, 
$\sigma^{i=1,2,3}_{p}$
represent 
the Pauli matrices 
at lattice cite $p$, and 
$g^{\mathrm{N}}(\vec{w})$ and $g^{\mathrm{S}}(\vec{w})$ are real-valued function given by
\begin{align}
g^{\mathrm{N}}(\vec{w})  = 
\begin{cases}w_4 & \left(0 \leq w_4 \leq 1\right), \\
0 & \left(-1 \leq w_4 \leq 0\right),\end{cases} 
\quad 
g^{\mathrm{S}}(\vec{w})  = 
\begin{cases}0 & \left(0 \leq w_4 \leq 1\right), \\
-w_4 & \left(-1 \leq w_4 \leq 0\right) .
\end{cases}
\end{align}
We call $\left\{\vec{w} \mid w_4 \geq 0\right\} \subset S^3$ as the North Hemisphere 
and $\left\{\vec{w} \mid w_4 \leq 0\right\} \subset S^3$ as the South Hemisphere. 
Note that the second term of the Hamiltonian 
\eqref{Wen model} is non-zero only 
in the North Hemisphere while 
the third term is non-zero only in the South Hemisphere. 
This model corresponds to half the process of Kitaev's canonical pump.

We construct explicit MPS matrices in Appendix 
\ref{Wen's model parameterized over X=S3}.
From the explicit MPS representations, 
we can show that
$b_{\alpha}'=0$
holds on all open sets in this model. Thus, we do not need to introduce a partition of unity.
We can easily check that the higher Berry curvature of 
the MPS matrices vanishes on the North hemisphere. 
Thus we compute the contribution from the South hemisphere.
The transfer matrix 
on the South hemisphere
$T_{S}(\vec{w})=\sum_{ij}\tilde{A}^{i,j\ast}_{S}(\vec{w})\otimes \tilde{A}^{i,j}_{S}(\vec{w})$ 
is given by
\begin{eqnarray}
T_{S}(\vec{w})=\frac{1}{2}\left(
\begin{array}{cccc}
  1+\sqrt{1-w_4^2} & 0 & 0 &  1-\sqrt{1-w_4^2} \\
 0 & 0 & 0 & 0 \\
 0 & 0 & 0 & 0 \\
  1+\sqrt{1-w_4^2} & 0 & 0 &  1-\sqrt{1-w_4^2} \\
\end{array}
\right)
\end{eqnarray}
The right eigenvector is $1_{2}$, and the left eigenvector satisfying 
the normalization condition Eq.\ (\ref{norm})
is given by
\begin{eqnarray}
\Lambda_{S}^{L}(\vec{w})=\frac{1}{\sqrt{2}}\begin{pmatrix}
    1+\sqrt{1-w_4^2}&0\\
    0&1-\sqrt{1-w_4^2}
\end{pmatrix}.
\end{eqnarray}
Remark that the vector representation of $1_{2}$ and $\Lambda_{S}^{L}(\vec{w})$ is
\begin{eqnarray}
1_{2}=\frac{1}{\sqrt{2}}\begin{pmatrix}
    1\\
    0\\
    0\\
    1
\end{pmatrix}, 
\quad
\Lambda_{S}^{L}(\vec{w})=\frac{1}{\sqrt{2}}\begin{pmatrix}
    1+\sqrt{1-w_4^2}\\
    0\\
    0\\
    1-\sqrt{1-w_4^2}
\end{pmatrix}.
\end{eqnarray}
Let's introduce a new coordinate 
$0\leq t\leq\pi$ such that $w_4=\cos(t)$. By an explicit calculation, we get
\begin{align}
B^{(2)}(\vec{w})= \begin{cases}0 & \text { on the north patch } U_N, \\ -\frac{1}{2} i \cos (t) \cos (\theta) d t \wedge d \phi & \text { on the south patche } U_S,\end{cases}
\end{align}
up to exact forms. 
Thus the 3-form curvature is
\begin{align}
H^{(3)}(\vec{w})
= \begin{cases}0 & \text { on the north patch } U_\alpha, \\ -\frac{1}{2} i \cos (t) \sin (\theta) d t \wedge d \theta \wedge d \phi & \text { on the south patches } U_\beta, U_\gamma \text { and } U_\delta .\end{cases}
\end{align}
Therefore, we obtain
\begin{align}
\int_{S^3} \frac{H^{(3)}}{2 \pi i}=-\frac{1}{4 \pi} \int_{\pi / 2}^\pi d t \int_0^\pi d \theta \int_0^{2 \pi} d \phi \cos (t) \sin (\theta)=1 .
\end{align}
This implies that the gerbe with the connection is non-trivial.

\subsection{Example 2: $X=S^2\times S^1$}\label{sec:example2}

Let's compute the higher Berry curvature for another model. 
To introduce the model, let us start by considering 
the following Su–Schrieffer–Heeger type Hamiltonian:
\begin{align}
\label{eq:Rice-Mele}
H(t^{(0)},t^{(1)},\mu)
&:=\sum_{i}t^{(0)}
\hat{c}^{\dagger}_{A,i}\hat{c}_{B,i} 
+t^{(0)}\hat{c}^{\dagger }_{B,i}\hat{c}^{\ }_{A,i}
+t^{(1)}\hat{c}^{\dagger }_{B,i}\hat{c}^{\ }_{A,i+1}
+t^{(1)}\hat{c}^{\dagger}_{A,i+1}\hat{c}_{B,i}
\nonumber \\
&
\quad
+
\sum_i \mu 
\hat{c}^{\dagger}_{A,i}\hat{c}^{\ }_{A,i}
-\mu \hat{c}^{\dagger}_{B,i}\hat{c}^{\ }_{B,i}
\end{align}
for $t^{(0)}, t^{(1)}, \mu\in\mathbb{R}$. This model has ${U}(1)$ symmetry 
\begin{eqnarray}
    \hat{c}_{A,i}\mapsto e^{i\phi}\hat{c}_{A,i}, 
    \quad \hat{c}_{B,i}\mapsto e^{i\phi}\hat{c}_{B,i}.
\end{eqnarray}
The continuous deformation along the path 
\begin{eqnarray}
    (t^{(0)}(t),t^{(1)}(t),\mu(t))=
    \begin{cases}
        (0,\sin(t),\cos(t))&(0\leq t\leq\pi)\\
        (-\sin(t),0,\cos(t))&(\pi\leq t\leq2\pi)\\
    \end{cases}
\end{eqnarray}
gives ${U}(1)$-charge pumping 
(see, e.g., Ref.\ \cite{HBSR20}).

By deforming this model, we construct a model parametrized by $S^2\times S^1$ with nontrivial higher Berry curvature. To introduce a parameter, we rewrite the degrees of freedom as follows:
\begin{eqnarray}
    c_{A,i}:=\hat{c}_{A,i}, 
    \quad c_{B,i}:=\hat{c}^{\dagger}_{B,i}.
\end{eqnarray}
In this notation, the Hamiltonian is written as 
\begin{align}
\label{eq:Rice-Mele}
H(t^{(0)},t^{(1)},\mu)
&:=\sum_{i}t^{(0)}c^{\dagger}_{A,i}c^{\dagger}_{B,i} 
+t^{(0)}c^{\ }_{B,i}c^{\ }_{A,i}
+t^{(1)}c^{\ }_{B,i}c^{\ }_{A,i+1}
+t^{(1)}c^{\dagger}_{A,i+1}c^{\dagger}_{B,i}
\nonumber \\
&
\quad
+
\sum_i \mu 
c^{\dagger}_{A,i}c^{\ }_{A,i}
+\mu c^{\dagger}_{B,i}c^{\ }_{B,i}
\end{align}
and the $U(1)$ symmetry becomes
\begin{eqnarray}
    c_{A,i}\mapsto e^{i\phi}c_{A,i}, 
    \quad c_{B,i}\mapsto e^{-i\phi}c_{B,i}.
\end{eqnarray}
%
Now,
let's take $\{\vec{z}\in\mathbb{C}^{2}\left.\right|\left|\vec{z}\right|=1\}\sim S^{3}$, and perform an unitary transformation
\begin{eqnarray}\label{eq:SU(2) rotation}
    \begin{pmatrix}
        c_{A,i}(\vec{z})\\
        c_{B,i}(\vec{z})
    \end{pmatrix}=
    \begin{pmatrix}
        z_{1}&z_{2}\\
        -z_{2}^{\ast}&z_{1}^{\ast}
    \end{pmatrix}
    \begin{pmatrix}
        c_{A,i}\\
        c_{B,i}
    \end{pmatrix}\Leftrightarrow
    \begin{cases}
        c_{A,i}(\vec{z})=z_{1}c_{A,i}+z_{2}c_{B,i},\\
        c_{B,i}(\vec{z})=-z_{2}^{\ast}c_{A,i}+z_{1}^{\ast}c_{B,i}.
    \end{cases}
\end{eqnarray}
Our model
$H(\vec{z},t)$ 
is obtained 
from 
$H(t):=H(t^{(0)}(t),t^{(1)}(t),\mu(t))$
by replacing
$c_{A,i}, c_{B,i}$
$\to$
$c_{A,i}(\vec{z}), c_{B,i}(\vec{z})$:
\begin{eqnarray}\label{eq:higher pump Rice-Mele}
H(\vec{z},t)&:=&\sum_{i}t^{(0)}(t)c^{\dagger}_{A,i}(\vec{z})c^{\dagger}_{B,i}(\vec{z})+t^{(0)}(t)c_{B,i}(\vec{z})c_{A,i}(\vec{z})+t^{(1)}(t)c_{B,i}(\vec{z})c_{A,i+1}(\vec{z})+t^{(1)}(t)c^{\dagger}_{A,i+1}(\vec{z})c^{\dagger}_{B,i}(\vec{z})\nonumber\\
&+&\sum_{i}\mu(t)c^{\dagger}_{A,i}(\vec{z})c_{A,i}(\vec{z})+\mu(t) c^{\dagger}_{B,i}(\vec{z})c_{B,i}(\vec{z}).
\end{eqnarray}
Remark that $c_{A,i}(z'\vec{z})=z'c_{A,i}(\vec{z})$ and $c_{B,i}(\vec{z})=z'^{\ast}c_{B,i}(\vec{z})$ for any $z'\in {U}(1)$. This is the symmetry of the original Hamiltonian 
Eq.\ \eqref{eq:Rice-Mele}, i.e.,
\begin{eqnarray}
H(z'\vec{z},t)=H(\vec{z},t),
\end{eqnarray}
for any $z'\in{U}(1)$. Therefore, $c_{A,i}(\vec{z})$ and $c_{B,i}(\vec{z})$ are parametrized by $S^{3}$, but $H(\vec{z},t)$ is parametrized by $S^{3}/S^1\times S^{1}\sim \mathbb{C}P^{1}\times S^1\sim S^{2}\times S^1$.

Let's compute the higher Berry curvature of this model. We construct explicit MPS matrices in Appendix 
\ref{A model parametrized S1xS2}\footnote{Although this model is a fermionic model, since the type of MPS is $(+)$\cite{BWHV17}, it can be regarded as a bosonic normal MPS by ignoring the $\zmod{2}$ grading.}. 
From the explicit MPS representations, 
we can show that
$b_{\alpha}'=0$
holds on all open sets in this model. Thus, we do not need to introduce a partition of unity.
In App.\ref{A model parametrized S1xS2}, we will divide $S^2\times S^1$ into two part: $U_+=S^2\times [0,\pi]$ and $U_-=S^2\times [\pi,2\pi]$. ince the transfer matrix $T_{-}(\vec{z},t)=\sum_{ij}\tilde{C}^{i,j\ast}_{-}(\vec{z},t)\otimes \tilde{C}^{i,j}_{-}(\vec{z},t)=1$ on $U_{-}$, the fixed point is trivial, and therefore the higher Berry curvature is also trivial. On the other hand, the transfer matrix  $T_{+}(\vec{z},t)=\sum_{ij}\tilde{C}^{i,j\ast}_{+}(\vec{z},t)\otimes \tilde{C}^{i,j}_{+}(\vec{z},t)$ on $U_{+}$ is given by
\begin{eqnarray}
T_{+}(\vec{z},t)=\begin{pmatrix}
    \cos^{2}(t/2)&0&0&\sin^{2}(t/2)\\
    0&0&0&0\\
    0&0&0&0\\
    \cos^{2}(t/2)&0&0&\sin^{2}(t/2)
\end{pmatrix}.
\end{eqnarray}
The right eigenvector is $1_{2}$, and the left eigenvector satisfying 
the normalization condition Eq.\ (\ref{norm}) is given by
\begin{eqnarray}
\Lambda_{+}^{L}(\vec{z},t)=\begin{pmatrix}
    \cos^{2}(t/2)&0\\
    0&\sin^{2}(t/2)
\end{pmatrix}.
\end{eqnarray}
Remark that the vector representations of $1_{2}$ and $\Lambda^L_{+}(\vec{z},t)$ is 
\begin{eqnarray}
1_{2}=\begin{pmatrix}
    1\\
    0\\
    0\\
    1
\end{pmatrix}, 
\quad
\Lambda_{+}^{L}(\vec{z},t)=\begin{pmatrix}
    \cos^{2}(t/2)\\
    0\\
    0\\
    \sin^{2}(t/2)
\end{pmatrix}.
\end{eqnarray}
By straightforward calculation,
we get 
\begin{eqnarray}
B^{(2)}(\vec{z},t)=\begin{cases}
-\frac{1}{2}i\sin(t)\sin^{2}(\theta/2)dt\wedge d\phi&(0\leq t\leq\pi),\\
0&(\pi\leq t\leq2\pi),
\end{cases}
\end{eqnarray}
up to an exact form. Thus the $3$-form curvature $H^{(3)}(\vec{z},t)$ is 
\begin{eqnarray}
H^{(3)}(\vec{z},t)=\begin{cases}
-\frac{1}{2}i\sin(t)\sin(\theta/2)\cos(\theta/2)dt\wedge d\theta\wedge d\phi&(0\leq t\leq\pi),\\
0&(\pi\leq t\leq2\pi).
\end{cases}
\end{eqnarray}
Therefore, we obtain
\begin{eqnarray}
\int_{S^{3}}\frac{H^{(3)}}{2\pi i}=-\frac{1}{4\pi}\int_{\pi/2}^{\pi}dt\int_{0}^{\pi}d\theta\int_{0}^{2\pi}d\phi\sin(t)\sin(\theta/2)\cos(\theta/2)=-1.
\end{eqnarray}
This implies that the gerbe with the connection is non-trivial.

We can regard the higher Berry curvature as a topological invariant for the Thouless pump phenomena in interacting systems. 
The original model Eq.\ \eqref{eq:Rice-Mele} has the $U(1)$ symmetry and $S^1$ parameter. 
In general, Thouless pump phenomena with a $G$-charge is classified by $\cohoZ{3}{BG\times S^1}$, where $BG$ is the classifying space of $G$. When $G=U(1)$, $BU(1)$ equals $S^\infty/U(1)$. Practically, $S^\infty$ can be regarded as a large enough dimensional sphere. This means we can introduce a sphere parameter, which is divided by $U(1)$ in the Hamiltonian. In the above construction, we introduced the $S^3$ parameter $\vec{z}$. However, due to the $U(1)$-symmetry of the model, the Hamiltonian is parameterized by $S^3/U(1)\sim S^2$. This implies that our model is parametrized by $S^3/S^1\times S^1$ in $BU(1)\times S^1$, and the higher Berry phase measures the nontriviality in $\cohoZ{3}{S^3/S^1\times S^1}\simeq\cohoZ{3}{BU(1)\times S^1}\simeq\mathbb{Z}$.


\section{Discussion}
\label{Discussion}

In this work, we undertook the task of constructing 
the higher Berry connection on an MPS gerbe.
We close by listing some open questions.



-- First and foremost,
it is important to 
apply our formula to more examples beyond the simple examples we studied.
It is interesting to apply our formalism to more "realistic" examples. 

-- Previous studies mainly focused on 
the higher Berry curvature,
$dB^{(2)}$, or its integral, 
$\int_X dB^{(2)}$. While our formula for the higher Berry connection can be used to calculate the higher Berry curvature, 
it is interesting to look for 
phenomena associated to 
the holonomy of $B^{(2)}$, rather than curvature. 
For the regular Berry phase in quantum mechanics, 
the loop integral
of the Berry connection $a^{(1)}$,
$\oint a^{(1)}$ or more precisely the Wilson loop 
$\exp [\oint a^{(1)}]$,
is a gauge invariant quantity, independent of 
the curvature $da^{(1)}$.
It appears in the Aharonov-Bohm effect 
\cite{aharonov_significance_1959}
or 
the theory of electric polarization in solids
\cite{zak_berrys_1989,
king-smith_theory_1993, resta_macroscopic_1994}.
Similarly, we could explore the role of 
$\exp \oint_{M_2} B^{(2)}$
in many-body quantum systems.

-- At the fundamental level, 
there is some room to develop our formalism further.
For example, it is good to have a better understanding 
of the role of $b_{\alpha}^{\prime}$ while it simply vanishes in the simple examples we studied.

--
Finally, it is important to establish 
a link between our higher Berry connection
and the triple inner product.
Similar to the case of the regular Berry phase,
we can consider
various ``overlaps'' of MPS.
It is convenient to use the noncommutative geometry 
notation introduced in Ref.\ \cite{OR23},
which uses the star product ($*$) and integration ($\int$).
The triple inner product 
$\int \Psi_{\alpha} * \Psi_{\beta} * \Psi_{\gamma}$
was shown to extract the Dixmier-Douady class,
$\int \Psi_{\alpha} * \Psi_{\beta} * \Psi_{\gamma}
  =
  c^{(0)}_{\alpha\beta\gamma}.
$
It is then natural to expect that
our one-form and two-form connections
can be obtained from the multi-wave function overlaps.

\begin{figure}[t]
\centering
\includegraphics[width=260pt]{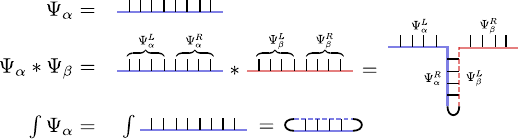}
\caption{
\label{NCG}
The noncommutative geometry notation from Ref.\ \cite{OR23}.
In this notation, 
we glue two MPS,
$\Psi_{\alpha}$ and $\Psi_{\beta}$, say,
by star operation,
$\Psi_{\alpha}*\Psi_{\beta}$.
Here, 
the right half of the first MPS
and the left half of the second MPS
are contracted.
In the integral $\int$, 
the left and right halves of MPS
are contracted.}
\end{figure}

Let us have a closer look at relevant wavefunction overlaps.
First, as a warmup, let us begin by discussing
the regular overlap of two MPS, such as $\int \Psi_{\alpha}* d\Psi_{\alpha}$.
(See Ref.\ \cite{green2016feynman} for the calculation of the regular Berry phase for MPS.)
This overlap can be computed as
\begin{align}
\includegraphics[scale=1.3]{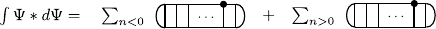}.
\end{align}
This inner product is extensive and 
hence divergent in the thermodynamic limit 
and hence ill-defined.
As we will see momentarily, 
the triple wave function overlap of three MPS,
has both divergent and non-divergent contributions.
We will also see that the non-divergent contributions are related to 
the higher Berry connection. 
When computing 
various multi-wave function overlaps,
it seems that contributions related to 
the regular inner product and the regular Berry phase are always divergent, while non-diverging contributions are relevant for 
the higher Berry connections.
We note that if we use tangent space MPS 
\cite{Vanderstraeten_2019},
such divergent contributions do not arise.
Tangent space MPS hence may allow us to focus on 
contributions relevant to the higher Berry phase.

%

Next, let us take a look
at $\int \Psi_{\alpha\beta}*d\Psi_{\alpha\beta}$.
Naively, one may interpret it
as a "regular Berry connection" associated 
to the mixed gauge MPS 
$\Psi_{\alpha\beta}$
and to the line bundle on $U_{\alpha\beta}$.
This would then give us $w^{(1)}_{\alpha\beta}$.
Explicitly, $\int \Psi_{\alpha\beta}*d\Psi_{\alpha\beta}$ can be
evaluated as
\begin{align}
    \includegraphics[scale=1.3]{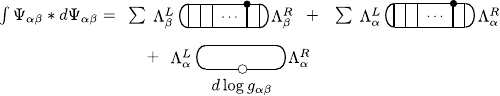}.
\end{align}
The first and second terms 
are divergent and correspond to the regular Berry connection. 
The last term is non-divergent 
and 
nothing but our one-form connection
$w^{(1)}_{\alpha\beta}$
\eqref{1-form conn}.
\begin{figure}[t]
\centering
\includegraphics[scale=1.2]{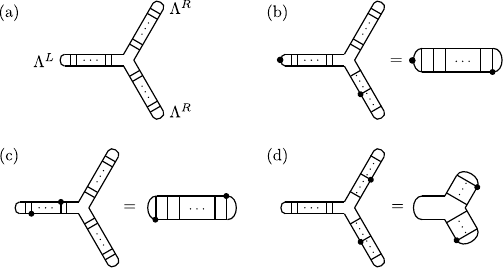}
\caption{\label{tri inner}
Various contributions (b-d) arising from the second derivative of the triple inner product (a).
Black dots represent $dA^s$ or $d\Lambda^L$.}
\end{figure}

Finally, let us discuss the triple inner product.
Here, we consider 
three MPS in the same patch but are at slightly different locations on $X$.
Taking the second derivative, 
we obtain diagrams as in Fig.\ \ref{tri inner}.
Here, we only show some representatives.
There are other diagrams that are related to the 
representatives by the identity
\eqref{deriv eig eq}.
The diagram (b) is nothing but our two-form connection 
\eqref{eq:def_of_b}.
\footnote{
Note that when we consider
triple inner products such 
as
$\int \Psi * d\Psi * d\Psi$,
$\int d\Psi * \Psi * d\Psi$
and
$\int \Psi * d\Psi * d\Psi$,
the derivative of $\Lambda^L$ does not appear directly. 
However,  
contributions of type (b) in Fig.\ \ref{tri inner}
can still arise
using the second identity
in Eq.\ \eqref{deriv eig eq}.}
We however also get different types of contributions, such as 
diagrams (c) and (d). The diagram (c) is extensive (divergent), when summed over possible positions of the derivative $dA^s$, and can be thought of as the overlap of 
two $d\Psi$. 
This contribution is an analog of $\langle d\psi| d\psi\rangle$ in quantum mechanics.
The diagram (d) is non-divergent and hence may have a physical significance.


\begin{acknowledgments}
  The authors thank Kiyonori Gomi and Ken Shiozaki for discussion and useful comments on the early stage of this project. We would also like to thank Yuya Kusuki and Bowei Liu for 
  collaboration in a related project.
  Special thanks go to Ophelia Evelyn Sommer,
  Xueda Wen, and Ashvin Vishwanath, who let us know of their related works, and also kindly agreed to coordinate submissions of our papers to arXiv.  
We thank the Yukawa Institute for Theoretical Physics at Kyoto University, where this work was initiated during the YITP-T-22-02 on "Novel Quantum States in Condensed Matter 2022".
S.O. is supported by JSPS KAKENHI Grant Number 23KJ1252 and 24K00522.
S.R. is supported 
by a Simons Investigator Grant from the Simons Foundation (Award No.~566116).
This work is supported by the Gordon and Betty Moore Foundation through Grant GBMF8685 toward the Princeton theory program. 
\end{acknowledgments}

\appendix

\section{A model parameterized over $X=S^3$}
\label{Wen's model parameterized over X=S3}


In this section, we discuss the derivation of 
the MPS representation \eqref{Wen model} 
for the model in Sec.\ \ref{sec:example1}. 
We also directly calculate the Dixmier-Douady class of the MPS gerbe, and compute a topological invariant proposed in \cite{OTS23,OR23}
(Sec.\ \ref{MPS gerbe and Dixmier-Douady class}).
This calculation of the topological invariant 
is 
expected to give  the same value as 
the calculation using the higher Berry curvature
presented in Sec.\ \ref{sec:example1}.
We will confirm the results of these calculations match.



For convenience, 
let us restate the model:
\begin{align}
H(\vec{w})=\sum_{p \in \mathbb{Z}} H_p(\vec{w})+\sum_{p \in 2 \mathbb{Z}+1} H_{p, p+1}^{\text {odd, } \mathrm{N}}(\vec{w})+\sum_{p \in 2 \mathbb{Z}} H_{p, p+1}^{\text {even,S }}(\vec{w}),
\end{align}
where $\vec{w}$ is a coordinate of $S^3=
\{\vec{w}=\left(w_1, w_2, w_3, w_4\right)| \sum_{\mu=1}^4 w_\mu^2=1\}$,
and each term is defined by
\begin{align}
&
H_p(\vec{w})  =(-1)^p\left(w_1 \sigma^1_p+w_2 \sigma^2_p+w_3 \sigma^3_p\right),
\nonumber \\
&
H_{p, p+1}^{\text {odd,N }}(\vec{w})  =g^{\mathrm{N}}(\vec{w}) \sum_{i=1}^3 \sigma_p^i \sigma_{p+1}^i,
\quad 
H_{p, p+1}^{\text {even,S }}(\vec{w})  =g^{\mathrm{S}}(\vec{w}) \sum_{i=1}^3 \sigma_p^i \sigma_{p+1}^i .
\end{align}
Here, $g^{\mathrm{N}}(\vec{w})$ and $g^{\mathrm{S}}(\vec{w})$ are real-valued functions given by
\begin{align}
g^{\mathrm{N}}(\vec{w}) 
& = \begin{cases}w_4 & \left(0 \leq w_4 \leq 1\right),  \\
0 & \left(-1 \leq w_4 \leq 0\right),\end{cases}
\quad
g^{\mathrm{S}}(\vec{w})  
= \begin{cases}
0  & \left(0 \leq w_4 \leq 1\right), \\
-w_4 & \left(-1 \leq w_4 \leq 0\right) .\end{cases}
\end{align}

\subsection{MPS representations}

Let's calculate the ground state of this model and determine the MPS representation. Since this model has rotational symmetry with respect to $(w_1, w_2, w_3)$, 
we assume $w_1=w_2=0$ without loss of generality. 
First, we regard sites $2 p-1$ and $2 p$ as unit cells and identify the ground state of the local Hamiltonian
\begin{align}
h_{p, p+1}\left(0,0, w_3, w_4\right):= \begin{cases}-w_3 \sigma_p^3+w_3 \sigma_{p+1}^3+w_4 \sum_{i=1}^3 \sigma_p^i \sigma_{p+1}^i & \left(p \in 2 \mathbb{Z}+1,0 \leq w_4 \leq 1\right), \\ w_3 \sigma_p^3-w_3 \sigma_{p+1}^3-w_4 \sum_{i=1}^3 \sigma_p^i \sigma_{p+1}^i & \left(p \in 2 \mathbb{Z}, -1 \leq w_4 \leq 0\right) .\end{cases}
\end{align}
When $0 \leq w_4 \leq 1$, the matrix representation of the local Hamiltonian 
for $p=2\mathbb{Z}+1$
is given by
\begin{align}
h_{p, p+1}\left(0,0, w_3, w_4\right)
=\left(\begin{array}{cccc}
w_4 & & & 0 \\
& -w_4-2 w_3 & 2 w_4 & \\
 & 2 w_4 & -w_4+2 w_3 & \\
0 & & & w_4
\end{array}\right),
\end{align}
with eigenvalues 
$0,0,-w_4+2$ and $-w_4-2$. 
Here, the basis of the local Hilbert space is
\begin{align}
\left(\begin{array}{c}
|\uparrow\rangle \\
|\downarrow\rangle
\end{array}\right)_p \otimes\left(\begin{array}{c}
|\uparrow\rangle \\
|\downarrow\rangle
\end{array}\right)_{p+1}=\left(\begin{array}{c}
|\uparrow \uparrow\rangle \\
|\uparrow \downarrow\rangle \\
|\downarrow \uparrow\rangle \\
|\downarrow \downarrow\rangle
\end{array}\right)_{p, p+1}
\end{align}
The minimum eigenvalue is $-w_4-2$ and its normalized eigenvector is given by
\begin{align}
&
(0, X_+, -X_-, 0)^T
\quad
\mbox{where}
\quad
X_{\pm}:= 
\frac{\sqrt{1+w_4}\pm \sqrt{1-w_4}}{2}. 
\end{align}
Note that this is an eigenvector even if $w_4=0$ and $w_4=1$. 
Thus the ground state of
$h_{p, p+1}(0,0, w_3, w_4)$
is
\begin{align}
\label{loc gs}
|\text{g.s.}\left(0,0, w_3, w_4\right)\rangle_{p, p+1}^{\mathrm{N}}:=
X_{+}
|\uparrow \downarrow\rangle_{p, p+1}
-X_-|\downarrow \uparrow\rangle_{p, p+1}
\end{align}
for any $p \in 2 \mathbb{Z}+1$. Therefore, when $0 \leq w_4 \leq 1$, 
the ground state of $H(\vec{w})$
is given by the tensor product
\begin{align}
\label{gs 1}
| \text{G.S.}(0,0, w_3, w_4)\rangle
:=\bigotimes_{p \in 2 \mathbb{Z}+1}
|\text{g.s.}(0,0, w_3, w_4)\rangle_{p, p+1}^{\mathrm{N}} .
\end{align}
For example, when $w_4=0$, 
$H(\vec{w})$ reduces to 
$H(0,0,1,0)=\sum_p(-1)^p \sigma^3_p$ and obviously, the ground state is a configuration with up arrows at odd sites and down arrows at even sites and this is consistent.

For $0 \leq w_4 \leq 1$, 
since the interactions between unit cells are trivial, the MPS representation of
Eq.\ \eqref{gs 1}
is given by
\begin{align}
A^{\uparrow \uparrow}=A^{\downarrow \downarrow}=0, \quad 
A^{\uparrow \downarrow}=
X_+,
\quad 
A^{\downarrow \uparrow}=
-X_-.
\end{align}
Note that this representation is in the canonical form of injective MPS, i.e., $\sum_{i, j=\uparrow, \downarrow} A^{i j} A^{i j}=1$.

When $-1 \leq w_4 \leq 0$, the matrix representation of the local Hamiltonian is given by
\begin{align}
h_{p, p+1}\left(0,0, w_3, w_4\right)=\left(\begin{array}{cccc}
\left(-w_4\right) & & & 0 \\
& -\left(-w_4\right)-2 w_3 & 2\left(-w_4\right) & \\
 & 2\left(-w_4\right) & -w_4+\left(-2 w_3\right) & \\
0& & & \left(-w_4\right)
\end{array}\right),
\end{align}
for $p \in 2 \mathbb{Z}$. We can obtain the local ground state by replacing $w_4$ with $-w_4$ and $w_3$ with $-w_3$ (i.e. flipping local spins;
we note that $w_3$ is nothing but a local Zeeman field at each site) 
in Eq.\ \eqref{loc gs}.
Explicitly, it is given by
\begin{align}
|\text{g.s.}\left(0,0, w_3, w_4\right)\rangle_{p, p+1}^{\mathrm{S}}:=
X_-
|\uparrow \downarrow\rangle_{p, p+1}+
X_+
|\downarrow \uparrow\rangle_{p, p+1},
\end{align}
for $p \in 2 \mathbb{Z}$. Therefore, when $-1 \leq w_4 \leq 0$, the ground state of $H(\vec{w})$ is the tensor product of these states,
\begin{align}
\label{gs 1}
| \text {G.S.}(0,0, w_3, w_4)
\rangle:=\bigotimes_{p \in 2 \mathbb{Z}} 
|\text{g.s.}\left(0,0, w_3, w_4\right)\rangle_{p, p+1}^{\mathrm{S}} .
\end{align}
To find an MPS representation of 
Eq.\ \eqref{gs 1},
let's expand the brackets of tensor products and examine the connections between unit cells:
\begin{align}
| \text{G.S.}\left(0,0, w_3, w_4\right)\rangle 
& :=\bigotimes_{p \in 2 \mathbb{Z}}\left(X_{-}|\uparrow \downarrow\rangle_{p, p+1}+X_{+}|\downarrow \uparrow\rangle_{p, p+1}\right) \nonumber \\
& =\bigotimes_{p \in 4 \mathbb{Z}}\left(X_{-}^2|\uparrow \downarrow \uparrow \downarrow\rangle+X_{-} X_{+}|\uparrow \downarrow \downarrow \uparrow\rangle+X_{+} X_{-}|\downarrow \uparrow \uparrow \downarrow\rangle+X_{+}^2|\downarrow \uparrow \downarrow \uparrow\rangle\right)_{p, p+1, p+2, p+3} .
\end{align}
From this expression, we read off the MPS matrices 
\begin{align}
A^{\uparrow \downarrow}=\left(\begin{array}{cc}
X_{+} & \\
& 0
\end{array}\right), 
\quad
A^{\downarrow \uparrow}=\left(\begin{array}{cc}
0 & \\
& X_{-}
\end{array}\right), 
\quad 
A^{\uparrow \uparrow}=\left(\begin{array}{cc}
 & X_{+} \\
0 & 
\end{array}\right), 
\quad 
A^{\downarrow \downarrow}=\left(
\begin{array}{cc}
0 & \\
 & X_{-}
\end{array}
\right) .
\end{align}
For example, when $w_4=0$, $H(\vec{w})$ reduces to $H(0,0,1,0)=\sum_p(-1)^p \sigma^3_p$ and, obviously, the ground state is given by the configuration with up arrows at odd sites and down arrows at even sites, and this is consistent. However, since the lower triangular part of the MPS is zero at $w_4=0$, the upper triangular matrix, $A^{\uparrow \uparrow}$, has no effect on the state. Consequently, the MPS is essentially represented as a $1 \times 1$ matrix. The change in the size of this matrix is essentially important for the non-triviality of the higher-order Berry phase.

We have determined the MPS representation for the case $w_1=w_2=0$. Using these results, let's find the MPS representation in the general parameter region. 
Instead of the coordinate $\left(w_1, w_2, w_3\right)$, we shall take the polar coordinate $(r, \theta, \phi)$. Since
$
\vec{w} \cdot \vec{\sigma}=e^{-i \frac{\phi}{2} \sigma^z} e^{-i \frac{\theta}{2} \sigma^y} \sigma^z e^{i \frac{\theta}{2} \sigma^y} e^{i \frac{\phi}{2} \sigma^z},
$
the Hamiltonian $H(\vec{w})$ is obtained  from $H(\vec{w})|_{w_1=w_2=0}$
by the unitary transformation, 
\begin{align}
H\left(r, \theta, \phi, w_4\right)=U(\theta, \phi) H\Big(\sqrt{1-w_4^2}, 0,0, w_4\Big) U(\theta, \phi)^{\dagger},
\end{align}
where $U(\theta, \phi)=e^{-i \frac{\phi}{2} \sigma^z} e^{-i \frac{\theta}{2} \sigma^y}$. We have already determined the MPS representation of the ground state for the case of $w_1=w_2=0$. Thus, the ground state is given by
\begin{align}
|\text{G.S.}(\vec{w})\rangle
=\sum_{\left\{i_k\right\}\left\{j_{k^{\prime}}\right\}} \operatorname{tr}\left(A^{i_1 j_1} \cdots A^{i_L j_L}\right)
\left|i_1 j_1(\theta, \phi)\right\rangle \otimes \cdots \otimes\left|i_L j_L(\theta, \phi)\right\rangle,
\end{align}
where $\left|i_1 j_1(\theta, \phi)\right\rangle=U(\theta, \phi)^{\otimes 2} \mid$ g.s. $\left.\left(r, 0,0, w_4\right)\right\rangle_{p, p+1}^{\mathrm{N} / \mathrm{S}}$. By pushing these dependences of the state on $\theta$ and $\phi$ onto the MPS matrices, we can obtain the MPS representation in the general parameter region. Explicitly, noting 
\begin{align}
U(\theta, \phi)|i\rangle= \begin{cases}e^{-i \frac{\phi}{2}} \cos \left(\frac{\theta}{2}\right)|\uparrow\rangle+e^{i \frac{\phi}{2}} \sin \left(\frac{\theta}{2}\right)|\downarrow\rangle & (i=\uparrow), \\ -e^{-i \frac{\phi}{2}} \sin \left(\frac{\theta}{2}\right)|\uparrow\rangle+e^{i \frac{\phi}{2}} \cos \left(\frac{\theta}{2}\right)|\downarrow\rangle & (i=\downarrow),\end{cases}
\end{align}
the ground state $|\text{G.S.}(\vec{w})\rangle$ can be recast into 
\begin{align}
|\mathrm{G.S.}
\left(r, \theta, \phi, w_4\right)\rangle & =\sum \operatorname{tr}\left(A^{\uparrow \uparrow} A^{i_2 j_2} \cdots A^{i_L j_L}\right)\left(e^{-} c|\uparrow\rangle+e^{+} s|\downarrow\rangle\right)\left(e^{-} c|\uparrow\rangle+e^{+} s|\downarrow\rangle\right)\left|i_2 j_2, \ldots, i_L j_L\right\rangle
\nonumber  \\
& +\sum \operatorname{tr}\left(A^{\downarrow \downarrow} A^{i_2 j_2} \cdots A^{i_L j_L}\right)\left(-e^{-} s|\uparrow\rangle+e^{+} c|\downarrow\rangle\right)\left(-e^{-} s|\uparrow\rangle+e^{+} c|\downarrow\rangle\right)\left|i_2 j_2, \ldots, i_L j_L\right\rangle
\nonumber  \\
& +\sum \operatorname{tr}\left(A^{\uparrow \downarrow} A^{i_2 j_2} \cdots A^{i_L j_L}\right)\left(e^{-} c|\uparrow\rangle+e^{+} s|\downarrow\rangle\right)\left(-e^{-} s|\uparrow\rangle+e^{+} c|\downarrow\rangle\right)\left|i_2 j_2, \ldots, i_L j_L\right\rangle
\nonumber \\
& +\sum \operatorname{tr}\left(A^{\downarrow \uparrow} A^{i_2 j_2} \cdots A^{i_L j_L}\right)\left(-e^{-} s|\uparrow\rangle+e^{+} c|\downarrow\rangle\right)\left(e^{-} c|\uparrow\rangle+e^{+} s|\downarrow\rangle\right)\left|i_2 j_2, \ldots, i_L j_L\right\rangle .
\end{align}
Here, we define $e^{ \pm}=e^{ \pm i \frac{\phi}{2}}, c=\cos \left(\frac{\theta}{2}\right)$ and $s=\sin \left(\frac{\theta}{2}\right)$. 
Therefore, MPS representation 
in the general parameter region is given by
\begin{align}
&
A^{\uparrow \uparrow}(r, \theta, \phi, w_4)
=e^{2-} c^2 A^{\uparrow \uparrow}+e^{2-} s^2 A^{\downarrow \downarrow}-e^{2-} c s A^{\uparrow \downarrow}-e^{2-} c s A^{\downarrow \uparrow}, 
\nonumber \\
&
A^{\downarrow \downarrow}
(r, \theta, \phi, w_4)=e^{2+} s^2 A^{\uparrow \uparrow}+e^{2+} c^2 A^{\downarrow \downarrow}+e^{2+} c s A^{\uparrow \downarrow}+e^{2+} c s A^{\downarrow \uparrow}, 
\nonumber \\
&
\textstyle
A^{\downarrow \uparrow}(
r, \theta, \phi, w_4)=c s A^{\uparrow \uparrow}-c s A^{\downarrow \downarrow}+c^2 A^{\uparrow \downarrow}-s^2 A^{\downarrow \uparrow}, 
\nonumber \\
&
A^{\downarrow \downarrow}
(r, \theta, \phi, w_4)=c s A^{\uparrow \uparrow}-c s A^{\downarrow \downarrow}-s^2 A^{\uparrow \downarrow}+c^2 A^{\downarrow \uparrow} .
\end{align}
Explicitly, for $0 \leq w_4 \leq 1$, the MPS matrices are given by
\begin{align}
& 
A_{\mathrm{N}}^{\uparrow \uparrow}
(r, \theta, \phi, w_4)=-e^{-i \phi} 
\cos \left(\frac{\theta}{2}\right) \sin \left(\frac{\theta}{2}\right) \sqrt{1-w_4},
\nonumber \\
&
A_{\mathrm{N}}^{\downarrow \downarrow}
(r, \theta, \phi, w_4)=e^{i \phi} \cos \left(\frac{\theta}{2}\right) \sin \left(\frac{\theta}{2}\right) \sqrt{1-w_4},
\nonumber \\
&
A_{\mathrm{N}}^{\uparrow \downarrow}
(r, \theta, \phi, w_4)=\frac{\sqrt{1+w_4}}{2}+\frac{\cos (\theta)}{2} \sqrt{1-w_4}, 
\nonumber \\
&
A_{\mathrm{N}}^{\downarrow \uparrow}
(r, \theta, \phi, w_4)=-\frac{\sqrt{1+w_4}}{2}+\frac{\cos (\theta)}{2} \sqrt{1-w_4},
\end{align}
and for $-1 \leq w_4 \leq 0$, the MPS matrices are given by
\begin{align}
& A_{\mathrm{S}}^{\uparrow \uparrow}
(r, \theta, \phi, w_4)=
e^{-i \phi}
\left(\begin{array}{cc}
- \cos \left(\frac{\theta}{2}\right) \sin \left(\frac{\theta}{2}\right) X_+ &  \cos ^2\left(\frac{\theta}{2}\right) X_- 
\nonumber \\
 \sin ^2\left(\frac{\theta}{2}\right) X_+ &
 - \cos \left(\frac{\theta}{2}\right) \sin \left(\frac{\theta}{2}\right) X_-
\end{array}\right),
\nonumber \\
& A_{\mathrm{S}}^{\downarrow \downarrow}
(r, \theta, \phi, w_4)=
e^{i \phi}
\left(\begin{array}{cc}
 \cos \left(\frac{\theta}{2}\right) \sin \left(\frac{\theta}{2}\right) X_+  &  \sin ^2\left(\frac{\theta}{2}\right) X_-  \\
 \cos ^2\left(\frac{\theta}{2}\right) X_+  &  \cos \left(\frac{\theta}{2}\right) \sin \left(\frac{\theta}{2}\right) X_- 
\end{array}\right), 
\nonumber \\
& A_{\mathrm{S}}^{\uparrow \downarrow}
(r, \theta, \phi, w_4)=\left(\begin{array}{cc}
\cos ^2\left(\frac{\theta}{2}\right) X_+ 
& \cos \left(\frac{\theta}{2}\right) \sin \left(\frac{\theta}{2}\right) X_- \\
-\cos \left(\frac{\theta}{2}\right) \sin \left(\frac{\theta}{2}\right) X_+ 
& -\sin ^2\left(\frac{\theta}{2}\right) X_-
\end{array}\right), 
\nonumber \\
& A_{\mathrm{S}}^{\downarrow \uparrow}
(r, \theta, \phi, w_4)=\left(\begin{array}{cc}
-\sin ^2\left(\frac{\theta}{2}\right) X_+ 
& \cos \left(\frac{\theta}{2}\right) \sin \left(\frac{\theta}{2}\right) X_-  \\
-\cos \left(\frac{\theta}{2}\right) \sin \left(\frac{\theta}{2}\right) X_+ 
& \cos ^2\left(\frac{\theta}{2}\right)
X_- 
\end{array}\right). 
\end{align}

\subsection{MPS gerbe and the Dixmier-Douady class}
\label{MPS gerbe and Dixmier-Douady class}

\begin{figure}[t]
\centering
\includegraphics[width=250pt]{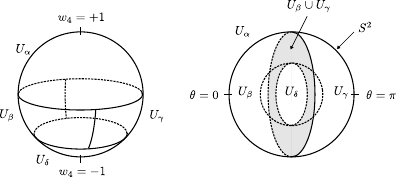}
\caption{
\label{atlas}
The atlas of $S^3$ with patches 
$U_{\alpha},
U_{\beta},
U_{\gamma},
U_{\delta}$
used in Appendix.}
\end{figure}


In this section, we construct 
an MPS gerbe 
of the model 
$H(\vec{w})$ over $S^3$ and determine its Dixmier-Douady class in 
$\mathrm{H}^3\left(S^3 ; \mathbb{Z}\right)$.
We note that a \v{C}ech representation of the Dixmier-Douady class is given by a lift of transition functions \cite{Picken03,Murray08}.
To this end, we first
note that the MPS matrices on the north hemisphere 
$\{A_{\mathrm{N}}^{i j}\}$ 
is global, i.e., parametrized by just the north hemisphere of $S^3$. 
In fact, when $w_4=1$,
$\{A_{\mathrm{N}}^{i j}\}$ is independent of $\theta$ and $\phi$, 
and when $\theta=0, \pi$,
$\{A_{\mathrm{N}}^{i j}\}$ is independent of $\phi$ for all $w_4$. 
On the other hand, the MPS matrices in the south hemisphere 
$\{A_{\mathrm{S}}^{i j}\}$ is not global. 
In fact, when $w_4=-1$,
$\{A_{\mathrm{S}}^{i j}\}$ 
is dependent on $\theta$ and $\phi$, and when $\theta=0, \pi$,
$\{A_{\mathrm{S}}^{i j}\}$ is dependent on $\phi$ for all $w_4$. 
In order to take a global gauge over each patch, it is necessary to divide the southern hemisphere into finer patches.
We consider finer patches 
$\left\{U_\alpha, U_\beta, U_\gamma, U_\delta\right\}$ in Fig.\ \ref{atlas}. 
On each patch, 
we take the gauge as follows.
\begin{itemize}
\item
On $U_\alpha$, since $A_N^{i j}$ is already global, we take
\begin{align}
A_\alpha^{i j}\left(r, \theta, \phi, w_4\right)=A_N^{i j}\left(r, \theta, \phi, w_4\right) .
\end{align}
\item 
On $U_\beta$, the $\phi$ dependence should vanish at $\theta=0$. Thus we take
\begin{align}
A_\beta^{i j}\left(r, \theta, \phi, w_4\right)=\left(\begin{array}{cc}e^{i \frac{\phi}{2}} & \\ & e^{-i \frac{\phi}{2}}\end{array}\right) A_{\mathrm{S}}^{i j}\left(r, \theta, \phi, w_4\right)\left(\begin{array}{ll}e^{-i \frac{\phi}{2}} & \\ & e^{i \frac{\phi}{2}}\end{array}\right).
\end{align}

\item 
On $U_\gamma$, the $\phi$ dependence should vanish at $\theta=\pi$. Thus we take
\begin{align}
A_\gamma^{i j}\left(r, \theta, \phi, w_4\right)=\left(\begin{array}{cc}
e^{-i \frac{\phi}{2}} & \\
& e^{i \frac{\phi}{2}}
\end{array}\right) A_{\mathrm{S}}^{i j}\left(r, \theta, \phi, w_4\right)\left(\begin{array}{ll}
e^{i \frac{\phi}{2}} & \\
& e^{-i \frac{\phi}{2}}
\end{array}\right) .
\end{align}

\item 
On $U_\delta$, the $\theta$ and $\phi$ dependence should vanish at $w_4=-1$. Thus we define
\begin{align}
U(\theta, \phi):=\left(\begin{array}{cc}
e^{i \frac{\phi}{2}} & \\
& e^{-i \frac{\phi}{2}}
\end{array}\right)\left(\begin{array}{cc}
\cos \left(\frac{\theta}{2}\right) & -\sin \left(\frac{\theta}{2}\right) \\
\sin \left(\frac{\theta}{2}\right) & \cos \left(\frac{\theta}{2}\right)
\end{array}\right),
\end{align}
and take our MPS matrices as
\begin{align}
A_\delta^{i j}\left(r, \theta, \phi, w_4\right)=U(\theta, \phi) A_{\mathrm{S}}^{i j} U(\theta, \phi)^{\dagger} .
\end{align}
\end{itemize}

Now, we take a global lift of the transition functions 
$\{g_{\alpha \beta} | A_\alpha^{i j}=g_{\alpha \beta} A_\beta^{i j} g_{\alpha \beta}^{\dagger}\}$ of this MPS gerbe. Note that on an intersection where the size of matrices changes, we take a projection of a larger transition function.
\begin{itemize}
\item
 On $U_{\alpha \beta}$, $A_\alpha^{i j}\left(r, \theta, \phi, w_4\right)$ and $A_\beta^{i j}\left(r, \theta, \phi, w_4\right)$ satisfy
 \begin{eqnarray}
     A_\alpha^{i j}\left(r, \theta, \phi, w_4\right)=
     \begin{pmatrix}
         1&0
     \end{pmatrix}
     A_\beta^{i j}\left(r, \theta, \phi, w_4\right)
     \begin{pmatrix}
         1\\
         0
     \end{pmatrix}.
 \end{eqnarray}
 Thus we take 
 \begin{align}
\tilde{g}_{\alpha \beta}=\begin{pmatrix}
         1&0
     \end{pmatrix}.
\end{align}
Here, we remind that the tilde implies the generalized transition function introduced in Sec.\ref{sec:MPSgerbe}.

\item  On $U_{\alpha \gamma}$, $A_\alpha^{i j}\left(r, \theta, \phi, w_4\right)$ and $A_\gamma^{i j}\left(r, \theta, \phi, w_4\right)$ satisfy
 \begin{eqnarray}
     A_\alpha^{i j}\left(r, \theta, \phi, w_4\right)=
     \begin{pmatrix}
         1&0
     \end{pmatrix}
     A_\gamma^{i j}\left(r, \theta, \phi, w_4\right)
     \begin{pmatrix}
         1\\
         0
     \end{pmatrix}.
 \end{eqnarray}
 Thus we take 
 \begin{align}
\tilde{g}_{\alpha \gamma}=\begin{pmatrix}
         1&0
     \end{pmatrix}.
\end{align}
Here, we remind that the tilde implies the generalized transition function introduced in Sec.\ref{sec:MPSgerbe}.

\item
 On $U_{\beta \gamma}$, we take
 \begin{align}
g_{\beta \gamma}=\left(\begin{array}{cc}
e^{i \frac{\phi}{2}} & \\
& e^{-i \frac{\phi}{2}}
\end{array}\right)\left(\begin{array}{ll}
e^{-i \frac{\phi}{2}} & \\
& e^{+i \frac{\phi}{2}}
\end{array}\right)^{\dagger}=\left(\begin{array}{ll}
e^{i \phi} & \\
& e^{-i \phi}
\end{array}\right) .
\end{align}

\item  On $U_{\beta \delta}$, we take
\begin{align}
g_{\beta \delta}=\left(\begin{array}{ll}
e^{i \frac{\phi}{2}} & \\
& e^{-i \frac{\phi}{2}}
\end{array}\right)\left\{\left(\begin{array}{cc}
e^{i \frac{\phi}{2}} & \\
& e^{-i \frac{\phi}{2}}
\end{array}\right)\left(\begin{array}{cc}
\cos \left(\frac{\theta}{2}\right) & -\sin \left(\frac{\theta}{2}\right) \\
\sin \left(\frac{\theta}{2}\right) & \cos \left(\frac{\theta}{2}\right)
\end{array}\right)\right\}^{\dagger}=\left(\begin{array}{cc}
\cos \left(\frac{\theta}{2}\right) & e^{i \phi} \sin \left(\frac{\theta}{2}\right) \\
-e^{-i \phi} \sin \left(\frac{\theta}{2}\right) & \cos \left(\frac{\theta}{2}\right)
\end{array}\right) .
\end{align}
Note that when $\theta=0, g_{\beta \delta}$ is independent of $\phi$.

\item 
On $U_{\gamma \delta}$, we take
\begin{align}
g_{\gamma \delta}=\left(\begin{array}{cc}
e^{-i \frac{\phi}{2}} & \\
& e^{i \frac{\phi}{2}}
\end{array}\right)\left\{\left(\begin{array}{cc}
e^{i \frac{\phi}{2}} & \\& e^{-i \frac{\phi}{2}}
\end{array}\right)\left(\begin{array}{cc}
\cos \left(\frac{\theta}{2}\right) & -\sin \left(\frac{\theta}{2}\right) \\
\sin \left(\frac{\theta}{2}\right) & \cos \left(\frac{\theta}{2}\right)
\end{array}\right)\right\}^{\dagger}=\left(\begin{array}{cc}
-e^{-i \phi} \cos \left(\frac{\theta}{2}\right) & \sin \left(\frac{\theta}{2}\right) \\
-\sin \left(\frac{\theta}{2}\right) & e^{i \phi} \cos \left(\frac{\theta}{2}\right)
\end{array}\right) .
\end{align}
Note that when $\theta=\pi, g_{\gamma \delta}$ is independent of $\phi$.
\end{itemize}

\begin{figure}[t]
\includegraphics[width=250pt]{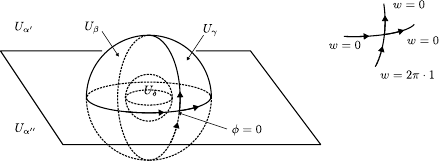}
\caption{
The calculation of the Dixmier-Douady class. The figure on the right is an enlargement around $\phi=0,\theta=\frac{\pi}{2}$. The $\mathbb{R}$-lift 
$w_{\alpha\beta\gamma\delta}$ corresponding to the horizontal and upward lines is $0$, but the 
$\mathbb{R}$-lift assigned to the line coming from below is $1$. This is a consequence of the non-trivial winding number.
\label{dd}}
\end{figure}

Finally, we identify the Dixmier-Douady class of this gerbe 
on each triple intersection.
\begin{align}
\label{eq: dd}
\tilde{g}_{\alpha \beta} p_\beta g_{\beta \gamma} p_\gamma=e^{i \phi}\cdot\begin{pmatrix}
    1,0
\end{pmatrix} =e^{i \phi} \tilde{g}_{\alpha \gamma},
\end{align}
Here, $\{p_\alpha\}$ is the projection onto the normal part.
On $U_{\beta \gamma \delta}$,
\begin{align}
g_{\beta \gamma} g_{\gamma \delta}=\left(\begin{array}{cc}
\cos \left(\frac{\theta}{2}\right) & e^{i \phi} \sin \left(\frac{\theta}{2}\right) \\
-e^{-i \phi} \sin \left(\frac{\theta}{2}\right) & \cos \left(\frac{\theta}{2}\right)
\end{array}\right)=g_{\beta \gamma} .
\end{align}
Thus $c_{\beta \gamma \delta}=1$. 
By taking a $\mathbb{R}$-lift 
$\left\{w_{\alpha \beta \gamma}\right\}$ of $\left\{c_{\alpha \beta \gamma}\right\}$,
i.e., taking $\left\{w_{\alpha \beta \gamma}\right\}$ 
such that $\left\{e^{2 \pi i w_{\alpha \beta \gamma}}=c_{\alpha \beta \gamma}\right\}$, the Dixmier-Douady class 
$\left\{d_{\alpha \beta \gamma \delta}\right\}$ is defined by $d_{\alpha \beta \gamma \delta}=(\delta w)_{\alpha \beta \gamma \delta} \in \mathbb{Z}$. In order to construct this class, we divide $U_\alpha$ into $U_{\alpha^{\prime}}$ and $U_{\alpha^{\prime \prime}}$ as in Fig.\ \ref{dd},
and take a trivial transition function on $U_{\alpha^{\prime} \alpha^{\prime \prime}}$. Then, $\left\{d_{\alpha \beta \gamma \delta}\right\}$ is only non-trivial on $U_{\alpha^{\prime} \alpha^{\prime \prime} \beta \gamma}$, and the value is $d_{\alpha^{\prime} \alpha^{\prime \prime} \beta \gamma}=1$. Therefore, $\left[d_{\alpha\beta \gamma\delta}\right]=1 \in \mathrm{H}^3(S^3 ; \mathbb{Z})\simeq\mathbb{Z}$\footnote{This quantity equivalent to the winding number 
$\int d\log c_{\alpha\beta\gamma}=1
$.}. 
This is consistent with the result in Sec.\ \ref{sec:example1}.

\subsection{Minimal canonical form}\label{sec:minimal}

For essentially normal MPSs, we can not connect them by a gauge transformation. However, 
By extending the gauge transformation group to the closure of gauge orbits, 
Ref.\ \cite{acuaviva2022minimal}
found that these can be connected by 
generalized
gauge transformations. In this section, we make a brief comment about the relation between their and our transition functions. 

First, we would like to extend the South patch by $\epsilon>0$, and make the intersection of the North and South patch open. For this purpose, we define a new MPS matrices $\bar{A}_{\mathrm{S}}^{i j}\left(r, \theta, \phi, w_4\right)$ as follows:
\begin{align}
\bar{A}_{\mathrm{S}}^{i j}\left(r, \theta, \phi, w_4\right)= \begin{cases}A_{\mathrm{S}}^{i j}\left(r, \theta, \phi, w_4\right) & \left(-1 \leq w_4 \leq 0\right), \\
\left(\begin{array}{ll}
A_{\mathrm{S}}^{i j}\left(r, \theta, \phi, w_4\right)_{00} \sqrt{1-w_4} & 0 \\
A_{\mathrm{S}}^{i j}\left(r, \theta, \phi, w_4\right)_{10} \sqrt{1-w_4} & 0
\end{array}\right) & \left(0 \leq w_4<\epsilon\right) .\end{cases}
\end{align}
Obviously, the MPS generated by $\bar{A}_{\mathrm{S}}^{i j}\left(r, \theta, \phi, w_4\right)$ on $w_4 \in[0, \epsilon)$ coincides with that of $A_{\mathrm{N}}^{i j}\left(r, \theta, \phi, w_4\right)$. Similarly, we define new MPS matrices $\bar{A}_{\mathrm{N}}^{i j}\left(r, \theta, \phi, w_4\right)$ as a restriction of $A_{\mathrm{N}}^{i j}\left(r, \theta, \phi, w_4\right)$ to $0<w_4 \leq 1$. Then, the intersection of these patches is $S^2 \times(0, \epsilon)$ and this is open.

What is the transition function? The difference between the two MPS representations lies only 
in their off-diagonal elements. 
Two such MPS give the same physical state, but cannot be connected by an ordinary gauge transformation. 

The idea proposed in \cite{acuaviva2022minimal} is to define the transition function as a convergent sequence of invertible matrices. In this case, if we take
\begin{align}
g_k=\left(\begin{array}{cc}
1 & \\
& \delta^{k}
\end{array}\right)
\end{align}
for $k \in \mathbb{N}$ and small positive number $\delta\in\mathbb{R}_{>0}$,
\begin{align}
g_k \bar{A}_{\mathrm{S}}^{i j}\left(r, \theta, \phi, w_4\right) g_k^{-1}=\left(\begin{array}{rl}
A_{\mathrm{S}}^{i j}\left(r, \theta, \phi, w_4\right)_{00} \sqrt{1-w_4} & 0 \\
\delta^{k} A_{\mathrm{S}}^{i j}\left(r, \theta, \phi, w_4\right)_{10} \sqrt{1-w_4} & 0
\end{array}\right),
\end{align}
for $0<w_4 \leq \epsilon$. Therefore,
\begin{align}
A_{\mathrm{N}}^{i j}\left(r, \theta, \phi, w_4\right)=\lim _{k \rightarrow \infty} g_k \bar{A}_{\mathrm{S}}^{i j}\left(r, \theta, \phi, w_4\right) g_k^{-1} .
\end{align}
If we use $\bar{A}_{ \mathrm{S}}^{i j}\left(r, \theta, \phi, w_4\right)$ instead of $A_{\mathrm{S}}^{i j}\left(r, \theta, \phi, w_4\right)$ and redo the calculations in Sec.\ 
\ref{MPS gerbe and Dixmier-Douady class},
$g_{\alpha \beta}$ and $g_{\alpha \gamma}$ are replaced by $g_{\alpha \beta}^{(k)}=g_{\alpha \gamma}^{(k)}=g_k$. Then, since
\begin{align}
g_{\alpha \beta}^{(k)} g_{\beta \gamma}=\left(\begin{array}{ll}
1 & \\
& \delta^{k}
\end{array}\right)\left(\begin{array}{cc}
e^{i \phi} & \\
& e^{-i \phi}
\end{array}\right)=\left(\begin{array}{cc}
e^{i \phi} & \\
& \delta^{k} e^{-i \phi}
\end{array}\right) \rightarrow\left(\begin{array}{cc}
e^{i \phi} & \\
& 0
\end{array}\right)
\end{align}
and
\begin{align}
g_{\alpha \gamma}^{(k)}=\left(\begin{array}{cc}
1 & \\
& \delta^{k}
\end{array}\right) \rightarrow\left(\begin{array}{ll}
1 & \\
& 0
\end{array}\right),
\end{align}
Eq.\ \eqref{eq: dd}
is naturally reproduced 
in the limit $k \rightarrow \infty$.

\section{A model parametrized $S^1\times S^2$}
\label{A model parametrized S1xS2}

In this section, 
we discuss the derivation of the MPS Eq.\ \eqref{eq:higher Rice-Mele mps matrices} 
for the model in Sec.\ \ref{sec:example2}. 
We also directly calculate the Dixmier-Douady class of the MPS gerbe, and compute a topological invariant proposed in \cite{OTS23,OR23} without using the higher Berry curvature. As a result, we confirm that it coincides with the invariant obtained using higher Berry curvature.

Here, we restate the model:
\begin{eqnarray}\label{eq:higher pump Rice-Mele}
H(\vec{z},t)&:=&\sum_{i}t^{(0)}(t)c^{\dagger}_{A,i}(\vec{z})c^{\dagger}_{B,i}(\vec{z})+t^{(0)}(t)c_{B,i}(\vec{z})c_{A,i}(\vec{z})+t^{(1)}(t)c_{B,i}(\vec{z})c_{A,i+1}(\vec{z})+t^{(1)}(t)c^{\dagger}_{A,i+1}(\vec{z})c^{\dagger}_{B,i}(\vec{z})\nonumber\\
&+&\sum_{i}\mu(t)c^{\dagger}_{A,i}(\vec{z})c_{A,i}(\vec{z})+\mu(t) c^{\dagger}_{B,i}(\vec{z})c_{B,i}(\vec{z}),
\end{eqnarray}
where $\vec{z}$ is a coordinate of $S^2\sim\{\vec{z}\in\mathbb{C}^{2}\left.\right|\left|\vec{z}\right|=1\}/(\vec{z}\sim z\vec{z})$ and the parameters are given by 
\begin{eqnarray}
    (t^{(0)}(t),t^{(1)}(t),\mu(t))=
    \begin{cases}
        (0,\sin(t),\cos(t))&(0\leq t\leq\pi)\\
        (-\sin(t),0,\cos(t))&(\pi\leq t\leq2\pi)\\
    \end{cases}
\end{eqnarray}
and 
\begin{eqnarray}\label{eq:SU(2) rotation}
    \begin{pmatrix}
        c_{A,i}(\vec{z})\\
        c_{B,i}(\vec{z})
    \end{pmatrix}=
    \begin{pmatrix}
        z_{1}&z_{2}\\
        -z_{2}^{\ast}&z_{1}^{\ast}
    \end{pmatrix}
    \begin{pmatrix}
        c_{A,i}\\
        c_{B,i}
    \end{pmatrix}\Leftrightarrow
    \begin{cases}
        c_{A,i}(\vec{z})=z_{1}c_{A,i}+z_{2}c_{B,i},\\
        c_{B,i}(\vec{z})=-z_{2}^{\ast}c_{A,i}+z_{1}^{\ast}c_{B,i}.
    \end{cases}
\end{eqnarray}

\subsection{MPS representations}

In Ref.\ \cite{HBSR20}, 
an MPS representation of the Hamiltonian $H(t)$ 
is given by 
\begin{eqnarray}\label{eq:Rice-Mele mps matrices}
A^0(t)=\begin{pmatrix}
    \gamma(t)&\\
    &0
\end{pmatrix},
\quad
A^1(t)=\begin{pmatrix}
    &\beta(t)\\
    -\alpha(t)&
\end{pmatrix},
\quad
B^0(t)=\begin{pmatrix}
    \gamma(t)&\\
    &0
\end{pmatrix},
\quad
B^1(t)=\begin{pmatrix}
    &\alpha(t)\\
    \beta(t)&
\end{pmatrix},
\end{eqnarray}
where
\begin{eqnarray}\label{eq:Rice-Mele path}
(\alpha(t),\beta(t),\gamma(t))=\begin{cases}
    (\sqrt{\sin(t/2)},0,\sqrt{\cos(t/2)})&(0\leq t\leq\pi),\\
    (0,\sqrt{\sin(t/2)},\sqrt{-\cos(t/2)})&(\pi\leq t\leq2\pi),
\end{cases}
\end{eqnarray}
and the ground state of $H(t)$ is given by 
\begin{eqnarray}
\ket{\mathrm{G.S.}(t)}=\sum_{\{i_{k}=0,1\},\{j_{k}=0,1\}}\tr{A^{i_1}(t)B^{j_1}(t)\cdots A^{i_L}(t)B^{j_L}(t)}\ket{i_{1}j_{1}\cdots i_{L}j_{L}}.
\end{eqnarray}
Here $\ket{00}$  and $\ket{ij}$ are defined by
\begin{eqnarray}
c_{A}\ket{00}=c_{B}\ket{00}=0,
\quad
\ket{ij}=(c^{\dagger}_{A})^{i}(c^{\dagger}_{B})^{j}\ket{00}.
\end{eqnarray}
Since Eq.\ \eqref{eq:SU(2) rotation} is a unitary transformation, the algebraic relation of $c_{A,i}(\vec{z})$ and $c_{B,i}(\vec{z})$ is the same as that of $c_{A,i}$ and $c_{B,i}$. Therefore the ground state of the Hamiltonian 
Eq.\ \eqref{eq:higher pump Rice-Mele} is given by the same MPS matrices under the basis of $c_{A,i}(\vec{z})$ and $c_{B,i}(\vec{z})$:
\begin{eqnarray}\label{eq:z dep gs}
\ket{\mathrm{G.S.}(\vec{z},t)}=\sum_{\{i_{k}=0,1\},\{j_{k}=0,1\}}\tr{A^{i_1}(t)B^{j_1}(t)\cdots A^{i_L}(t)B^{j_L}(t)}\ket{i_{1}(\vec{z})j_{1}(\vec{z})\cdots i_{L}(\vec{z})j_{L}(\vec{z})},
\end{eqnarray}
where $\ket{0(\vec{z})0(\vec{z})}$
and 
$\ket{i(\vec{z})j(\vec{z})}$
are defined by
\begin{eqnarray}
\ket{0(\vec{z})0(\vec{z})}=\ket{00},
\quad
\ket{i(\vec{z})j(\vec{z})}=(c^{\dagger}_{A}(\vec{z}))^{i}(c^{\dagger}_{B}(\vec{z}))^{j}\ket{00}.
\end{eqnarray}
More explicitly, 
\begin{eqnarray}\label{eq:z dep basis}
\ket{0(\vec{z})0(\vec{z})}&=&\ket{00},
\nonumber \\
\ket{1(\vec{z})0(\vec{z})}&=&z_{1}^{\ast}\ket{10}+z_{2}^{\ast}\ket{01},
\nonumber \\
\ket{0(\vec{z})0(\vec{z})}&=&-z_{2}\ket{10}+z_{1}\ket{01},
\nonumber \\
\ket{1(\vec{z})1(\vec{z})}&=&\ket{11}.
\end{eqnarray}
By substituting 
Eq.\ \eqref{eq:z dep basis} in Eq.\ \eqref{eq:z dep gs}, 
we can easily check that
\begin{eqnarray}\label{eq:z dep gs 2}
\ket{\mathrm{G.S.}(\vec{z},t)}=\sum_{\{i_{k}=0,1\},\{j_{k}=0,1\}}\tr{C^{i_1,j_1}(\vec{z},t)\cdots C^{i_L,j_L}(\vec{z},t)}\ket{i_{1}j_{1}\cdots i_{L}j_{L}},
\end{eqnarray}
where
\begin{eqnarray}\label{eq:higher Rice-Mele mps matrices}
C^{0,0}(\vec{z},t)&=&A^{0}(t)B^{0}(t),
\nonumber \\
C^{1,0}(\vec{z},t)&=&z_{1}^{\ast}A^{1}(t)B^{0}(t)-z_{2}A^{0}(t)B^{1}(t),
\nonumber \\
C^{0,1}(\vec{z},t)&=&z_{2}^{\ast}A^{1}(t)B^{0}(t)+z_{1}A^{0}(t)B^{1}(t),
\nonumber \\
C^{1,1}(\vec{z},t)&=&A^{1}(t)B^{1}(t).
\end{eqnarray}
Unfortunately, these matrices are not in the canonical form. In the following, we compute the matrices explicitly. 

Let $C^{i,j}_{+}(\vec{z},t)$ be the MPS matrices in $0\leq t\leq\pi$. By substituting 
Eqs.\ \eqref{eq:Rice-Mele mps matrices} 
and \eqref{eq:Rice-Mele path} in 
Eq.\ \eqref{eq:higher Rice-Mele mps matrices}, we obtain that
\begin{eqnarray}\label{eq:Rice-Mele mps matrices plus}
C^{0,0}_{+}(\vec{z},t)&=&\begin{pmatrix}
    \cos(t/2)&\\
    &0
\end{pmatrix},
\nonumber \\
C^{1,0}_{+}(\vec{z},t)&=&\begin{pmatrix}
    &z_{2}\sqrt{\sin(t/2)\cos(t/2)}\\
    -z_{1}^{\ast}\sqrt{\sin(t/2)\cos(t/2)}&
\end{pmatrix},
\nonumber \\
C^{0,1}_{+}(\vec{z},t)&=&\begin{pmatrix}
    &-z_{1}\sqrt{\sin(t/2)\cos(t/2)}\\
    -z_{2}^{\ast}\sqrt{\sin(t/2)\cos(t/2)}&
\end{pmatrix},
\nonumber \\
C^{1,1}_{+}(\vec{z},t)&=&\begin{pmatrix}
    0&\\
    &\sin(t/2)
\end{pmatrix}.
\end{eqnarray}
To convert Eq.\ \eqref{eq:Rice-Mele mps matrices plus} into a canonical form, take
\begin{eqnarray}
X(t):=\begin{pmatrix}
    \sqrt{\sin(t/2)}&\\
    &\sqrt{\cos(t/2)}
\end{pmatrix}
\end{eqnarray}
and perform a similar transformation
\begin{eqnarray}
C^{i,j}_{+}(\vec{z},t)\mapsto \tilde{C}^{i,j}_{+}:=XC^{i,j}_{+}X^{-1}.
\end{eqnarray}
Note that $\tilde{C}^{i,j}_{+}$ satisfies 
\begin{eqnarray}
\sum_{i}\tilde{C}^{i,j}_{+}(\vec{z},t)(\tilde{C}^{i,j}_{+}(\vec{z},t))^{\dagger}=1_{2},
\end{eqnarray}
i.e., $\tilde{C}^{i,j}_{+}$ is in the right canonical form\footnote{$X(t)^{-1}$ is ill-defined at $t=0,\pi$ but $C^{i,j}_{+}(\vec{z},t)$ gives correct ground state of $H(\vec{z},t)$.}. 
Consequently, the MPS matrices in the right canonical form are given by
\begin{eqnarray}\label{eq:Rice-Mele mps matrices plus, canonical}
\tilde{C}^{0,0}_{+}(\vec{z},t)&=&\begin{pmatrix}
    \cos(t/2)&\\
    &0
\end{pmatrix},
\nonumber \\
\tilde{C}^{1,0}_{+}(\vec{z},t)&=&\begin{pmatrix}
    &z_{2}\sin(t/2)\\
    -z_{1}^{\ast}\cos(t/2)&
\end{pmatrix},
\nonumber \\
\tilde{C}^{0,1}_{+}(\vec{z},t)&=&\begin{pmatrix}
    &-z_{1}\sin(t/2)\\
    -z_{2}^{\ast}\cos(t/2)&
\end{pmatrix},
\nonumber \\
\tilde{C}^{1,1}_{+}(\vec{z},t)&=&\begin{pmatrix}
    0&\\
    &\sin(t/2)
\end{pmatrix}.
\end{eqnarray}
We use two patches to cover the $S^2$. We take $z_{2}\in\mathbb{R}$ in the North hemisphere, and take $z_{1}\in\mathbb{R}$ in the South hemisphere:
\begin{eqnarray}
(z_{1},z_{2})=\begin{cases}
    (e^{i\phi}\sin(\theta/2),\cos(\theta/2))&(0\leq\theta<\pi, 0\leq\phi<2\pi),\\
    (\sin(\theta/2),e^{-i\phi}\cos(\theta/2))&(0<\theta\leq\pi, 0\leq\phi<2\pi).
\end{cases}
\end{eqnarray}
Let $\tilde{C}^{i,j}_{+,N}(\vec{z},t)$ and $\tilde{C}^{i,j}_{+,S}(\vec{z},t)$ be the MPS matrices on the North and South hemisphere,
\begin{eqnarray}\label{eq:Rice-Mele mps matrices plus, canonical, north}
\tilde{C}^{0,0}_{+,N}(\vec{z},t)&=&\begin{pmatrix}
    \cos(t/2)&\\
    &0
\end{pmatrix},\\
\tilde{C}^{1,0}_{+,N}(\vec{z},t)&=&\begin{pmatrix}
    &\cos(\theta/2)\sin(t/2)\\
    -e^{-i\phi}\sin(\theta/2)\cos(t/2)&
\end{pmatrix},\\
\tilde{C}^{0,1}_{+,N}(\vec{z},t)&=&\begin{pmatrix}
    &-e^{i\phi}\sin(\theta/2)\sin(t/2)\\
    -\cos(\theta/2)\cos(t/2)&
\end{pmatrix},\\
\tilde{C}^{1,1}_{+,N}(\vec{z},t)&=&\begin{pmatrix}
    0&\\
    &\sin(t/2)
\end{pmatrix},
\end{eqnarray}
and 
\begin{eqnarray}\label{eq:Rice-Mele mps matrices plus, canonical, south}
\tilde{C}^{0,0}_{+,S}(\vec{z},t)&=&\begin{pmatrix}
    \cos(t/2)&\\
    &0
\end{pmatrix},\\
\tilde{C}^{1,0}_{+,S}(\vec{z},t)&=&\begin{pmatrix}
    &e^{-i\phi}\cos(\theta/2)\sin(t/2)\\
    -\sin(\theta/2)\cos(t/2)&
\end{pmatrix},\\
\tilde{C}^{0,1}_{+,S}(\vec{z},t)&=&\begin{pmatrix}
    &-\sin(\theta/2)\sin(t/2)\\
    -e^{i\phi}\cos(\theta/2)\cos(t/2)&
\end{pmatrix},\\
\tilde{C}^{1,1}_{+,S}(\vec{z},t)&=&\begin{pmatrix}
    0&\\
    &\sin(t/2)
\end{pmatrix}.
\end{eqnarray}
Remark that
\begin{eqnarray}
\tilde{C}^{i,j}_{+,N}(\vec{z},t)=\begin{pmatrix}
    1&\\
    &e^{-i\phi}
\end{pmatrix}
\tilde{C}^{i,j}_{+,S}(\vec{z},t)\begin{pmatrix}
    1&\\
    &e^{i\phi}
\end{pmatrix}
\end{eqnarray}
on $0<\theta<\pi$.

Let $C^{i,j}_{-}(\vec{z},t)$ be the MPS matrices in $\pi\leq t\leq 2\pi$. 
By substituting Eqs.\ \eqref{eq:Rice-Mele mps matrices} and 
\eqref{eq:Rice-Mele path} 
in Eq.\ \eqref{eq:higher Rice-Mele mps matrices}, 
we obtain 
\begin{eqnarray}\label{eq:Rice-Mele mps matrices minus}
C^{0,0}_{-}(\vec{z},t)=\begin{pmatrix}
    -\cos(t/2)&\\
    &0
\end{pmatrix},
\quad
C^{1,0}_{-}(\vec{z},t)=C^{0,1}_{-}(\vec{z},t)=0_2,
\quad
C^{1,1}_{-}(\vec{z},t)=\begin{pmatrix}
    -\sin(t/2)&\\
    &0
\end{pmatrix}.
\end{eqnarray}
These matrices are in the canonical form.

\subsection{MPS gerbe and the Dixmier-Douady class}

\begin{figure}[t]
\centering
\includegraphics[width=100pt]{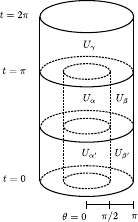}
\caption{
\label{atlas2}
The atlas of $S^1\times S^2$ with patches 
$U_{\alpha},
U_{\beta},
U_{\gamma},
$
used in Appendix. In the vertical direction, a periodic boundary condition is imposed, representing the $S^1$ direction. The boundary of the horizontal circle is compactified to one point, representing the $S^2$ direction.}
\end{figure}

In this section, we construct an MPS bundle of the model 
$H(\vec{w})$ over $S^1\times S^2$ and determine its Dixmier-Douady class in 
$\mathrm{H}^3\left(S^1\times S^2 ; \mathbb{Z}\right)$.
We note that
a \v{C}ech representation of the Dixmier-Douady class is given by a lift of transition functions. 
To this end, we first
note that the MPS matrices on the north hemisphere 
$\{A_{\mathrm{N}}^{i j}\}$ 
is not global for $0\leq t\leq\pi$. 
In order to take a global gauge,
we consider finer patches 
$\left\{U_\alpha, U_\beta, U_\gamma \right\}$ in Fig.\ \ref{atlas2}. 
On each patch, 
we take the gauge as follows.

\begin{itemize}
\item
On $U_\alpha$ and $U_{\alpha^{\prime}}$, the $\phi$ dependence should vanish at $\theta=0$. Thus we take
$
(z_{1},z_{2})=(e^{i\phi}\sin(\theta/2),\cos(\theta/2)).
$
Then, the MPS representation is
\begin{eqnarray}\label{eq:Rice-Mele mps matrices plus, canonical, north}
C^{i,j}_{\alpha}(\vec{z},t)=C^{i,j}_{\alpha^{\prime}}(\vec{z},t)=\tilde{C}^{i,j}_{+,N}(\vec{z},t).
\end{eqnarray}
\item 
On $U_\beta$ and $U_{\beta^{\prime}}$, the $\phi$ dependence should vanish at $\theta=\pi$. Thus we take
$
(z_{1},z_{2})=(\sin(\theta/2),e^{-i\phi}\cos(\theta/2)).
$
Then, the MPS representation is
\begin{eqnarray}\label{eq:Rice-Mele mps matrices plus, canonical, south}
C^{i,j}_{\beta}(\vec{z},t)=C^{i,j}_{\beta^{\prime}}(\vec{z},t)=\tilde{C}^{i,j}_{+,S}(\vec{z},t)
\end{eqnarray}
\item 
On $U_\gamma$, since $C_-^{i j}\left(\vec{z},t\right)$ is already global, we take
\begin{align}
C_\gamma^{i j}\left(\vec{z},t\right)=C_-^{i j}\left(\vec{z},t\right).
\end{align}
\end{itemize}

Now, we take a global lift of the transition functions 
$\{g_{\alpha \beta}  | C_\alpha^{i j}=g_{\alpha \beta} C_\beta^{i j} g_{\alpha \beta}^{\dagger}\}$ of this MPS gerbe. Note that on an intersection where the size of matrices changes, we take a projection of a larger transition function.
\begin{itemize}
\item
 On $U_{\alpha \beta}$, $C_\alpha^{i j}(\vec{z},t)$ and $C_\beta^{i j}(\vec{z},t)$ satisfy
\begin{eqnarray}
C^{i,j}_{\alpha}(\vec{z},t)=\begin{pmatrix}
    1&\\
    &e^{-i\phi}
\end{pmatrix}
C^{i,j}_{\beta}(\vec{z},t)\begin{pmatrix}
    1&\\
    &e^{i\phi}
\end{pmatrix}
\end{eqnarray}
on $0<\theta<\pi$\footnote{
The original model $H(t)$
has ${U}(1)$ symmetry, and the unitary matrix $\begin{pmatrix}
    1&\\
    &e^{-i\phi}
\end{pmatrix}$ is the symmetry of the MPS matrices.}.
 Thus we take 
 \begin{align}
g_{\alpha \beta}=\begin{pmatrix}
    1&\\
    &e^{-i\phi}
\end{pmatrix}
\end{align}
as a lift of the transition function.
\item
 On $U_{\beta^{\prime}\gamma}$, $C_{\beta^{\prime}}^{i j}(\vec{z},t)$ and $C_\gamma^{i j}(\vec{z},t)$ satisfy
\begin{eqnarray}
C^{i,j}_{\beta^{\prime}}(\vec{z},t)=\begin{pmatrix}
    1\\
    0
\end{pmatrix}
C^{i,j}_{\gamma}(\vec{z},t)\begin{pmatrix}
    1&0
\end{pmatrix}.
\end{eqnarray}
 Thus we take 
 \begin{align}
g_{\beta^{\prime}\gamma}=\begin{pmatrix}
    1\\
    0
\end{pmatrix}
\end{align}
as a lift of the transition function.
\item
 On $U_{\alpha^{\prime} \gamma}$, $C_{\alpha^{\prime}}^{i j}(\vec{z},t)$ and $C_\gamma^{i j}(\vec{z},t)$ satisfy
\begin{eqnarray}
C^{i,j}_{\alpha^{\prime}}(\vec{z},t)=\begin{pmatrix}
    1\\
    0
\end{pmatrix}
C^{i,j}_{\gamma}(\vec{z},t)\begin{pmatrix}
    1&0
\end{pmatrix}
\end{eqnarray}
 Thus we take 
 \begin{align}
g_{\alpha^{\prime} \gamma}=\begin{pmatrix}
    1\\
    0
\end{pmatrix}
\end{align}
as a lift of the transition function.
\item
 On $U_{\beta\gamma}$, $C_\beta^{i j}(\vec{z},t)$ and $C_\gamma^{i j}(\vec{z},t)$ satisfy
\begin{eqnarray}
p_\beta C^{i,j}_{\beta}(\vec{z},t)p_\beta=\begin{pmatrix}
    0\\
    1
\end{pmatrix}
C^{i,j}_{\gamma}(\vec{z},t)\begin{pmatrix}
    0&1
\end{pmatrix}.
\end{eqnarray}
 Thus we take 
 \begin{align}
g_{\beta\gamma}=\begin{pmatrix}
    0\\
    1
\end{pmatrix}
\end{align}
as a lift of the transition function.
\item
 On $U_{\alpha\gamma}$, $C_\alpha^{i j}(\vec{z},t)$ and $C_\gamma^{i j}(\vec{z},t)$ satisfy
\begin{eqnarray}
p_\alpha C^{i,j}_{\alpha}(\vec{z},t)p_\alpha=\begin{pmatrix}
    0\\
    1
\end{pmatrix}
C^{i,j}_{\gamma}(\vec{z},t)\begin{pmatrix}
    0&1
\end{pmatrix}.
\end{eqnarray}
 Thus we take 
 \begin{align}
g_{\alpha\gamma}=\begin{pmatrix}
    0\\
    1
\end{pmatrix}
\end{align}
as a lift of the transition function.
\item On $U_{\alpha\alpha^{\prime}}$ and $U_{\beta\beta^{\prime}}$, the MPS matrices are glued trivially. Thus we can take
\begin{eqnarray}
    g_{\alpha\alpha^{\prime}}=g_{\beta\beta^{\prime}}=g_{\alpha\beta^{\prime}}=g_{\alpha^\prime\beta}=1_2.
\end{eqnarray}
\end{itemize}

\begin{figure}[t]
\includegraphics[width=250pt]{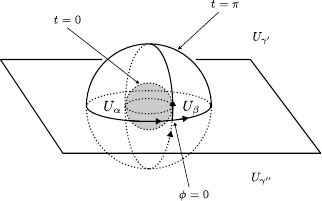}
\caption{
The calculation of the Dixmier-Douady class. The gray area represents the holes, and the surface represents the constant plane at $t=0$. The outer spherical shell represents the constant plane at $t=\pi$, and outside of it, the patch $U_\gamma$ extends. Although not depicted in the figure, there is an even larger spherical shell outside that corresponds to $t=2\pi$, and reflecting the periodicity in the $t$ direction, this shell connecting to the inner spherical shell of the hole.
\label{dd2}}
\end{figure}

Finally, we identify the Dixmier-Douady class of this bundle. On $U_{\alpha^{\prime} \beta^{\prime} \gamma}$,
\begin{align}
\label{eq: dd2}
g_{\alpha^{\prime} \beta^{\prime}} g_{\beta^{\prime} \gamma} = g_{\alpha^{\prime} \gamma}.
\end{align}
Thus $c_{\alpha^{\prime} \beta^{\prime} \gamma}=1$. 
On $U_{\alpha \beta\gamma}$ around $t=\pi$,
\begin{align}
p_\alpha g_{\alpha \beta}p_\beta \tilde{g}_{\beta \gamma} = e^{-i\phi}\tilde{g}_{\alpha \gamma}.
\end{align}
Here, $\{p_\alpha\}$ is the projection onto the normal mart. Thus $c_{\alpha \beta \gamma}=e^{-i\phi}$ on $U_{\alpha\beta\gamma}$. 
By taking a $\mathbb{R}$-lift 
$\left\{w_{\alpha \beta \gamma}\right\}$ of $\left\{c_{\alpha \beta \gamma}\right\}$,
i.e., taking $\left\{w_{\alpha \beta \gamma}\right\}$ 
such that $\left\{e^{2 \pi i w_{\alpha \beta \gamma}}=c_{\alpha \beta \gamma}\right\}$, the Dixmier-Douady class 
$\left\{d_{\alpha \beta \gamma \delta}\right\}$ is defined by $d_{\alpha \beta \gamma \delta}=(\delta w)_{\alpha \beta \gamma \delta} \in \mathbb{Z}$. In order to construct this class, we divide $U_\gamma$ into $U_{\gamma^{\prime}}$ and $U_{\gamma^{\prime \prime}}$ as in Fig.\ref{dd2}.
and take a trivial transition function on $U_{\gamma^{\prime} \gamma^{\prime \prime}}$. Then, $\left\{d_{\alpha \beta \gamma \delta}\right\}$ is only non-trivial on $U_{\alpha \beta \gamma^{\prime} \gamma^{\prime \prime}}$, and the value is $d_{\alpha \beta \gamma^{\prime} \gamma^{\prime \prime}}=-1$. Therefore, $\left[d_{\alpha\beta \gamma\delta}\right]=-1 \in \mathrm{H}^3(S^1\times S^2 ; \mathbb{Z})\simeq\mathbb{Z}$\footnote{This quantity equivalent to the winding number 
$\int d\log c_{\alpha\beta\gamma}=-1$.}. 
This is consistent with the result in Sec.\ \ref{sec:example2}.

\bibliography{ref}
\end{document}